\documentclass[a4paper,11pt]{article}
\makeatletter
\RequirePackage{etoolbox}
\providebool{true}\booltrue{true}
\providebool{false}\boolfalse{false}
\newbool{paper@anon}
\newbool{paper@draft}
\newbool{paper@libertinus}
\newbool{paper@newtx}
\newbool{paper@logo}\booltrue{paper@logo}

\RequirePackage[T1]{fontenc}
\RequirePackage[utf8]{inputenc}
\ifbool{paper@libertinus}{\RequirePackage{libertinus}}{\ifbool{paper@newtx}{\RequirePackage{newtxtext}}{\RequirePackage{lmodern}}}
\RequirePackage{microtype}

\RequirePackage{amsmath,amsthm,mathtools}
\allowdisplaybreaks

\RequirePackage[dvipsnames,svgnames,table]{xcolor}
\definecolor{accent}{HTML}{000000}
\definecolor{accentdark}{HTML}{000000}
\definecolor{accentsoft}{HTML}{EFEFEF}
\RequirePackage[final]{graphicx}
\RequirePackage{xspace}
\RequirePackage{geometry}
\RequirePackage{titlesec}
\titleformat{\section}{\Large\sffamily\bfseries\color{accentdark}}{\thesection.}{0.6em}{}
\titleformat{\subsection}{\large\sffamily\bfseries\color{accentdark}}{\thesubsection.}{0.5em}{}
\titleformat{\subsubsection}{\normalsize\sffamily\bfseries}{\thesubsubsection.}{0.5em}{}
\titleformat{\paragraph}[runin]{\normalfont\normalsize\bfseries}{}{0pt}{}[.]
\titlespacing*{\paragraph}{0pt}{0.7\baselineskip}{0.6em}
\titlespacing*{\section}{0pt}{1.3\baselineskip}{0.4\baselineskip}
\titlespacing*{\subsection}{0pt}{1.0\baselineskip}{0.3\baselineskip}

\RequirePackage{enumitem}
\setlist[enumerate,1]{label={(\roman*)},leftmargin=*}
\setlist[itemize,1]{leftmargin=*}

\RequirePackage{caption,subcaption}
\RequirePackage{mdframed}
\newmdenv[hidealllines=true,backgroundcolor=blue!5,skipabove=0.6\baselineskip,
  skipbelow=0.6\baselineskip,innertopmargin=8pt,innerbottommargin=8pt,
  innerleftmargin=10pt,innerrightmargin=10pt]{highlightbox}

\RequirePackage[hyphens]{url}
\RequirePackage[final]{hyperref}
\hypersetup{
  colorlinks=true,
  linkcolor=accentdark,
  citecolor=accentdark,
  urlcolor=accent,
  linktoc=all,
  breaklinks=true,
  pdfborder={0 0 0},
}
\RequirePackage[capitalize,noabbrev]{cleveref}

\newtheoremstyle{paperplain}{0.8\baselineskip}{0.8\baselineskip}{\itshape}{}{\bfseries}{.}{0.5em}{}
\newtheoremstyle{paperdefn}{0.8\baselineskip}{0.8\baselineskip}{\normalfont}{}{\bfseries}{.}{0.5em}{}
\newtheoremstyle{paperremark}{0.6\baselineskip}{0.6\baselineskip}{\normalfont}{}{\itshape}{.}{0.5em}{}

\theoremstyle{paperplain}

\crefname{informaltheorem}{Informal statement}{Informal statements}
\Crefname{informaltheorem}{Informal statement}{Informal statements}

\theoremstyle{paperdefn}

\crefname{openquestion}{Open question}{Open questions}
\Crefname{openquestion}{Open question}{Open questions}

\theoremstyle{paperremark}
\newtheorem{remarkx}{Remark}[section]
\newenvironment{remark}{\pushQED{\qed}\remarkx}{\popQED\endremarkx}
\crefname{remarkx}{Remark}{Remarks}\Crefname{remarkx}{Remark}{Remarks}

\crefname{examplex}{Example}{Examples}\Crefname{examplex}{Example}{Examples}

\crefname{counterexamplex}{Counterexample}{Counterexamples}\Crefname{counterexamplex}{Counterexample}{Counterexamples}

\RequirePackage{fancyhdr}
\newcommand{\@logofile}{assets/logo}
\newcommand{\logofile}[1]{\renewcommand{\@logofile}{#1}}
\renewcommand{\headrule}{\color{accent}\hrule\@height\headrulewidth\@width\headwidth\vskip-\headrulewidth}

\fancypagestyle{paper@first}{\fancyhf{}\renewcommand{\headrulewidth}{0.8pt}\fancyhead[L]{\footnotesize\sffamily\color{accentdark}\@date}\fancyhead[R]{\ifbool{paper@anon}{}{{\footnotesize\sffamily\color{accentdark}\href{https://activefluidcontrol.com}{activefluidcontrol.com}}\ifbool{paper@logo}{\hspace{0.7em}\raisebox{-0.35\height}{\includegraphics[height=22pt]{\@logofile}}}{}}}\fancyfoot[R]{\footnotesize\sffamily\color{accentdark}\thepage}}
\fancypagestyle{paper@running}{\fancyhf{}\renewcommand{\headrulewidth}{0.4pt}\fancyhead[L]{\footnotesize\sffamily\color{accentdark}\nouppercase{\@title}}\fancyhead[R]{\footnotesize\sffamily\color{accentdark}\thepage}}
\newcommand{\@correspondence}{}
\newcommand{\correspondence}[1]{\renewcommand{\@correspondence}{#1}}
\newcommand{\@keywords}{}
\newcommand{\keywords}[1]{\renewcommand{\@keywords}{#1}}

\renewcommand{\maketitle}{\par\thispagestyle{paper@first}\begingroup
    \renewcommand\thefootnote{\fnsymbol{footnote}}\begin{center}{\LARGE\sffamily\bfseries\color{accentdark}\@title\par}\vskip 1.1em\ifbool{paper@anon}{{\large Anonymous Author(s)}\par\vskip 0.2em{\normalsize Affiliation withheld for review}\par}{{\large\@author\par}}\ifdefempty{\@correspondence}{}{\ifbool{paper@anon}{}{\vskip 0.6em{\small Correspondence to: \@correspondence}\par}}\@thanks
    \end{center}\endgroup
  \setcounter{footnote}{0}\global\let\@thanks\@empty
  \vskip 0.4em}
\renewenvironment{abstract}{{\centering\sffamily\bfseries\large\color{accentdark}Abstract\par}\vskip 0.1em\begin{quote}\small\noindent\ignorespaces}{\ifdefempty{\@keywords}{}{\par\vskip 0.6em\noindent{\small\textit{Keywords:} \@keywords}}\end{quote}\vskip 0.8em}

\newbool{draftmode}
\newbool{anonmode}
\ifbool{paper@anon}{\booltrue{anonmode}}{}
\ifbool{paper@draft}{\booltrue{draftmode}\setlength{\overfullrule}{4pt}}{}

\RequirePackage{booktabs}
\makeatother

\newbool{proofInMain}
\newbool{proofInAppendix}
\newbool{duplicateStatementInAppendix}

\newbool{showLinks}
\newbool{showBoxes}

\boolfalse{proofInMain}
\booltrue{proofInAppendix}
\boolfalse{duplicateStatementInAppendix}
\booltrue{showLinks}
\booltrue{showBoxes}

\usepackage{amssymb,amsfonts,bbm,dsfont,stmaryrd}
\usepackage{aligned-overset}
\usepackage{bm}
\usepackage{empheq}
\usepackage{cancel}
\usepackage{centernot}
\usepackage{physics}
\allowdisplaybreaks

\usepackage{multirow,multicol}
\usepackage{wrapfig}
\usepackage{floatrow}
\usepackage{adjustbox}
\usepackage{pdfpages}
\usepackage{xurl}

\usepackage{algorithm}
\usepackage{algpseudocode}

\usepackage[acronym,nopostdot,nogroupskip,nonumberlist]{glossaries}
\glsdisablehyper

\newacronym{acr:aepe}{AEPE}{average end-point error}
\newacronym{acr:aid}{AID}{approximate implicit differentiation}
\newacronym{acr:cnn}{CNN}{convolutional neural network}
\newacronym{acr:dro}{DRO}{distributionally robust optimization}
\newacronym{acr:epe}{EPE}{end-point error}
\newacronym{acr:ift}{IFT}{implicit function theorem}
\newacronym{acr:lhs}{LHS}{left-hand side}
\newacronym{acr:lp}{LP}{linear program}
\newacronym{acr:lsc}{lsc}{lower semicontinuous}
\newacronym{acr:mlp}{MLP}{multilayer perceptron}
\newacronym{acr:mse}{MSE}{mean squared error}
\newacronym{acr:nn}{NN}{neural network}
\newacronym{acr:ode}{ODE}{ordinary differential equation}
\newacronym{acr:of}{OF}{optical flow}
\newacronym{acr:ot}{OT}{optimal transport}
\newacronym{acr:pde}{PDE}{partial differential equation}
\newacronym{acr:piv}{PIV}{particle image velocimetry}
\newacronym{acr:qp}{QP}{quadratic program}
\newacronym{acr:raepe}{rAEPE}{relative average end-point error reduction}
\newacronym{acr:rhs}{RHS}{right-hand side}
\newacronym{acr:rl}{RL}{reinforcement learning}
\newacronym{acr:sde}{SDE}{stochastic differential equation}
\newacronym{acr:sgd}{SGD}{stochastic gradient descent}
\newacronym{acr:sota}{SOTA}{state of the art}
\newacronym{acr:usc}{usc}{upper semicontinuous}
\newacronym{acr:wlog}{WLOG}{without loss of generality}

\usepackage[authoryear,round]{natbib}

\ifbool{draftmode}{
  \usepackage[textsize=small,linecolor=red,backgroundcolor=red!20,bordercolor=red,prependcaption]{todonotes}
}{
  \usepackage[disable]{todonotes}
}
\usepackage{comment}
\usepackage{lipsum}

\usepackage[normalem]{ulem}
\definecolor{placeholdernum}{HTML}{B7352D}
\newcommand{\pnum}[1]{\textcolor{placeholdernum}{\uwave{#1}}}

\newcommand{\bibtoverify}[1][]{\ifbool{draftmode}{\textcolor{placeholdernum}{\textbf{[verify\ifstrempty{#1}{}{: #1}]}}\xspace}{}}

\ifbool{draftmode}{\hypersetup{citecolor=red}}{}

\DeclareGraphicsExtensions{.pdf,.png,.jpg}
\graphicspath{{figures/}{figures/build/}}

\let\abs\relax

\newcommand{\nrange}[1]               {\llbracket 1, #1 \rrbracket}

\newcommand{\set}[1]                  {\left\{#1\right\}}

\newcommand{\landau}[1]               {O\!\left(#1\right)}

\DeclareMathOperator*{\argmin}        {arg\,min}
\DeclareMathOperator*{\argmax}        {arg\,max}

\newcommand{\abs}[1]                  {\left\lvert #1\right\rvert}

\newcommand{\expectedValue}[2][]      {\mathbb{E}\ifstrempty{#1}{}{_{#1}}\!\left[#2\right]}

\newcommand{\indicatorOf}[1]          {\mathbbm{1}_{\left\{#1\right\}}}

\newcommand{\emphBox}[2]{\ifbool{showBoxes}{\begin{highlightbox}\textbf{#1.}\quad #2\end{highlightbox}}{#2}}

\newcommand{\externalLink}[3]{\ifbool{showLinks}{\href{#1}{\textcolor{#3}{#2}}}{#2}}

\DeclareRobustCommand{\githubPackage}[2]{\texorpdfstring{\ifbool{showLinks}{\href{https://github.com/#1}{\mbox{\texttt{#2}\hspace{0.18em}\raisebox{-0.08ex}{\includegraphics[height=1.4ex]{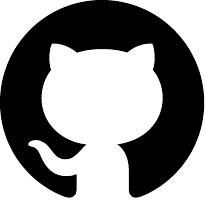}}}}}{\texttt{#2}}}{#2}}

\newif\ifmargincomments \margincommentstrue

\newcommand{\eg}{e.g.,\xspace}
\newcommand{\ie}{i.e.,\xspace}

\newcommand{\imageDomain}    {\Omega}

\newcommand{\framePrev}      {I_0}
\newcommand{\frameNext}      {I_1}
\newcommand{\imagePair}      {(\framePrev,\frameNext)}
\newcommand{\flow}           {u}

\newcommand{\flowGT}         {\flow^{\star}}

\newcommand{\library}        {\mathcal{A}}
\newcommand{\pool}           {\mathcal{P}}
\newcommand{\numEstimators}  {N}
\newcommand{\estimator}[1]   {A_{#1}}
\newcommand{\estimate}[1]    {\hat{\flow}_{#1}}
\newcommand{\speedBudget}    {B}
\newcommand{\estimatorRateUnit}{\text{image pair/s}}
\newcommand{\cache}          {\mathcal{C}}

\newcommand{\selectorParams} {\theta}
\newcommand{\selector}       {\pi_{\selectorParams}}
\newcommand{\score}          {q_{\selectorParams}}
\newcommand{\scoreTarget}    {z}
\newcommand{\selectionLabel} {y}
\newcommand{\huberLoss}      {\rho}
\newcommand{\meanPairError}  {\bar e}
\newcommand{\targetLogOffset}{\delta_{\log}}

\newcommand{\action}         {a}
\newcommand{\reward}         {r}
\newcommand{\rewardBaseline} {r_0}
\newcommand{\explore}        {\varepsilon}

\DeclareMathOperator{\epeOp} {EPE}\newcommand{\epe}{\epeOp}
\DeclareMathOperator{\aepeOp}{AEPE}\NewDocumentCommand{\aepe}{mo}{\aepeOp(#1\IfValueT{#2}{;#2})}
\newcommand{\raepe}[2]       {\mathrm{rAEPE}(#1,#2)}

\newcommand{\ourmethod}      {\text{ART of PIV}\xspace}
\DeclareRobustCommand{\flowgym}{\githubPackage{antonioterpin/flowgym}{FlowGym}\xspace}
\DeclareRobustCommand{\flowgames}{\githubPackage{antonioterpin/flowgames}{FlowGames}\xspace}
\DeclareRobustCommand{\synthpix}{\githubPackage{antonioterpin/synthpix}{SynthPix}\xspace}
\NewDocumentCommand{\goggles}{s}{\IfBooleanTF{#1}{\githubPackage{antonioterpin/goggles}{Goggles}}{\texttt{Goggles}}\xspace}
\newcommand{\dis}            {\texttt{DIS}\xspace}
\NewDocumentCommand{\dreamerVThree}{s}{\IfBooleanTF{#1}{\githubPackage{danijar/dreamerv3}{DreamerV3}}{\texttt{DreamerV3}}\xspace}
\newcommand{\raft}           {\texttt{RAFT}\xspace}
\newcommand{\raftpiv}        {\texttt{RAFT32-PIV}\xspace}
\newcommand{\lima}           {\texttt{LIMA}\xspace}

\newcommand{\openpiv}        {\texttt{OpenPIV}\xspace}
\newcommand{\jax}            {\texttt{JAX}\xspace}

\newcommand{\quietPosition}     {\boldsymbol{x}}
\newcommand{\quietVelocity}     {v}
\newcommand{\quietReference}    {v_{\mathrm{ref}}}
\newcommand{\quietBroadForce}   {f_b}
\newcommand{\quietFineForce}    {f_f}
\newcommand{\quietReferenceForce}{f_{\mathrm{ref}}}
\newcommand{\quietBroad}        {\phi_b}
\newcommand{\quietFine}         {\phi_f}
\newcommand{\quietTolerance}    {\epsilon}
\newcommand{\quietEstimatedError}{\widehat m}
\newcommand{\quietIntegral}     {I}
\newcommand{\quietUnclippedCommand}{\widetilde a}
\newcommand{\quietCost}         {\bar J}
\newcommand{\quietUsableTime} {\mathcal U}
\newcommand{\quietCostDelta}    {\Delta J}
\newcommand{\quietUsableTimeDelta}{\Delta\mathcal U}

\newcommand{\channelActuationLimit}{\bar{A}}

\newcommand{\channelEstimatedFlow}{\widehat{\boldsymbol u}}
\newcommand{\channelTargetFlow}{\boldsymbol u^{\star}}

\newcommand{\quietBrakingTimePrecise}{0.760}
\newcommand{\quietBrakingTimeRounded}{0.76}
\newcommand{\quietDiffusionRatio}{8.57}

\newcommand{\quietNoEffortCostDelta}{-12.33}
\newcommand{\quietNumericsCostDeltaMax}{-12.22}
\newcommand{\quietNumericsCostDeltaMin}{-13.23}
\newcommand{\quietObjectiveSettings}{13}
\newcommand{\quietOriginalCostDelta}{-12.46}
\newcommand{\quietOriginalCount}{30}
\newcommand{\quietOriginalUsableDelta}{+12.33}
\newcommand{\quietParamAmplitude}{2.1}
\newcommand{\quietParamBitDepth}{8}
\newcommand{\quietParamBroadBound}{0.06}
\newcommand{\quietParamCoupling}{0.3}
\newcommand{\quietParamDensity}{0.04}
\newcommand{\quietParamDerivativeTau}{0.48}
\newcommand{\quietParamDiameter}{2.5}
\newcommand{\quietParamEffortWeight}{0.01}
\newcommand{\quietParamFineAmplitude}{0.01}
\newcommand{\quietParamFineBound}{0.04}
\newcommand{\quietParamFineMode}{16}
\newcommand{\quietParamGrid}{256}
\newcommand{\quietParamHalo}{16}
\newcommand{\quietParamHold}{0.72}
\newcommand{\quietParamHorizon}{45.6}
\newcommand{\quietParamImageSize}{512}
\newcommand{\quietParamIntensity}{65}
\newcommand{\quietParamInterval}{0.12}
\newcommand{\quietParamKd}{0.02}
\newcommand{\quietParamKi}{0}
\newcommand{\quietParamKp}{0.5}
\newcommand{\quietParamQuadrature}{64}
\newcommand{\quietParamReference}{0.02}
\newcommand{\quietParamStep}{0.00125}
\newcommand{\quietParamTolerance}{0.04}
\newcommand{\quietParamViscosity}{10^{-5}}
\newcommand{\quietParamWidth}{0.3}
\newcommand{\quietPulseAmplitude}{2}
\newcommand{\quietPulseDuration}{6}
\newcommand{\quietPulseJitter}{2}
\newcommand{\quietScreenCostDelta}{-12.73}
\newcommand{\quietScreenCount}{3}
\newcommand{\quietScreenUsableDelta}{+12.54}
\newcommand{\quietSensitivityExclusions}{}
\newcommand{\quietTableLibraryRowAColA}{Specialist 0}
\newcommand{\quietTableLibraryRowAColB}{15}
\newcommand{\quietTableLibraryRowAColC}{7}
\newcommand{\quietTableLibraryRowAColD}{16}
\newcommand{\quietTableLibraryRowAColE}{4}
\newcommand{\quietTableLibraryRowAColF}{1}
\newcommand{\quietTableLibraryRowBColA}{Specialist 1}
\newcommand{\quietTableLibraryRowBColB}{9}
\newcommand{\quietTableLibraryRowBColC}{8}
\newcommand{\quietTableLibraryRowBColD}{12}
\newcommand{\quietTableLibraryRowBColE}{4}
\newcommand{\quietTableLibraryRowBColF}{0}
\newcommand{\quietTableLibraryRowCColB}{13}
\newcommand{\quietTableLibraryRowCColC}{10}
\newcommand{\quietTableLibraryRowCColD}{8}
\newcommand{\quietTableLibraryRowCColE}{5}
\newcommand{\quietTableLibraryRowCColF}{0}
\newcommand{\quietTableOutcomesRowAColB}{$+4.83$}
\newcommand{\quietTableOutcomesRowAColC}{$-6.01$}
\newcommand{\quietTableOutcomesRowAColD}{0/30}
\newcommand{\quietTableOutcomesRowBColA}{Fixed specialist 0}
\newcommand{\quietTableOutcomesRowBColB}{$+5.78$}
\newcommand{\quietTableOutcomesRowBColC}{$-6.01$}
\newcommand{\quietTableOutcomesRowBColD}{0/30}
\newcommand{\quietTableOutcomesRowCColA}{Fixed specialist 1}
\newcommand{\quietTableOutcomesRowCColB}{$+4.05$}
\newcommand{\quietTableOutcomesRowCColC}{$-3.71$}
\newcommand{\quietTableOutcomesRowCColD}{7/30}
\newcommand{\quietTableOutcomesRowDColB}{$0.00$}
\newcommand{\quietTableOutcomesRowDColC}{$0.00$}
\newcommand{\quietTableOutcomesRowDColD}{15/30}
\newcommand{\quietTableOutcomesRowEColB}{$-12.46$}
\newcommand{\quietTableOutcomesRowEColC}{$+12.33$}
\newcommand{\quietTableOutcomesRowEColD}{29/30}
\newcommand{\quietTableOutcomesRowFColB}{$-13.98$}
\newcommand{\quietTableOutcomesRowFColC}{$+13.81$}
\newcommand{\quietTableOutcomesRowFColD}{30/30}
\newcommand{\quietTableOutcomesRowGColB}{$-16.20$}
\newcommand{\quietTableOutcomesRowGColC}{$+15.91$}
\newcommand{\quietTableOutcomesRowGColD}{30/30}
\newcommand{\quietTableSensitivityControllerRowAColA}{$K_p$ ($\mathrm{s}^{-1}$)}
\newcommand{\quietTableSensitivityControllerRowAColB}{$0.5$}
\newcommand{\quietTableSensitivityControllerRowAColC}{$0.25,\,1$}
\newcommand{\quietTableSensitivityControllerRowAColD}{$[-15.14, -1.88]$}
\newcommand{\quietTableSensitivityControllerRowAColE}{$[+1.71, +14.89]$}
\newcommand{\quietTableSensitivityControllerRowBColA}{$K_i$ ($\mathrm{s}^{-2}$)}
\newcommand{\quietTableSensitivityControllerRowBColB}{$0$}
\newcommand{\quietTableSensitivityControllerRowBColC}{$0.005,\,0.02$}
\newcommand{\quietTableSensitivityControllerRowBColD}{$[-15.13, -12.73]$}
\newcommand{\quietTableSensitivityControllerRowBColE}{$[+12.54, +14.88]$}
\newcommand{\quietTableSensitivityControllerRowCColB}{$0.02$}
\newcommand{\quietTableSensitivityControllerRowCColC}{$0,\,0.04$}
\newcommand{\quietTableSensitivityControllerRowCColD}{$[-13.12, -11.86]$}
\newcommand{\quietTableSensitivityControllerRowCColE}{$[+11.67, +12.90]$}
\newcommand{\quietTableSensitivityControllerRowDColB}{$0.48$}
\newcommand{\quietTableSensitivityControllerRowDColC}{$0,\,0.96$}
\newcommand{\quietTableSensitivityControllerRowDColD}{$[-13.00, -10.41]$}
\newcommand{\quietTableSensitivityControllerRowDColE}{$[+10.21, +12.78]$}
\newcommand{\quietTableSensitivityControllerRowEColA}{$b_b$ ($\ell/\mathrm{s}^2$)}
\newcommand{\quietTableSensitivityControllerRowEColB}{$0.06$}
\newcommand{\quietTableSensitivityControllerRowEColC}{$0.045,\,0.075$}
\newcommand{\quietTableSensitivityControllerRowEColD}{$[-20.74, -0.10]$}
\newcommand{\quietTableSensitivityControllerRowEColE}{$[0.00, +20.50]$}
\newcommand{\quietTableSensitivityControllerRowFColA}{$b_f$ ($\ell/\mathrm{s}^2$)}
\newcommand{\quietTableSensitivityControllerRowFColB}{$0.04$}
\newcommand{\quietTableSensitivityControllerRowFColC}{$0.024,\,0.06$}
\newcommand{\quietTableSensitivityControllerRowFColD}{$[-12.73, -5.47]$}
\newcommand{\quietTableSensitivityControllerRowFColE}{$[+5.25, +12.54]$}
\newcommand{\quietTableSensitivityNumericsRowAColB}{$256$}
\newcommand{\quietTableSensitivityNumericsRowAColC}{$128,\,512$}
\newcommand{\quietTableSensitivityNumericsRowAColD}{$[-13.23, -12.22]$}
\newcommand{\quietTableSensitivityNumericsRowAColE}{$[+12.02, +13.01]$}
\newcommand{\quietTableSensitivityNumericsRowBColB}{$0.00125$}
\newcommand{\quietTableSensitivityNumericsRowBColC}{$0.000625,\,0.0025$}
\newcommand{\quietTableSensitivityNumericsRowBColD}{$[-12.86, -12.66]$}
\newcommand{\quietTableSensitivityNumericsRowBColE}{$[+12.46, +12.66]$}
\newcommand{\quietTableSensitivityNumericsRowCColB}{$45.6$}
\newcommand{\quietTableSensitivityNumericsRowCColC}{$57.6$}
\newcommand{\quietTableSensitivityNumericsRowCColD}{$[-16.79, -12.73]$}
\newcommand{\quietTableSensitivityNumericsRowCColE}{$[+12.54, +16.59]$}
\newcommand{\quietTableSensitivityNumericsRowDColB}{$16$}
\newcommand{\quietTableSensitivityNumericsRowDColC}{$8,\,32$}
\newcommand{\quietTableSensitivityNumericsRowDColD}{$[-12.95, -12.73]$}
\newcommand{\quietTableSensitivityNumericsRowDColE}{$[+12.54, +12.74]$}
\newcommand{\quietTableSensitivityNumericsRowEColC}{$1/3,\,2/3$}
\newcommand{\quietTableSensitivityNumericsRowEColD}{$[-12.73, -9.36]$}
\newcommand{\quietTableSensitivityNumericsRowEColE}{$[+9.18, +12.54]$}
\newcommand{\quietTableSensitivityObjectiveRowAColB}{$0.04$}
\newcommand{\quietTableSensitivityObjectiveRowAColC}{$0.02,\,0.03,\,0.06,\,0.08$}
\newcommand{\quietTableSensitivityObjectiveRowAColD}{$[-13.18, -10.68]$}
\newcommand{\quietTableSensitivityObjectiveRowAColE}{$[+10.55, +13.05]$}
\newcommand{\quietTableSensitivityObjectiveRowBColB}{$0.01$}
\newcommand{\quietTableSensitivityObjectiveRowBColC}{$0,\,0.001,\,0.1$}
\newcommand{\quietTableSensitivityObjectiveRowBColD}{$[-13.64, -12.33]$}
\newcommand{\quietTableSensitivityObjectiveRowBColE}{$+12.33$}
\newcommand{\quietTableSensitivityObjectiveRowCColB}{$0.72$}
\newcommand{\quietTableSensitivityObjectiveRowCColC}{$0.36,\,1.44,\,2.88$}
\newcommand{\quietTableSensitivityObjectiveRowCColD}{$-12.46$}
\newcommand{\quietTableSensitivityObjectiveRowCColE}{$+12.33$}
\newcommand{\quietTableSensitivityObjectiveRowDColB}{$45.6$}
\newcommand{\quietTableSensitivityObjectiveRowDColC}{$22.8,\,34.2$}
\newcommand{\quietTableSensitivityObjectiveRowDColD}{$[-12.46, -0.02]$}
\newcommand{\quietTableSensitivityObjectiveRowDColE}{$[0.00, +12.33]$}
\newcommand{\quietTableSensitivityObservationRowAColB}{$64$}
\newcommand{\quietTableSensitivityObservationRowAColC}{$32,\,128$}
\newcommand{\quietTableSensitivityObservationRowAColD}{$[-12.73, -11.88]$}
\newcommand{\quietTableSensitivityObservationRowAColE}{$[+11.68, +12.54]$}
\newcommand{\quietTableSensitivityObservationRowBColB}{$\pi/2$}
\newcommand{\quietTableSensitivityObservationRowBColC}{$\pi/8$}
\newcommand{\quietTableSensitivityObservationRowBColD}{$[-12.73, -11.57]$}
\newcommand{\quietTableSensitivityObservationRowBColE}{$[+11.50, +12.54]$}
\newcommand{\quietTableSensitivityObservationRowCColB}{$0.12$}
\newcommand{\quietTableSensitivityObservationRowCColC}{$0.06,\,0.24$}
\newcommand{\quietTableSensitivityObservationRowCColD}{$[-13.79, -0.06]$}
\newcommand{\quietTableSensitivityObservationRowCColE}{$[0.00, +13.45]$}
\newcommand{\quietTableSensitivityObservationRowDColB}{$0.04$}
\newcommand{\quietTableSensitivityObservationRowDColC}{$0.02,\,0.08$}
\newcommand{\quietTableSensitivityObservationRowDColD}{$[-12.73, -11.89]$}
\newcommand{\quietTableSensitivityObservationRowDColE}{$[+11.72, +12.54]$}
\newcommand{\quietTableSensitivityObservationRowEColB}{$2.5$}
\newcommand{\quietTableSensitivityObservationRowEColC}{$1.5,\,3.5$}
\newcommand{\quietTableSensitivityObservationRowEColD}{$[-12.73, -0.80]$}
\newcommand{\quietTableSensitivityObservationRowEColE}{$[+0.62, +12.54]$}
\newcommand{\quietTableSensitivityObservationRowFColB}{$65$}
\newcommand{\quietTableSensitivityObservationRowFColC}{$32.5,\,130$}
\newcommand{\quietTableSensitivityObservationRowFColD}{$[-13.19, -11.53]$}
\newcommand{\quietTableSensitivityObservationRowFColE}{$[+11.35, +12.98]$}
\newcommand{\quietTableSensitivityObservationRowGColB}{$512$}
\newcommand{\quietTableSensitivityObservationRowGColC}{$256,\,384$}
\newcommand{\quietTableSensitivityObservationRowGColD}{$[-12.73, -0.16]$}
\newcommand{\quietTableSensitivityObservationRowGColE}{$[0.00, +12.54]$}
\newcommand{\quietTableSensitivityObservationRowHColB}{$64$}
\newcommand{\quietTableSensitivityObservationRowHColC}{$24,\,32,\,128$}
\newcommand{\quietTableSensitivityObservationRowHColD}{$[-12.75, -12.33]$}
\newcommand{\quietTableSensitivityObservationRowHColE}{$[+12.13, +12.55]$}
\newcommand{\quietTableSensitivityPhysicsRowAColB}{$0.3$}
\newcommand{\quietTableSensitivityPhysicsRowAColC}{$0,\,0.15,\,0.45$}
\newcommand{\quietTableSensitivityPhysicsRowAColD}{$[-14.15, -2.04]$}
\newcommand{\quietTableSensitivityPhysicsRowAColE}{$[+1.87, +13.82]$}
\newcommand{\quietTableSensitivityPhysicsRowBColB}{$16$}
\newcommand{\quietTableSensitivityPhysicsRowBColC}{$8,\,24$}
\newcommand{\quietTableSensitivityPhysicsRowBColD}{$[-12.73, -8.42]$}
\newcommand{\quietTableSensitivityPhysicsRowBColE}{$[+8.21, +12.54]$}
\newcommand{\quietTableSensitivityPhysicsRowCColB}{$0.3$}
\newcommand{\quietTableSensitivityPhysicsRowCColC}{$0.15,\,0.6$}
\newcommand{\quietTableSensitivityPhysicsRowCColD}{$[-12.73, -7.45]$}
\newcommand{\quietTableSensitivityPhysicsRowCColE}{$[+7.22, +12.54]$}
\newcommand{\quietTableSensitivityPhysicsRowDColA}{$\nu$ ($\ell^2/\mathrm{s}$)}
\newcommand{\quietTableSensitivityPhysicsRowDColB}{$10^{-5}$}
\newcommand{\quietTableSensitivityPhysicsRowDColC}{$5\times10^{-6},\,2\times10^{-5}$}
\newcommand{\quietTableSensitivityPhysicsRowDColD}{$[-12.73, -12.37]$}
\newcommand{\quietTableSensitivityPhysicsRowDColE}{$[+12.16, +12.54]$}
\newcommand{\quietTableSensitivityPhysicsRowEColB}{$2.1$}
\newcommand{\quietTableSensitivityPhysicsRowEColC}{$1.575,\,2.625$}
\newcommand{\quietTableSensitivityPhysicsRowEColD}{$[-13.08, -0.23]$}
\newcommand{\quietTableSensitivityPhysicsRowEColE}{$[0.00, +12.93]$}
\newcommand{\quietTableSensitivityPhysicsRowFColB}{$0.01$}
\newcommand{\quietTableSensitivityPhysicsRowFColC}{$0,\,0.05$}
\newcommand{\quietTableSensitivityPhysicsRowFColD}{$[-12.73, -12.27]$}
\newcommand{\quietTableSensitivityPhysicsRowFColE}{$[+12.08, +12.54]$}
\newcommand{\quietTableSensitivityPhysicsRowGColB}{$0.02$}
\newcommand{\quietTableSensitivityPhysicsRowGColC}{$0,\,0.04$}
\newcommand{\quietTableSensitivityPhysicsRowGColD}{$[-13.76, -11.97]$}
\newcommand{\quietTableSensitivityPhysicsRowGColE}{$[+11.78, +13.53]$}
\newcommand{\quietTableSensitivityPhysicsRowHColB}{$1$}
\newcommand{\quietTableSensitivityPhysicsRowHColC}{$0,\,2$}
\newcommand{\quietTableSensitivityPhysicsRowHColD}{$[-13.77, -5.16]$}
\newcommand{\quietTableSensitivityPhysicsRowHColE}{$[+5.02, +13.63]$}
\newcommand{\quietWindowCenterY}{\pi/2}
\newcommand{\quietWindowCentersX}{\pi/2,\pi,3\pi/2}

\newcommand{\quietWindowSide}{\pi\ell/4}

\newcommand{\quietArtAmplitudeVariationPercent}{15}
\newcommand{\quietArtBatch}{16}
\newcommand{\quietArtChannels}{16,32}
\newcommand{\quietArtConsistentEpisodeCount}{30}
\newcommand{\quietArtCostDeltaMax}{-11.75}
\newcommand{\quietArtCostDeltaMin}{-12.06}
\newcommand{\quietArtEmbedding}{32}
\newcommand{\quietArtFixedOneCostDelta}{+3.52}
\newcommand{\quietArtFixedZeroCostDelta}{+7.16}
\newcommand{\quietArtHidden}{64}
\newcommand{\quietArtHoldsMin}{30}
\newcommand{\quietArtLearningRate}{0.001}

\newcommand{\quietArtOracleCostDelta}{-12.54}

\newcommand{\quietArtReferenceHolds}{18}
\newcommand{\quietArtSeedCount}{3}
\newcommand{\quietArtSetupChoice}{0}
\newcommand{\quietArtSetupTime}{0.52}
\newcommand{\quietArtSteps}{5,000}
\newcommand{\quietArtStride}{4}
\newcommand{\quietArtTableOutcomesRowAColA}{Fixed specialist 0}
\newcommand{\quietArtTableOutcomesRowAColB}{$+7.16$}
\newcommand{\quietArtTableOutcomesRowAColC}{$-7.38$}
\newcommand{\quietArtTableOutcomesRowAColD}{0/30}
\newcommand{\quietArtTableOutcomesRowBColA}{Fixed specialist 1}
\newcommand{\quietArtTableOutcomesRowBColB}{$+3.52$}
\newcommand{\quietArtTableOutcomesRowBColC}{$-3.24$}
\newcommand{\quietArtTableOutcomesRowBColD}{11/30}
\newcommand{\quietArtTableOutcomesRowCColB}{$0.00$}
\newcommand{\quietArtTableOutcomesRowCColC}{$0.00$}
\newcommand{\quietArtTableOutcomesRowCColD}{18/30}
\newcommand{\quietArtTableOutcomesRowDColB}{$[-12.06, -11.75]$}
\newcommand{\quietArtTableOutcomesRowDColC}{$[+11.62, +11.92]$}
\newcommand{\quietArtTableOutcomesRowDColD}{30/30}
\newcommand{\quietArtTableOutcomesRowEColB}{$-12.54$}
\newcommand{\quietArtTableOutcomesRowEColC}{$+12.41$}
\newcommand{\quietArtTableOutcomesRowEColD}{30/30}
\newcommand{\quietArtTableOutcomesRowFColB}{$-16.08$}
\newcommand{\quietArtTableOutcomesRowFColC}{$+15.79$}
\newcommand{\quietArtTableOutcomesRowFColD}{30/30}
\newcommand{\quietArtTeacherCount}{5}
\newcommand{\quietArtTestCount}{30}
\newcommand{\quietArtTrainCount}{12}
\newcommand{\quietArtTrainPairs}{22,800}
\newcommand{\quietArtUsableDeltaMax}{+11.92}
\newcommand{\quietArtUsableDeltaMin}{+11.62}
\newcommand{\quietArtValidationCount}{6}
\newcommand{\quietArtValidationPairs}{11,400}
\newcommand{\quietArtViscosityVariationPercent}{10}

\RequirePackage{tikz}
\RequirePackage{pgfplots}
\RequirePackage{pgfplotstable}
\pgfplotsset{compat=1.18}
\usetikzlibrary{
  calc, intersections, through, positioning, fit, backgrounds,
  arrows.meta, decorations.pathreplacing, decorations.markings,
  patterns, patterns.meta, shapes.geometric, shapes.misc, shapes.arrows,
  hobby,
}
\usepgfplotslibrary{groupplots, fillbetween, statistics, colormaps, dateplot}

\newcommand{\figureFontScale}{1}
\newcommand{\figureFont}[2]{\pgfmathsetmacro{\figureFontSize}{#1*\figureFontScale}\pgfmathsetmacro{\figureFontBaseline}{#2*\figureFontScale}\fontsize{\figureFontSize}{\figureFontBaseline}\selectfont}
\newcommand{\figureTitleFont}{\figureFont{9}{11}}
\newcommand{\figureLabelFont}{\figureFont{9}{11}}
\newcommand{\figureTextFont}{\figureFont{8}{10}}

\definecolor{ink}{HTML}{1A1A1A}
\definecolor{paperbg}{HTML}{FFFFFF}
\definecolor{seriesA}{HTML}{215CAF}
\definecolor{seriesB}{HTML}{B7352D}
\definecolor{seriesC}{HTML}{627313}
\definecolor{seriesD}{HTML}{8E6713}
\definecolor{seriesE}{HTML}{A3076C}
\definecolor{seriesF}{HTML}{007894}
\definecolor{seriesG}{HTML}{6F6F6F}
\colorlet{seriesAlight}{seriesA!22}
\colorlet{seriesBlight}{seriesB!22}
\colorlet{seriesClight}{seriesC!22}
\colorlet{seriesDlight}{seriesD!22}
\colorlet{seriesElight}{seriesE!22}
\colorlet{seriesFlight}{seriesF!22}
\pgfplotscreateplotcyclelist{paper}{
  {seriesA, mark=*},      {seriesB, mark=square*},  {seriesC, mark=triangle*},
  {seriesD, mark=diamond*},{seriesE, mark=pentagon*},{seriesF, mark=oplus*},
  {seriesG, mark=x},
}

\pgfplotsset{
  every axis/.append style={
    line width=0.7pt, tick style={line width=0.6pt, color=ink},
    axis line style={color=ink}, label style={font=\figureLabelFont},
    tick label style={font=\figureTextFont}, title style={font=\figureTitleFont},
    legend cell align=left, legend style={font=\figureTextFont, draw=none, fill=none},
    grid=major, grid style={dashed, color=ink!12},
  },
  cycle list name=paper,
  /pgf/number format/.cd, dec sep={.}, 1000 sep={\,},
}

\pgfplotsset{
  legend on top/.style={
    legend cell align=left,
    legend columns=-1,
    legend style={
      font=\figureTextFont, draw=ink!30, fill=white,
      at={(0.5,1.04)}, anchor=south,
      /tikz/every even column/.append style={column sep=6pt},
    },
  },
}

\tikzset{
  every picture/.append style={font=\figureTextFont},
  >=Latex,
  node/.style={circle, draw=ink, fill=white, inner sep=1.6pt, minimum size=5pt},
  point/.style={circle, fill=seriesA, inner sep=0pt, minimum size=4pt},
  arrow/.style={-{Latex[length=2.2mm]}, line width=0.7pt},
  thinarrow/.style={-{Latex[length=1.6mm]}, line width=0.5pt, color=ink!70},
  box/.style={rounded corners=2pt, draw=ink, fill=accentsoft, inner sep=5pt},
  emph/.style={color=seriesB},
  annot/.style={font=\figureTextFont, color=ink!75},
  flow/.style={->, decoration={markings, mark=at position 0.5 with {\arrow{Latex}}}, postaction={decorate}},
}

\pgfmathdeclarefunction{gauss}{2}{\pgfmathparse{1/(#2*sqrt(2*pi))*exp(-((x-#1)^2)/(2*#2^2))}}

\definecolor{ours}{HTML}{215CAF}
\colorlet{ourslight}{ours!20}
\definecolor{baselineOne}{HTML}{B7352D}
\definecolor{baselineTwo}{HTML}{627313}
\definecolor{baselineThree}{HTML}{8E6713}

\definecolor{famCorrelation}{HTML}{8E6713}
\definecolor{famOpticalFlow}{HTML}{627313}
\definecolor{famNeural}{HTML}{A3076C}
\colorlet{famCorrelationlight}{famCorrelation!20}
\colorlet{famOpticalFlowlight}{famOpticalFlow!20}
\colorlet{famNeurallight}{famNeural!20}

\definecolor{positive}{HTML}{215CAF}
\definecolor{negative}{HTML}{B7352D}
\definecolor{neutral}{HTML}{6F6F6F}

\definecolor{quietOracle}{HTML}{007F82}
\definecolor{quietFixed}{HTML}{B7352D}
\definecolor{quietGroundTruth}{HTML}{303840}
\definecolor{quietZero}{HTML}{8E6713}
\definecolor{quietOne}{HTML}{8663A8}

\definecolor{pinballRegulationBg}{HTML}{E8EDF5}
\definecolor{pinballAcquisitionBg}{HTML}{DDF0FF}
\definecolor{pinballEstimationBg}{HTML}{FFF0D0}
\definecolor{pinballControlBg}{HTML}{DDF5E8}
\definecolor{pinballActuationBg}{HTML}{EBE1FF}
\definecolor{pinballTrackingError}{HTML}{B35A16}

\newcommand{\disTwentyHzArtofpivAepe}{0.1268}

\newcommand{\disTwentyHzBestSingleAepe}{0.1498}

\newcommand{\disTwentyHzRaepePct}{15.3}

\newcommand{\disTwentyHzNominalRateValue}{20}
\newcommand{\disTwentyHzNominalRate}{20\,\estimatorRateUnit}
\newcommand{\disOneKHzArtofpivAepe}{0.1385}

\newcommand{\disOneKHzBestSingleAepe}{0.1577}

\newcommand{\disOneKHzRaepePct}{12.2}

\newcommand{\disOneKHzNominalRateValue}{1000}

\newcommand{\openpivOneKHzArtofpivAepe}{0.1979}

\newcommand{\openpivOneKHzBestSingleAepe}{0.2544}

\newcommand{\openpivOneKHzRaepePct}{22.2}

\newcommand{\openpivOneKHzNominalRateValue}{1000}
\newcommand{\openpivOneKHzNominalRate}{1000\,\estimatorRateUnit}
\newcommand{\openpivFourKHzArtofpivAepe}{0.1974}

\newcommand{\openpivFourKHzBestSingleAepe}{0.2544}

\newcommand{\openpivFourKHzRaepePct}{22.4}

\newcommand{\openpivFourKHzNominalRateValue}{4000}

\newcommand{\raftThirtyTwoArtofpivAepe}{0.0096}

\newcommand{\raftThirtyTwoBestSingleAepe}{0.0120}

\newcommand{\raftThirtyTwoRaepePct}{20.3}

\newcommand{\raftThirtyTwoNominalRateValue}{1.7}
\newcommand{\raftThirtyTwoNominalRate}{1.7\,\estimatorRateUnit}
\newcommand{\archDisOneKHzArtofpivAepe}{0.1279}

\newcommand{\archDisOneKHzRaepePct}{15.1}
\newcommand{\archDisOneKHzLargeArtofpivAepe}{0.1272}

\newcommand{\archDisOneKHzLargeRaepePct}{15.6}
\newcommand{\archOpenpivOneKHzArtofpivAepe}{0.1827}

\newcommand{\archOpenpivOneKHzRaepePct}{21.0}
\newcommand{\archOpenpivOneKHzLargeArtofpivAepe}{0.1823}

\newcommand{\archOpenpivOneKHzLargeRaepePct}{21.2}
\newcommand{\archOpenpivFourKHzArtofpivAepe}{0.1828}

\newcommand{\archOpenpivFourKHzRaepePct}{21.0}
\newcommand{\archOpenpivFourKHzLargeArtofpivAepe}{0.1824}

\newcommand{\archOpenpivFourKHzLargeRaepePct}{21.2}
\newcommand{\agentOverheadDisDeployedMs}{0.16}
\newcommand{\agentOverheadDisLargeMs}{0.27}
\newcommand{\agentOverheadOpenpivDeployedMs}{0.11}
\newcommand{\agentOverheadOpenpivLargeMs}{0.21}
\newcommand{\librarySelectionGreedyGapTwoPct}{6.7}
\newcommand{\librarySelectionGreedyGapFromThreeMaxPct}{0.4}
\newcommand{\librarySelectionMilpMinS}{6}
\newcommand{\librarySelectionMilpMaxS}{17}
\newcommand{\epeStatsDisTwentyHzRouterMeanPx}{0.1268}
\newcommand{\epeStatsDisTwentyHzRouterStdPx}{0.0967}
\newcommand{\epeStatsDisTwentyHzRouterMedianPx}{0.0944}
\newcommand{\epeStatsDisTwentyHzRouterNinetyPx}{0.2384}
\newcommand{\epeStatsDisTwentyHzRouterNinetyFivePx}{0.3210}
\newcommand{\epeStatsDisTwentyHzRouterNinetyNinePx}{0.5130}
\newcommand{\epeStatsDisTwentyHzRouterMaxPx}{0.6733}
\newcommand{\epeStatsDisTwentyHzBestMeanPx}{0.1498}
\newcommand{\epeStatsDisTwentyHzBestStdPx}{0.1270}
\newcommand{\epeStatsDisTwentyHzBestMedianPx}{0.1055}
\newcommand{\epeStatsDisTwentyHzBestNinetyPx}{0.2767}
\newcommand{\epeStatsDisTwentyHzBestNinetyFivePx}{0.3793}
\newcommand{\epeStatsDisTwentyHzBestNinetyNinePx}{0.6275}
\newcommand{\epeStatsDisTwentyHzBestMaxPx}{1.4324}
\newcommand{\epeStatsDisTwentyHzBetterPct}{68.9}
\newcommand{\epeStatsDisTwentyHzTiePct}{13.7}
\newcommand{\epeStatsDisTwentyHzWorsePct}{17.4}
\newcommand{\epeStatsDisOneKHzRouterMeanPx}{0.1385}
\newcommand{\epeStatsDisOneKHzRouterStdPx}{0.1058}
\newcommand{\epeStatsDisOneKHzRouterMedianPx}{0.1023}
\newcommand{\epeStatsDisOneKHzRouterNinetyPx}{0.2648}
\newcommand{\epeStatsDisOneKHzRouterNinetyFivePx}{0.3499}
\newcommand{\epeStatsDisOneKHzRouterNinetyNinePx}{0.5973}
\newcommand{\epeStatsDisOneKHzRouterMaxPx}{0.7854}
\newcommand{\epeStatsDisOneKHzBestMeanPx}{0.1577}
\newcommand{\epeStatsDisOneKHzBestStdPx}{0.1219}
\newcommand{\epeStatsDisOneKHzBestMedianPx}{0.1133}
\newcommand{\epeStatsDisOneKHzBestNinetyPx}{0.3016}
\newcommand{\epeStatsDisOneKHzBestNinetyFivePx}{0.3882}
\newcommand{\epeStatsDisOneKHzBestNinetyNinePx}{0.6307}
\newcommand{\epeStatsDisOneKHzBestMaxPx}{1.0053}
\newcommand{\epeStatsDisOneKHzBetterPct}{78.5}
\newcommand{\epeStatsDisOneKHzTiePct}{0.2}
\newcommand{\epeStatsDisOneKHzWorsePct}{21.3}
\newcommand{\epeStatsOpenpivOneKHzRouterMeanPx}{0.1979}
\newcommand{\epeStatsOpenpivOneKHzRouterStdPx}{0.1575}
\newcommand{\epeStatsOpenpivOneKHzRouterMedianPx}{0.1423}
\newcommand{\epeStatsOpenpivOneKHzRouterNinetyPx}{0.3910}
\newcommand{\epeStatsOpenpivOneKHzRouterNinetyFivePx}{0.4905}
\newcommand{\epeStatsOpenpivOneKHzRouterNinetyNinePx}{0.8317}
\newcommand{\epeStatsOpenpivOneKHzRouterMaxPx}{1.0057}
\newcommand{\epeStatsOpenpivOneKHzBestMeanPx}{0.2544}
\newcommand{\epeStatsOpenpivOneKHzBestStdPx}{0.2559}
\newcommand{\epeStatsOpenpivOneKHzBestMedianPx}{0.1656}
\newcommand{\epeStatsOpenpivOneKHzBestNinetyPx}{0.4993}
\newcommand{\epeStatsOpenpivOneKHzBestNinetyFivePx}{0.8557}
\newcommand{\epeStatsOpenpivOneKHzBestNinetyNinePx}{1.4207}
\newcommand{\epeStatsOpenpivOneKHzBestMaxPx}{1.9545}
\newcommand{\epeStatsOpenpivOneKHzBetterPct}{65.6}
\newcommand{\epeStatsOpenpivOneKHzTiePct}{21.9}
\newcommand{\epeStatsOpenpivOneKHzWorsePct}{12.4}
\newcommand{\epeStatsOpenpivFourKHzRouterMeanPx}{0.1974}
\newcommand{\epeStatsOpenpivFourKHzRouterStdPx}{0.1558}
\newcommand{\epeStatsOpenpivFourKHzRouterMedianPx}{0.1428}
\newcommand{\epeStatsOpenpivFourKHzRouterNinetyPx}{0.3899}
\newcommand{\epeStatsOpenpivFourKHzRouterNinetyFivePx}{0.4882}
\newcommand{\epeStatsOpenpivFourKHzRouterNinetyNinePx}{0.8223}
\newcommand{\epeStatsOpenpivFourKHzRouterMaxPx}{0.9795}
\newcommand{\epeStatsOpenpivFourKHzBestMeanPx}{0.2544}
\newcommand{\epeStatsOpenpivFourKHzBestStdPx}{0.2559}
\newcommand{\epeStatsOpenpivFourKHzBestMedianPx}{0.1656}
\newcommand{\epeStatsOpenpivFourKHzBestNinetyPx}{0.4993}
\newcommand{\epeStatsOpenpivFourKHzBestNinetyFivePx}{0.8557}
\newcommand{\epeStatsOpenpivFourKHzBestNinetyNinePx}{1.4207}
\newcommand{\epeStatsOpenpivFourKHzBestMaxPx}{1.9545}
\newcommand{\epeStatsOpenpivFourKHzBetterPct}{61.5}
\newcommand{\epeStatsOpenpivFourKHzTiePct}{27.3}
\newcommand{\epeStatsOpenpivFourKHzWorsePct}{11.2}
\newcommand{\epeStatsRaftThirtyTwoRouterMeanPx}{0.0096}
\newcommand{\epeStatsRaftThirtyTwoRouterStdPx}{0.0073}
\newcommand{\epeStatsRaftThirtyTwoRouterMedianPx}{0.0077}
\newcommand{\epeStatsRaftThirtyTwoRouterNinetyPx}{0.0179}
\newcommand{\epeStatsRaftThirtyTwoRouterNinetyFivePx}{0.0227}
\newcommand{\epeStatsRaftThirtyTwoRouterNinetyNinePx}{0.0365}
\newcommand{\epeStatsRaftThirtyTwoRouterMaxPx}{0.0424}
\newcommand{\epeStatsRaftThirtyTwoBestMeanPx}{0.0120}
\newcommand{\epeStatsRaftThirtyTwoBestStdPx}{0.0090}
\newcommand{\epeStatsRaftThirtyTwoBestMedianPx}{0.0103}
\newcommand{\epeStatsRaftThirtyTwoBestNinetyPx}{0.0200}
\newcommand{\epeStatsRaftThirtyTwoBestNinetyFivePx}{0.0270}
\newcommand{\epeStatsRaftThirtyTwoBestNinetyNinePx}{0.0451}
\newcommand{\epeStatsRaftThirtyTwoBestMaxPx}{0.0790}
\newcommand{\epeStatsRaftThirtyTwoBetterPct}{98.1}
\newcommand{\epeStatsRaftThirtyTwoTiePct}{0.0}
\newcommand{\epeStatsRaftThirtyTwoWorsePct}{1.9}

\newcommand{\mainRaepePctRange}{12.2--22.4}
\newcommand{\mainMinNominalRate}{1.7\,\estimatorRateUnit}
\newcommand{\mainMaxNominalRate}{4000\,\estimatorRateUnit}
\newcommand{\mainHighRateRange}{1000--4000}
\newcommand{\ablationBaselineAepe}{0.1370}

\newcommand{\ablationClassificationRefAepe}{0.1076}

\newcommand{\ablationOracleRefAepe}{0.1034}

\newcommand{\ablationSignalRowACount}{1}
\newcommand{\ablationSignalRowAOracle}{0.1370}
\newcommand{\ablationSignalRowBCount}{2}
\newcommand{\ablationSignalRowBOracle}{0.1196}
\newcommand{\ablationSignalRowBClassification}{0.1208}
\newcommand{\ablationSignalRowBRegression}{0.1211}
\newcommand{\ablationSignalRowBBandits}{0.1219}
\newcommand{\ablationSignalRowCCount}{3}
\newcommand{\ablationSignalRowCOracle}{0.1073}
\newcommand{\ablationSignalRowCClassification}{0.1086}
\newcommand{\ablationSignalRowCRegression}{0.1090}
\newcommand{\ablationSignalRowCBandits}{0.1096}
\newcommand{\ablationSignalRowDCount}{4}
\newcommand{\ablationSignalRowDOracle}{0.1064}
\newcommand{\ablationSignalRowDClassification}{0.1081}
\newcommand{\ablationSignalRowDRegression}{0.1081}
\newcommand{\ablationSignalRowDBandits}{0.1101}
\newcommand{\ablationSignalRowECount}{5}
\newcommand{\ablationSignalRowEOracle}{0.1056}
\newcommand{\ablationSignalRowEClassification}{0.1079}
\newcommand{\ablationSignalRowERegression}{0.1080}
\newcommand{\ablationSignalRowEBandits}{0.1102}
\newcommand{\ablationSignalRowFCount}{10}
\newcommand{\ablationSignalRowFOracle}{0.1041}
\newcommand{\ablationSignalRowFClassification}{0.1075}
\newcommand{\ablationSignalRowFRegression}{0.1068}
\newcommand{\ablationSignalRowFBandits}{0.1101}
\newcommand{\ablationSignalRowGCount}{15}
\newcommand{\ablationSignalRowGOracle}{0.1036}
\newcommand{\ablationSignalRowGClassification}{0.1076}
\newcommand{\ablationSignalRowGRegression}{0.1066}
\newcommand{\ablationSignalRowGBandits}{0.1097}
\newcommand{\ablationSignalRowHCount}{20}
\newcommand{\ablationSignalRowHOracle}{0.1034}
\newcommand{\ablationSignalRowHClassification}{0.1076}
\newcommand{\ablationSignalRowHRegression}{0.1067}
\newcommand{\ablationSignalRowHBandits}{0.1106}
\newcommand{\ablationSignalRows}{
\ablationSignalRowACount & \ablationBaselineAepe & \ablationSignalRowAOracle & --- & --- & --- \\
\ablationSignalRowBCount & \ablationBaselineAepe & \ablationSignalRowBOracle & \textbf{\ablationSignalRowBClassification} & \ablationSignalRowBRegression & \ablationSignalRowBBandits \\
\ablationSignalRowCCount & \ablationBaselineAepe & \ablationSignalRowCOracle & \textbf{\ablationSignalRowCClassification} & \ablationSignalRowCRegression & \ablationSignalRowCBandits \\
\ablationSignalRowDCount & \ablationBaselineAepe & \ablationSignalRowDOracle & \textbf{\ablationSignalRowDClassification} & \ablationSignalRowDRegression & \ablationSignalRowDBandits \\
\ablationSignalRowECount & \ablationBaselineAepe & \ablationSignalRowEOracle & \textbf{\ablationSignalRowEClassification} & \ablationSignalRowERegression & \ablationSignalRowEBandits \\
\ablationSignalRowFCount & \ablationBaselineAepe & \ablationSignalRowFOracle & \ablationSignalRowFClassification & \textbf{\ablationSignalRowFRegression} & \ablationSignalRowFBandits \\
\ablationSignalRowGCount & \ablationBaselineAepe & \ablationSignalRowGOracle & \ablationSignalRowGClassification & \textbf{\ablationSignalRowGRegression} & \ablationSignalRowGBandits \\
\ablationSignalRowHCount & \ablationBaselineAepe & \ablationSignalRowHOracle & \ablationSignalRowHClassification & \textbf{\ablationSignalRowHRegression} & \ablationSignalRowHBandits \\
}
\newcommand{\ablationRegressionRawAepe}{0.1084}
\newcommand{\ablationRegressionMeanAepe}{0.1072}
\newcommand{\ablationRegressionLogAepe}{0.1067}
\newcommand{\raftThirtyTwoBestSingleAepeTable}{\raftThirtyTwoBestSingleAepe}
\newcommand{\raftThirtyTwoArtofpivAepeTable}{\textbf{\raftThirtyTwoArtofpivAepe}}
\newcommand{\disTwentyHzBestSingleAepeTable}{\disTwentyHzBestSingleAepe}
\newcommand{\disTwentyHzArtofpivAepeTable}{\textbf{\disTwentyHzArtofpivAepe}}
\newcommand{\disOneKHzBestSingleAepeTable}{\disOneKHzBestSingleAepe}
\newcommand{\disOneKHzArtofpivAepeTable}{\textbf{\disOneKHzArtofpivAepe}}
\newcommand{\openpivOneKHzBestSingleAepeTable}{\openpivOneKHzBestSingleAepe}
\newcommand{\openpivOneKHzArtofpivAepeTable}{\textbf{\openpivOneKHzArtofpivAepe}}
\newcommand{\openpivFourKHzBestSingleAepeTable}{\openpivFourKHzBestSingleAepe}
\newcommand{\openpivFourKHzArtofpivAepeTable}{\textbf{\openpivFourKHzArtofpivAepe}}
\newcommand{\archDisOneKHzArtofpivAepeTable}{\archDisOneKHzArtofpivAepe}
\newcommand{\archDisOneKHzLargeArtofpivAepeTable}{\textbf{\archDisOneKHzLargeArtofpivAepe}}
\newcommand{\archDisOneKHzRaepePctTable}{\archDisOneKHzRaepePct}
\newcommand{\archDisOneKHzLargeRaepePctTable}{\textbf{\archDisOneKHzLargeRaepePct}}
\newcommand{\archOpenpivOneKHzArtofpivAepeTable}{\archOpenpivOneKHzArtofpivAepe}
\newcommand{\archOpenpivOneKHzLargeArtofpivAepeTable}{\textbf{\archOpenpivOneKHzLargeArtofpivAepe}}
\newcommand{\archOpenpivOneKHzRaepePctTable}{\archOpenpivOneKHzRaepePct}
\newcommand{\archOpenpivOneKHzLargeRaepePctTable}{\textbf{\archOpenpivOneKHzLargeRaepePct}}
\newcommand{\archOpenpivFourKHzArtofpivAepeTable}{\archOpenpivFourKHzArtofpivAepe}
\newcommand{\archOpenpivFourKHzLargeArtofpivAepeTable}{\textbf{\archOpenpivFourKHzLargeArtofpivAepe}}
\newcommand{\archOpenpivFourKHzRaepePctTable}{\archOpenpivFourKHzRaepePct}
\newcommand{\archOpenpivFourKHzLargeRaepePctTable}{\textbf{\archOpenpivFourKHzLargeRaepePct}}
\newcommand{\agentOverheadDisDeployedMsTable}{\textbf{\agentOverheadDisDeployedMs}}
\newcommand{\agentOverheadDisLargeMsTable}{\agentOverheadDisLargeMs}
\newcommand{\agentOverheadOpenpivDeployedMsTable}{\textbf{\agentOverheadOpenpivDeployedMs}}
\newcommand{\agentOverheadOpenpivLargeMsTable}{\agentOverheadOpenpivLargeMs}
\newcommand{\epeStatsDisTwentyHzBestMeanPxTable}{\epeStatsDisTwentyHzBestMeanPx}
\newcommand{\epeStatsDisTwentyHzRouterMeanPxTable}{\textbf{\epeStatsDisTwentyHzRouterMeanPx}}
\newcommand{\epeStatsDisTwentyHzBestMedianPxTable}{\epeStatsDisTwentyHzBestMedianPx}
\newcommand{\epeStatsDisTwentyHzRouterMedianPxTable}{\textbf{\epeStatsDisTwentyHzRouterMedianPx}}
\newcommand{\epeStatsDisTwentyHzBestNinetyPxTable}{\epeStatsDisTwentyHzBestNinetyPx}
\newcommand{\epeStatsDisTwentyHzRouterNinetyPxTable}{\textbf{\epeStatsDisTwentyHzRouterNinetyPx}}
\newcommand{\epeStatsDisTwentyHzBestNinetyFivePxTable}{\epeStatsDisTwentyHzBestNinetyFivePx}
\newcommand{\epeStatsDisTwentyHzRouterNinetyFivePxTable}{\textbf{\epeStatsDisTwentyHzRouterNinetyFivePx}}
\newcommand{\epeStatsDisTwentyHzBestNinetyNinePxTable}{\epeStatsDisTwentyHzBestNinetyNinePx}
\newcommand{\epeStatsDisTwentyHzRouterNinetyNinePxTable}{\textbf{\epeStatsDisTwentyHzRouterNinetyNinePx}}
\newcommand{\epeStatsDisTwentyHzBestMaxPxTable}{\epeStatsDisTwentyHzBestMaxPx}
\newcommand{\epeStatsDisTwentyHzRouterMaxPxTable}{\textbf{\epeStatsDisTwentyHzRouterMaxPx}}
\newcommand{\epeStatsDisOneKHzBestMeanPxTable}{\epeStatsDisOneKHzBestMeanPx}
\newcommand{\epeStatsDisOneKHzRouterMeanPxTable}{\textbf{\epeStatsDisOneKHzRouterMeanPx}}
\newcommand{\epeStatsDisOneKHzBestMedianPxTable}{\epeStatsDisOneKHzBestMedianPx}
\newcommand{\epeStatsDisOneKHzRouterMedianPxTable}{\textbf{\epeStatsDisOneKHzRouterMedianPx}}
\newcommand{\epeStatsDisOneKHzBestNinetyPxTable}{\epeStatsDisOneKHzBestNinetyPx}
\newcommand{\epeStatsDisOneKHzRouterNinetyPxTable}{\textbf{\epeStatsDisOneKHzRouterNinetyPx}}
\newcommand{\epeStatsDisOneKHzBestNinetyFivePxTable}{\epeStatsDisOneKHzBestNinetyFivePx}
\newcommand{\epeStatsDisOneKHzRouterNinetyFivePxTable}{\textbf{\epeStatsDisOneKHzRouterNinetyFivePx}}
\newcommand{\epeStatsDisOneKHzBestNinetyNinePxTable}{\epeStatsDisOneKHzBestNinetyNinePx}
\newcommand{\epeStatsDisOneKHzRouterNinetyNinePxTable}{\textbf{\epeStatsDisOneKHzRouterNinetyNinePx}}
\newcommand{\epeStatsDisOneKHzBestMaxPxTable}{\epeStatsDisOneKHzBestMaxPx}
\newcommand{\epeStatsDisOneKHzRouterMaxPxTable}{\textbf{\epeStatsDisOneKHzRouterMaxPx}}
\newcommand{\epeStatsOpenpivOneKHzBestMeanPxTable}{\epeStatsOpenpivOneKHzBestMeanPx}
\newcommand{\epeStatsOpenpivOneKHzRouterMeanPxTable}{\textbf{\epeStatsOpenpivOneKHzRouterMeanPx}}
\newcommand{\epeStatsOpenpivOneKHzBestMedianPxTable}{\epeStatsOpenpivOneKHzBestMedianPx}
\newcommand{\epeStatsOpenpivOneKHzRouterMedianPxTable}{\textbf{\epeStatsOpenpivOneKHzRouterMedianPx}}
\newcommand{\epeStatsOpenpivOneKHzBestNinetyPxTable}{\epeStatsOpenpivOneKHzBestNinetyPx}
\newcommand{\epeStatsOpenpivOneKHzRouterNinetyPxTable}{\textbf{\epeStatsOpenpivOneKHzRouterNinetyPx}}
\newcommand{\epeStatsOpenpivOneKHzBestNinetyFivePxTable}{\epeStatsOpenpivOneKHzBestNinetyFivePx}
\newcommand{\epeStatsOpenpivOneKHzRouterNinetyFivePxTable}{\textbf{\epeStatsOpenpivOneKHzRouterNinetyFivePx}}
\newcommand{\epeStatsOpenpivOneKHzBestNinetyNinePxTable}{\epeStatsOpenpivOneKHzBestNinetyNinePx}
\newcommand{\epeStatsOpenpivOneKHzRouterNinetyNinePxTable}{\textbf{\epeStatsOpenpivOneKHzRouterNinetyNinePx}}
\newcommand{\epeStatsOpenpivOneKHzBestMaxPxTable}{\epeStatsOpenpivOneKHzBestMaxPx}
\newcommand{\epeStatsOpenpivOneKHzRouterMaxPxTable}{\textbf{\epeStatsOpenpivOneKHzRouterMaxPx}}
\newcommand{\epeStatsOpenpivFourKHzBestMeanPxTable}{\epeStatsOpenpivFourKHzBestMeanPx}
\newcommand{\epeStatsOpenpivFourKHzRouterMeanPxTable}{\textbf{\epeStatsOpenpivFourKHzRouterMeanPx}}
\newcommand{\epeStatsOpenpivFourKHzBestMedianPxTable}{\epeStatsOpenpivFourKHzBestMedianPx}
\newcommand{\epeStatsOpenpivFourKHzRouterMedianPxTable}{\textbf{\epeStatsOpenpivFourKHzRouterMedianPx}}
\newcommand{\epeStatsOpenpivFourKHzBestNinetyPxTable}{\epeStatsOpenpivFourKHzBestNinetyPx}
\newcommand{\epeStatsOpenpivFourKHzRouterNinetyPxTable}{\textbf{\epeStatsOpenpivFourKHzRouterNinetyPx}}
\newcommand{\epeStatsOpenpivFourKHzBestNinetyFivePxTable}{\epeStatsOpenpivFourKHzBestNinetyFivePx}
\newcommand{\epeStatsOpenpivFourKHzRouterNinetyFivePxTable}{\textbf{\epeStatsOpenpivFourKHzRouterNinetyFivePx}}
\newcommand{\epeStatsOpenpivFourKHzBestNinetyNinePxTable}{\epeStatsOpenpivFourKHzBestNinetyNinePx}
\newcommand{\epeStatsOpenpivFourKHzRouterNinetyNinePxTable}{\textbf{\epeStatsOpenpivFourKHzRouterNinetyNinePx}}
\newcommand{\epeStatsOpenpivFourKHzBestMaxPxTable}{\epeStatsOpenpivFourKHzBestMaxPx}
\newcommand{\epeStatsOpenpivFourKHzRouterMaxPxTable}{\textbf{\epeStatsOpenpivFourKHzRouterMaxPx}}
\newcommand{\epeStatsRaftThirtyTwoBestMeanPxTable}{\epeStatsRaftThirtyTwoBestMeanPx}
\newcommand{\epeStatsRaftThirtyTwoRouterMeanPxTable}{\textbf{\epeStatsRaftThirtyTwoRouterMeanPx}}
\newcommand{\epeStatsRaftThirtyTwoBestMedianPxTable}{\epeStatsRaftThirtyTwoBestMedianPx}
\newcommand{\epeStatsRaftThirtyTwoRouterMedianPxTable}{\textbf{\epeStatsRaftThirtyTwoRouterMedianPx}}
\newcommand{\epeStatsRaftThirtyTwoBestNinetyPxTable}{\epeStatsRaftThirtyTwoBestNinetyPx}
\newcommand{\epeStatsRaftThirtyTwoRouterNinetyPxTable}{\textbf{\epeStatsRaftThirtyTwoRouterNinetyPx}}
\newcommand{\epeStatsRaftThirtyTwoBestNinetyFivePxTable}{\epeStatsRaftThirtyTwoBestNinetyFivePx}
\newcommand{\epeStatsRaftThirtyTwoRouterNinetyFivePxTable}{\textbf{\epeStatsRaftThirtyTwoRouterNinetyFivePx}}
\newcommand{\epeStatsRaftThirtyTwoBestNinetyNinePxTable}{\epeStatsRaftThirtyTwoBestNinetyNinePx}
\newcommand{\epeStatsRaftThirtyTwoRouterNinetyNinePxTable}{\textbf{\epeStatsRaftThirtyTwoRouterNinetyNinePx}}
\newcommand{\epeStatsRaftThirtyTwoBestMaxPxTable}{\epeStatsRaftThirtyTwoBestMaxPx}
\newcommand{\epeStatsRaftThirtyTwoRouterMaxPxTable}{\textbf{\epeStatsRaftThirtyTwoRouterMaxPx}}
\newcommand{\ablationRegressionRawAepeTable}{\ablationRegressionRawAepe}
\newcommand{\ablationRegressionMeanAepeTable}{\ablationRegressionMeanAepe}
\newcommand{\ablationRegressionLogAepeTable}{\textbf{\ablationRegressionLogAepe}}
\newcommand{\quietArtCostReductionRange}{13.24--13.59}

\newcommand{\channelUnknown}{\ifmmode\text{\pnum{...}}\else\pnum{...}\fi}
\newcommand{\channelOperatingSystem}{Ubuntu 22.04}
\newcommand{\channelCpuModel}{AMD Ryzen Threadripper PRO 5995WX}
\newcommand{\channelGpuModel}{NVIDIA RTX 4090}
\newcommand{\channelGpuCount}{4}
\newcommand{\channelAquariumModel}{Olibetta Long Aquarium}
\newcommand{\channelAquariumCapacityL}{90}
\newcommand{\channelAquariumLengthM}{0.90}
\newcommand{\channelAquariumWidthM}{0.30}
\newcommand{\channelAquariumHeightM}{0.30}

\newcommand{\channelInletSpeedMS}{0.13}
\newcommand{\channelCylinderCount}{3}
\newcommand{\channelPropellerCount}{2}
\newcommand{\channelMotorModel}{MyActuator L4010-2}
\newcommand{\channelCanInterface}{Kvaser}
\newcommand{\channelCylinderRadiusM}{0.03}

\newcommand{\channelCameraRateHz}{60}

\newcommand{\channelAcquisitionHeightPx}{896}
\newcommand{\channelAcquisitionWidthPx}{1028}
\newcommand{\channelCropHeightPx}{720}
\newcommand{\channelCropWidthPx}{768}

\newcommand{\channelObservationHeight}{16}
\newcommand{\channelObservationWidth}{16}

\newcommand{\channelDisLatencyBudgetMs}{1}
\newcommand{\channelFlowAcquisitionRateHz}{30}

\newcommand{\channelPreprocessIntensityCappingN}{3.0}

\newcommand{\channelPreprocessStretchContrastScale}{255.0}
\newcommand{\channelMotorCount}{5}

\newcommand{\channelTrackingVelocitySpanPx}{2}

\newcommand{\channelRotationRateLimitRadS}{15.708}
\newcommand{\channelSurfaceSpeedMS}{0.471}
\newcommand{\channelSurfaceSpeedRatio}{3.62}
\newcommand{\channelCropTopPx}{156}
\newcommand{\channelCropLeftPx}{255}
\newcommand{\channelObservationCellCount}{256}

\newcommand{\channelRoiSidePx}{256}
\newcommand{\channelSelectorRecordingDurationS}{60}
\newcommand{\channelSelectorSettlingDurationS}{30}
\newcommand{\channelSelectorTrainingRecordings}{3}
\newcommand{\channelSelectorValidationRecordings}{1}
\newcommand{\channelSelectorPairsPerRecording}{1799}
\newcommand{\channelPairIntervalMs}{2.5}
\newcommand{\channelPropellerSpeedRpm}{200}

\newcommand{\channelSelectorBatchSize}{16}
\newcommand{\channelSelectorTrainingSteps}{3000}
\newcommand{\channelSelectorLearningRate}{0.001}
\newcommand{\channelRoiSelectorLearningRate}{0.0001}
\newcommand{\channelSelectorTrainPairsPerClass}{5\,397}
\newcommand{\channelSelectorValidationPairsPerClass}{1\,799}
\newcommand{\channelDisTrainPairs}{10\,794}
\newcommand{\channelDisValidationPairs}{3\,598}
\newcommand{\channelRoiTrainPairs}{16\,191}
\newcommand{\channelRoiValidationPairs}{5\,397}
\newcommand{\channelFrozenDisRows}{A & 0 & 1 & 19 & 11 & 4\\
B & 1 & 3 & 13 & 9 & 4\\
C & 0 & 2 & 13 & 7 & 6\\}
\newcommand{\channelFrozenRegimeRows}{Tuning & A & 150 & -150 & -150\\
Tuning & B & 0 & 150 & -150\\
Region & R1 & 150 & 150 & 150\\
Region & R2 & 150 & 150 & -150\\
Region & R3 & 150 & -150 & 150\\}
\newcommand{\channelRoiRows}{R1 & 256 & 64\\
R2 & 0 & 256\\
R3 & 512 & 384\\}

\newcommand{\pinballTrainingEpisode}{17}

\newcommand{\pinballDisValPairs}{3\,598}
\newcommand{\pinballDisTestPairs}{3\,598}

\newcommand{\pinballTestSeparationMinutes}{35.8}
\newcommand{\pinballDisValidationAccuracyPct}{100.00}

\newcommand{\pinballDisSelectedStep}{500}

\newcommand{\pinballRoiValidationAccuracyPct}{99.93}

\newcommand{\pinballRoiSelectedStep}{3000}

\newcommand{\pinballSelectorValidationRows}{Tuning & A & 1\,799/1\,799 & 100.00\\
Tuning & B & 1\,799/1\,799 & 100.00\\
Region & R1 & 1\,795/1\,799 & 99.78\\
Region & R2 & 1\,799/1\,799 & 100.00\\
Region & R3 & 1\,799/1\,799 & 100.00\\}
\newcommand{\pinballAugmentationWithCompletedSeeds}{5}

\newcommand{\pinballAugmentationWithAccuracyPct}{90.00}

\newcommand{\pinballAugmentationWithStdPct}{20.00}

\newcommand{\pinballAugmentationWithoutAccuracyPct}{80.00}

\newcommand{\pinballAugmentationWithoutStdPct}{24.49}
\newcommand{\pinballAugmentationRows}{With augmentation & 5/5 & $\mathbf{90.00} \pm 20.00$\\
Without augmentation & 5/5 & $80.00 \pm 24.49$\\}
\newcommand{\pinballAugmentationCompletionNote}{All five paired seeds per variant are complete.}

\newcommand{\pinballSettingAASixteenEpe}{0.4609}

\newcommand{\pinballSettingABSixteenEpe}{0.4861}

\newcommand{\pinballSettingBASixteenEpe}{0.9951}

\newcommand{\pinballSettingBBSixteenEpe}{0.5359}

\newcommand{\pinballSettingASixteenPreferredWinPct}{55.5}

\newcommand{\pinballSettingBSixteenPreferredWinPct}{98.8}

\newcommand{\pinballPreferenceReferenceRows}{A & A & \textbf{0.3666} & \textbf{0.3248} & \textbf{0.4609}\\
A & B & 0.4585 & 0.4511 & 0.4861\\
B & A & 0.6170 & 0.6095 & 0.9951\\
B & B & \textbf{0.4864} & \textbf{0.4765} & \textbf{0.5359}\\}
\newcommand{\pinballControlModelPreset}{size25m}
\newcommand{\pinballControlTrainRatio}{-1}
\newcommand{\pinballControlTrainMinutes}{20}
\newcommand{\pinballControlTrainSteps}{36\,000}
\newcommand{\pinballControlReplayCapacity}{1\,000\,000}
\newcommand{\pinballControlConvChannels}{48, 72}

\title{ART of PIV: Agentic Real-Time Particle Image Velocimetry for Active Flow Control}
\author{Antonio Terpin$^{1,2,\ast}$\quad Francesco Banelli$^{1,2}$\quad
  Alan Bonomi$^{1,2}$\quad Raffaello D'Andrea$^{1}$\\[0.5em]
  {\small
    $^{1}$Institute for Dynamic Systems and Control, ETH Zurich,
    8092 Zurich, Switzerland\\[0.35em]
    $^{2}$These authors contributed equally: Antonio Terpin, Francesco Banelli, Alan Bonomi.\\
    $^{\ast}$Correspondence and requests for materials: Antonio Terpin
    (\href{mailto:aterpin@ethz.ch}{\texttt{aterpin@ethz.ch}}).
  }}
\date{Preprint --- October 30, 2026}

\begin{document}
\maketitle

\begin{abstract}
  \noindent
\Gls{acr:piv} measures fluid motion from images of tracer particles, but its
accuracy depends on choices throughout the estimation pipeline, including
image preprocessing, flow estimation, and postprocessing. These choices are
typically left to expert tuning, and a fixed configuration can become less
accurate as conditions change. This matters
especially when \gls{acr:piv} supplies feedback for real-time control, where
measurements must be both accurate and fast, and flow conditions vary.
We introduce \ourmethod---short
for Agentic Real-Time and also a nod to the art of tuning a \gls{acr:piv}
pipeline---which automates pipeline choice for each image pair.
A learned agent selects a configuration according to a chosen measurement
objective from a library assembled for a specified speed budget.
Across cross-correlation,
classical optical-flow, and learning-based estimators, \ourmethod reduces
\gls{acr:aepe} by \mainRaepePctRange\,\%
relative to the best single estimator at matched speed, with gains extending
to \mainHighRateRange\,\estimatorRateUnit. In a simulated local velocity regulation task, it reduces
mean normalized closed-loop cost by
\quietArtCostReductionRange{} percentage
points relative to a tuned fixed estimator, with the same classical controller
in both cases. In a tabletop water channel, we illustrate two ways to tune
\ourmethod for a fluidic pinball of three rotating cylinders: selecting
between complementary optical-flow tunings for accuracy, and selecting
regions of interest with coherent rotation or shear using a common flow
estimator. For the first variant, we train a \dreamerVThree reinforcement
learning controller directly in the physical world and demonstrate tracking
of a prescribed flow field in the wake of a fluidic pinball.
\glsreset{acr:piv}
\glsreset{acr:aepe}

\end{abstract}

\vfill
\begin{figure}[H]\centering
  \includegraphics[width=.9\linewidth]{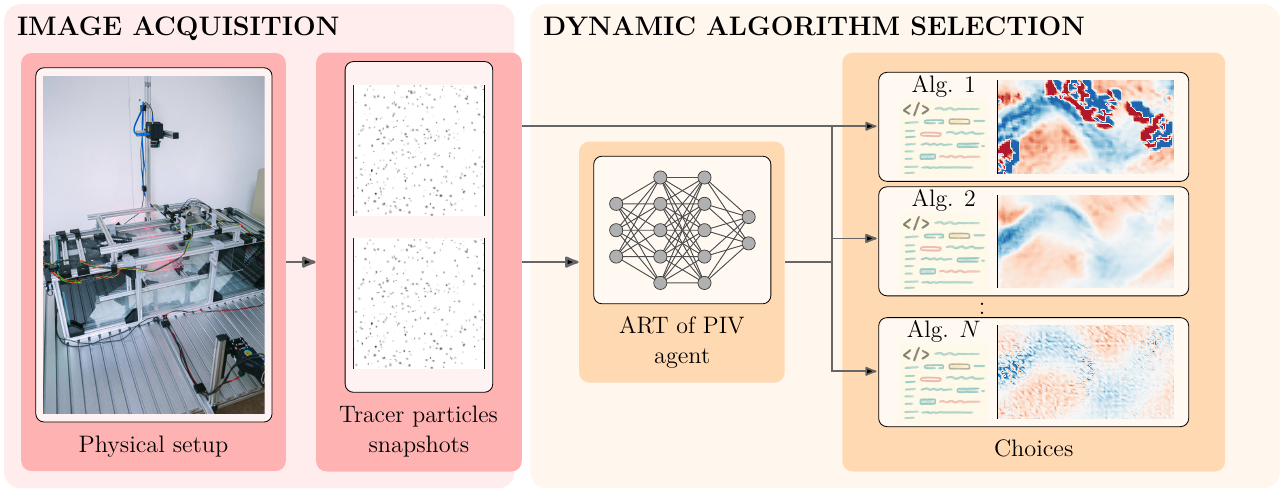}
  \caption{\ourmethod casts flow quantification as an agentic decision: given
    a pair of tracer-particle snapshots, a lightweight agent selects one
    estimator from a library $\estimator{1},\dots,\estimator{\numEstimators}$ and
    returns the selected estimator's flow.}
  \label{fig:teaser}
\end{figure}
\clearpage

\section{Introduction}
\label{sec:introduction}

\Gls{acr:piv} \citep{adrian1991piv,raffel2018piv,westerweel2013piv,scharnowski2020piv}
and related image-based flow-quantification methods
\citep{horn1981determining,liu2008fluid} are among the central tools for
measuring flow fields, with applications spanning laboratory
\citep{adrian1991piv}, industrial \citep{raffel2018piv}, microfluidic
\citep{santiago1998micropiv}, and environmental \citep{fujita2007lspiv}
settings. Traditionally, their role has been observational: to reconstruct a
velocity field for analysis, validation, and diagnostics. Performance has
accordingly been measured almost exclusively through accuracy. For an image
pair, the \gls{acr:epe} is the $\ell_2$ distance between the estimated and
ground-truth flow, averaged over pixels. Averaging these errors over the image
pairs in a dataset gives the \gls{acr:aepe}.
Throughout, we use \emph{algorithm} and \emph{estimator} broadly to include
the configured measurement pipeline: image preprocessing, flow estimation,
and postprocessing, following the organization of \flowgym
\citep{banelli2026flowgym}. Two configurations may therefore differ in their
crop, image conditioning, or output resolution while sharing the same
underlying flow-estimation method.

Recent results in active flow control invite a broader perspective.
\citet{terpin2025ff} showed that, on a real-world flow-control task,
an agent equipped with dense flow feedback discovers high-performance strategies
within tens of minutes of real-world interaction, whereas an agent without that
feedback fails to learn a policy that increases drag. Once flow quantification is embedded in a real-time
learning or control loop, latency becomes as consequential as accuracy: a
highly accurate but slow estimator simply cannot run online, while a fast but
crude one risks yielding feedback too noisy to learn from. With accuracy as
the measurement objective, this leads to the \emph{speed--accuracy trade-off}:
the best accuracy attainable at a given inference speed.
Three estimator families span this trade-off: classical
cross-correlation \citep{scarano2002widim,liberzon2021openpiv}, optical flow
\citep{kroeger2016fast,horn1981determining,farneback2003two}, and learning-based
estimators \citep{cai2019dense,huang2022flowformer,luo2024flowdiffuser}. No
single method dominates. For instance, using \flowgym \citep{banelli2026flowgym}, we find that the
accurate \raftpiv \citep{lagemann2021raft} runs at only
\raftThirtyTwoNominalRate{} on an RTX 5090, while fast estimators such as \dis
\citep{kroeger2016fast} and \openpiv run from tens to thousands of
image pairs per second at lower accuracy.
For representative operating points in these families, \cref{tab:main}
compares fixed estimators with our approach, which selects an estimator for
each image pair.

\begin{table}[t]\centering
  \caption{Selecting an estimator for each image pair improves accuracy at
    matched speed across three estimator families. The library spans flow-class
    specialists or hyperparameter tunings of the listed estimator. Speeds are
    in \estimatorRateUnit{} and include the agent in the sequential reference
    implementation. Errors are \gls{acr:aepe} (px; lower is better);
    \acrshort{acr:raepe} is the percentage decrease in \gls{acr:aepe} relative
    to the best single estimator (higher is better). The fixed baseline is the
    best budget-feasible tuning for \dis and \openpiv, and the pretrained
    network for \raftpiv. Results are means over three training seeds on the
    held-out test set of \citet{cai2019dense}; experimental details are in
    \cref{sec:results:accuracy}. Bold marks the lower error at each operating point.}
  \label{tab:main}
  \small
  \begin{tabular}{llcccc}
    \toprule
    & & & \multicolumn{2}{c}{\gls{acr:aepe} (px)} & \\
    \cmidrule(lr){4-5}
    Estimator & Library spans & \shortstack{Speed\\(\estimatorRateUnit)} & Best single & \ourmethod & \acrshort{acr:raepe} (\%) \\
    \midrule
    \multicolumn{6}{l}{\emph{Learned}}\\
    \raftpiv    & flow classes      & $\approx$\raftThirtyTwoNominalRateValue & \raftThirtyTwoBestSingleAepeTable & \raftThirtyTwoArtofpivAepeTable & \raftThirtyTwoRaepePct \\
    \midrule
    \multicolumn{6}{l}{\emph{Classical optical flow}}\\
    \dis        & hyperparameters   & $\approx$\disTwentyHzNominalRateValue  & \disTwentyHzBestSingleAepeTable & \disTwentyHzArtofpivAepeTable & \disTwentyHzRaepePct \\
    \dis        & hyperparameters   & $\approx$\disOneKHzNominalRateValue  & \disOneKHzBestSingleAepeTable & \disOneKHzArtofpivAepeTable & \disOneKHzRaepePct \\
    \midrule
    \multicolumn{6}{l}{\emph{Cross-correlation}}\\
    \openpiv    & hyperparameters   & $\approx$\openpivOneKHzNominalRateValue  & \openpivOneKHzBestSingleAepeTable & \openpivOneKHzArtofpivAepeTable & \openpivOneKHzRaepePct \\
    \openpiv    & hyperparameters   & $\approx$\openpivFourKHzNominalRateValue  & \openpivFourKHzBestSingleAepeTable & \openpivFourKHzArtofpivAepeTable & \openpivFourKHzRaepePct \\
    \bottomrule
  \end{tabular}
\end{table}

In this work, we ask how a flow-measurement pipeline can adapt to changing
conditions within a given inference-time budget. The preferred configuration
can depend on both the flow and what the user wants to measure. Even when
accuracy is the objective, the best-performing estimator or parameter setting
can vary across flows and regions of a single image
pair~\citep{bonomi2025consensus}. Tuning a \gls{acr:piv} pipeline is therefore an
\emph{art}: as the 2014 \gls{acr:piv} Challenge put it, ``the largest
uncertainties are introduced by the user who selects the evaluation parameters
based on knowledge, experience and intuition''~\citep{kahler2016main}. One line of
work responds by adapting a single estimator's parameters to the data: the
interrogation window in cross-correlation
\citep{scarano1999multigrid,theunissen2007adaptive,theunissen2010ensemble}, or
the regularization in optical flow \citep{jassal2025optical}. We build a library
of configurations suited to different conditions and learn when to use each.
The motivation is practical: tuning a pipeline for a particular regime can be
easier than finding a common configuration for all regimes.
Human expertise can guide library design and training labels. Alternatively,
labels can follow a quantitative criterion, such as minimum estimation error
when the true flow is known. The agent learns to predict the preferred
configuration for new image pairs.

\paragraph{Our contributions}
We introduce \ourmethod, an agent that learns to select a measurement pipeline
for each image pair within a specified speed budget. A small \gls{acr:cnn}
predicts the preferred configuration from a supplied library and returns its
flow; see \cref{fig:teaser}. The library and supervision encode the user's
measurement preferences, including choices of preprocessing, estimator tuning,
and postprocessing. Selection can overlap with estimator execution to keep
its additional latency small. Adapting the choice on every image pair is
particularly relevant in active flow control, where the controller continually
changes the flow conditions.

We evaluate selection for accuracy using estimators from three paradigms:
classical optical flow, learning-based estimation, and cross-correlation,
including \dis, \raftpiv, and \openpiv. Relative to the best single configuration
at matched speed, \ourmethod reduces \gls{acr:aepe} by
\mainRaepePctRange\,\%. This gain persists at
\mainHighRateRange\,\estimatorRateUnit, where the resulting accuracy could enable control with flow feedback in systems with very fast dynamics, such as
combustion instabilities \citep{mcmanus1993review}, aerodynamic
flow separation \citep{greenblatt2000control}, and gust
load alleviation \citep{regan2012survey}.

To assess whether these estimation gains translate to active flow control,
we first evaluate \ourmethod on a simulated local velocity regulation task.
With the same classical controller, \ourmethod increases the time a work area
stays near its target velocity and reduces mean normalized closed-loop cost by
\quietArtCostReductionRange{} percentage points relative to a separately tuned
fixed \dis reference (see \cref{sec:afc:quieting}).
We then deploy \ourmethod on the tabletop water-channel setup of
\citet{terpin2025ff}, available at \url{https://activefluidcontrol.com},
using a fluidic pinball of three rotating cylinders to illustrate two ways
of learning our measurement preferences. The first variant selects between \dis~A and
\dis~B, complementary tunings chosen for accuracy in different flow regimes.
The second selects regions of interest with coherent rotation or shear using
the common flow estimator \dis~C. Using flow feedback from the first variant,
we train a \dreamerVThree reinforcement learning controller directly on the
physical setup and demonstrate successful flow tracking
(see \cref{sec:afc:channel}).
The method and evaluation infrastructure are available in
\flowgym~\citep{banelli2026flowgym},
and the deployment code is available in \flowgames.

\section{Our method}
\label{sec:method}

We first formalize pipeline selection under a speed budget
(\cref{sec:method:formulation}), then describe the agent, its execution
schedule, and training from scores or labels (\cref{sec:method:learning}).
We close with library construction (\cref{sec:method:library}).

\subsection{Flow quantification as a selection problem}
\label{sec:method:formulation}
We are given a library
$\library=\set{\estimator{1},\dots,\estimator{\numEstimators}}$ of flow
estimators, each including its preprocessing and postprocessing. Members can
be different \dis~\citep{kroeger2016fast} tunings, fine-tuned \raftpiv variants
\citep{lagemann2021raft}, or a common estimator applied to different regions.
Each maps an image pair $\imagePair$ to a candidate flow
$\estimate{i}=\estimator{i}\imagePair$ on its configured spatial domain.
A policy $\selector$, with learnable parameters $\selectorParams$, maps the
pair to an action $\action\in\nrange{\numEstimators}$, yielding the flow
$\estimate{\action}$.

Let $\reward(\framePrev,\frameNext,\action)$ quantify how well that choice
serves the measurement objective, with larger values preferred. We seek to maximize
its expectation over the image pairs encountered at deployment:
\begin{equation}\label{eq:selection}
  \max_{\selectorParams}\;
  \expectedValue[\imagePair]{\reward(\framePrev,\frameNext,\selector\imagePair)}.
\end{equation}
The library is restricted a priori to pipelines that meet the speed budget
$\speedBudget$ (see \cref{sec:method:library}). We treat each configured
pipeline's runtime, including its preprocessing and postprocessing, as a
fixed, deterministic cost. Under the overlapping execution schedule of
\cref{sec:method:execution}, meeting the budget is then a property of the
pipeline alone. This is a contextual bandit formulation
\citep{sutton2018reinforcement}: one measurement decision per image pair.
The selection reward specifies which measurement to return; a downstream
controller can have a separate task reward, such as tracking a prescribed flow field.

\begin{remark}[Numerical objectives and expert labels]
\label{rem:method:objectives}
The selection reward can be specified through a numerical criterion or expert
labels. In the first approach, a numerical score directly defines $\reward$.
For example, when a ground-truth flow $\flowGT$ is available, we measure accuracy
using
$\reward(\framePrev,\frameNext,\action)=-\epe(\estimate{\action},\flowGT)$,
where \gls{acr:epe} is the pixel-averaged $\ell_2$ error on a common domain
$\imageDomain$, with estimates expressed in the same coordinates and units.
We use this objective in the accuracy benchmarks
(\cref{sec:results:accuracy}) and the simulated local velocity regulation task
(\cref{sec:afc:quieting}). Other numerical criteria, such as scores for desired
flow structures, can define $\reward$ in the same way when available.

Alternatively, an expert can supply a preferred member
$\selectionLabel\imagePair$ for each image pair $\imagePair$. Setting
$\reward(\framePrev,\frameNext,\action)=\indicatorOf{\action=\selectionLabel\imagePair}$
then makes \eqref{eq:selection} maximize agreement with those labels, without
requiring ground-truth flow. We showcase this use with labels that encode
our choices between \dis tunings in the fluidic-pinball experiment of
\cref{sec:afc:channel}, and between regions of interest with a common estimator in
\cref{sec:appendix:water-channel:regions}.
\end{remark}

\subsection{The selection agent}
\label{sec:method:learning}
The agent uses a small \gls{acr:cnn}. Given the image pair $\imagePair$,
it produces in a single forward pass a vector of $\numEstimators$ scores
$\score\imagePair$, one per library member. A higher score indicates a
stronger predicted preference under the chosen supervision; the scores are
classification logits in our default training scheme. The agent's standalone
execution time is reported in \cref{sec:appendix:ablations:arch}.
At deployment, we return the flow of the highest-scoring member:
$\selector\imagePair=\argmax_i\score\imagePair_i$.

\paragraph{Sequential and speculative execution}
\label{sec:method:execution}
In sequential execution, we first evaluate the agent and then run only the
selected estimator. The total latency is the sum of their runtimes.
The agent and the estimators consume the same image pair, so estimator execution
need not wait for the selection decision. A deployment can use \emph{speculative
execution}, analogous to executing instructions ahead of branch resolution in a
processor: start the library estimators concurrently with the agent, then retain
the selected branch and cancel the others once the decision is available
(\cref{alg:inference}). With sufficient parallel resources to avoid slowing the
selected estimator, and with the agent finishing before that estimator, the
decision is hidden entirely within estimator execution.
A very fast estimator may finish before the agent; in that case the
decision determines when the output can be returned, motivating a smaller agent
or a downsampled agent input (\cref{sec:appendix:ablations:arch}).

\begin{algorithm}[H]
\caption{\ourmethod: deployment with optional speculative execution.}
\label{alg:inference}
\begin{algorithmic}[1]
\Require image pair $\imagePair$, library $\library=\set{\estimator{1},\dots,\estimator{\numEstimators}}$, trained score network $\score$
\Require execution mode: sequential or speculative
\If{execution mode is speculative}
  \For{each $i\in\nrange{\numEstimators}$}
    \State Launch $\estimator{i}\imagePair$ asynchronously
  \EndFor
\EndIf
\State Evaluate the scores $\score\imagePair$
\State $\action \gets \argmax_{i\in\nrange{\numEstimators}} \score\imagePair_i$
\If{execution mode is speculative}
  \State Cancel estimator branches $i\ne\action$
  \State Wait for $\estimator{\action}$ to finish, yielding $\estimate{\action}$
\Else
  \State $\estimate{\action} \gets \estimator{\action}\imagePair$
\EndIf
\State \Return $\estimate{\action}$
\end{algorithmic}
\end{algorithm}

\paragraph{Training from scores or labels}
We use \emph{classification} as a training surrogate for \eqref{eq:selection}.
For each image pair $\imagePair$, let
$\selectionLabel\imagePair$ denote the desired library member. This label is
either supplied directly, as in \cref{rem:method:objectives}, or obtained by selecting
$\selectionLabel\imagePair\in\argmax_k\reward(\framePrev,\frameNext,k)$,
with ties resolved consistently.
For negative \gls{acr:epe}, this recovers the minimum-error label
$k^\star\imagePair=\argmin_k e_k\imagePair$, where $e_k\imagePair$ is the
error of estimator $\estimator{k}$ on $\imagePair$. A cross-entropy loss drives
$p_{\selectorParams}(k\mid\framePrev,\frameNext)=\mathrm{softmax}(\score\imagePair)_k$
toward
$\selectionLabel\imagePair$; \cref{alg:training} gives the procedure. Classification
uses the preferred member as its target and discards the size of reward
differences between members. Regression and bandit alternatives retain
numerical reward information; we compare these training paradigms for the
accuracy objective in \cref{sec:appendix:ablations:signal}.

We train only the agent: the library and the supervision criterion remain
fixed. For a fixed image pair, numerical rewards can therefore be computed
once and cached, together with the resulting label. Later epochs reuse that
label without rerunning the estimators. Directly supplied labels bypass
reward evaluation altogether. The cache is valid for unchanged inputs and
supervision; transformations that alter the preferred choice require new
labels or reward evaluations. We use Adam for the agent updates
(\cref{sec:appendix:training:hardware}).

\begin{algorithm}[H]
\caption{Training the agent from numerical rewards or supplied labels.}
\label{alg:training}
\begin{algorithmic}[1]
\Require image pairs $\set{\imagePair}$, library $\library$, optimizer; supplied labels $\selectionLabel\imagePair$ or a reward evaluator $\reward(\framePrev,\frameNext,k)$; label cache $\cache\gets\emptyset$
\For{each minibatch $\mathcal B$}
  \For{$\imagePair \in \mathcal B$}
    \If{$\imagePair\notin\cache$}
      \If{a label $\selectionLabel\imagePair$ is supplied}
        \State $\cache[\imagePair] \gets \selectionLabel\imagePair$
      \Else
        \State Evaluate and cache $\reward(\framePrev,\frameNext,k)$ for every library member $k$
        \State $\cache[\imagePair] \gets \argmax_k\reward(\framePrev,\frameNext,k)$ \Comment{Resolve ties consistently}
      \EndIf
    \EndIf
  \EndFor
  \State Update $\selectorParams$ on $-\frac{1}{|\mathcal B|}\sum_{\imagePair\in\mathcal B}\log p_{\selectorParams}(\cache[\imagePair]\mid\framePrev,\frameNext)$
\EndFor
\State \Return $\selectorParams$
\end{algorithmic}
\end{algorithm}

\paragraph{Learning with limited labeled data}
\label{sec:method:limited-data}
When ground-truth flow fields are available from computational fluid dynamics
simulations representative of the intended deployment, tools such as
\synthpix \citep{terpin2025synthpix} can generate arbitrarily many particle-image
pairs from those fields, without manual annotation. Training or fine-tuning
directly on a physical setup with manually assigned labels, as in our fluidic-pinball
experiment (\cref{sec:afc:channel}), instead relies on a limited set of
recordings and annotations. We reduce annotation effort by assigning
one label to each recording collected under a fixed actuation setting.
Augmentation offers a way to reuse labeled examples
\citep{cubuk2020randaugment}, varying illumination and contrast while preserving
the intended preference. Validation on separate recordings, or on disjoint
time blocks with a gap, helps assess generalization beyond neighboring frames.
The collection, augmentation, and validation procedure for our physical
experiment is detailed in \cref{sec:appendix:water-channel:selection}.

\subsection{Building the library}
\label{sec:method:library}
It remains to choose the library itself. Fix a pool $\pool$ of candidate
pipelines and a speed budget $\speedBudget$, and write
$\pool_{\speedBudget}\subseteq\pool$ for the candidates fast enough to meet it.
When numerical rewards are available for the pool, the best value attainable
on image pair $\imagePair$ by selecting from $\library$ is the \emph{oracle} reward
$\max_{\estimator{k}\in\library}\reward(\framePrev,\frameNext,k)$. We choose the library to maximize
this value in expectation, subject to the budget and a cap $\numEstimators$
on its size:
\begin{equation}\label{eq:library}
  \library^\star \in
  \argmax_{\substack{\library\subseteq\pool_{\speedBudget}\\[1pt] 1\le\abs{\library}\le\numEstimators}}\;
  \expectedValue[\imagePair]{\max_{\estimator{k}\in\library}\reward(\framePrev,\frameNext,k)}.
\end{equation}
For negative \gls{acr:epe}, this minimizes the expected oracle error.
We estimate the expectation on tuning data.
The budget keeps every member---and hence whichever flow the agent
returns---within the latency target under the schedule of
\cref{sec:method:execution}, and the cap
$\numEstimators$ keeps the agent's decision and the speculative work small.

For fixed, finite cached rewards independent of library composition, the
gain above a common lower baseline is monotone and submodular (\cref{sec:appendix:library}). We therefore initialize
the library greedily by adding the candidate with the largest marginal reward
gain until the cap $\numEstimators$ is reached. This attains at least
$(1-1/e)$ of the optimal gain above that baseline \citep{nemhauser1978}.
The greedy library can be refined to the exact optimum with a combinatorial
solver. For our negative-\gls{acr:epe} experiments, this takes
\librarySelectionMilpMinS--\librarySelectionMilpMaxS\,s on the \dis pool;
greedy lands within \librarySelectionGreedyGapFromThreeMaxPct\,\% of the optimal
\gls{acr:aepe} from three models on, and the libraries of \cref{tab:main} are
exact. We perform this selection offline using cached rewards. Sweeping
$\speedBudget$ traces the objective attained by the corresponding libraries;
for accuracy, \cref{tab:main} reports the resulting speed--accuracy trade-off
at the tested operating points.

\paragraph{Forming the pool}
The candidates in $\pool$ come from varying the flow estimator, its
parameters, or its preprocessing and postprocessing. When this is cheap we
over-generate and let \eqref{eq:library} prune: the patch size, stride, iteration
count, and pyramid depth of \dis, for instance, yield a large pool at negligible
cost. When it is expensive we cannot screen many, and instead build each
candidate deliberately; for example, by fine-tuning a pretrained \raftpiv
\citep{lagemann2021raft} into one specialist per flow regime. We give the \dis pool
and hyperparameters and the \raftpiv fine-tuning protocol in
\cref{sec:appendix:training}, and release both in the \flowgym repository
\citep{banelli2026flowgym}, together with the generic per-algorithm scripts that
generate the candidate pools, so a pool can be formed for any new estimator.

An expert can also form a small library directly, retaining configurations
preferred in different observed regimes without a numerical reward for
screening the pool. Provided every member meets the speed budget, the library
can be paired with expert labels to train the same selection agent. The
fluidic-pinball experiment in \cref{sec:afc:channel} illustrates this approach
with manually tuned estimators and labels encoding our choices;
\cref{sec:appendix:water-channel:regions}
gives a second example with a library of regions of interest.

\section{Results and discussion}
\label{sec:results}

We first evaluate \ourmethod in terms of \gls{acr:aepe} across estimators and
speed budgets (\cref{sec:results:accuracy}). We then test its use in active
flow control (\cref{sec:afc}).

\subsection{Flow estimation accuracy}
\label{sec:results:accuracy}

We compare \ourmethod with fixed estimators at the matched-speed operating
points in \cref{tab:main}, spanning three estimator families and two
specialization strategies.

\paragraph{Estimators}
We evaluate three estimators drawn from three paradigms. \raftpiv
\citep{lagemann2021raft} is a \emph{learned} optical-flow network;
\dis \citep{kroeger2016fast} is a classical \emph{optical-flow} method; \openpiv
\citep{liberzon2021openpiv} is a \emph{cross-correlation} method. Each estimator
contributes one or more operating points, and the library \ourmethod selects from
(\cref{sec:method:library}) is built by specializing that estimator along a
\emph{different} axis:
\begin{itemize}[leftmargin=*, itemsep=1pt, topsep=2pt]
  \item \raftpiv: one fine-tuned specialist \emph{per flow class}
    of the benchmark, so the library spans flow regimes
    (\cref{sec:appendix:training:raft}).
  \item \dis and \openpiv: a pool of \emph{hyperparameter} tunings (interrogation
    window, overlap, and the \dis parameters), from which \eqref{eq:library} selects the
    library (\cref{sec:appendix:training:dis,sec:appendix:training:openpiv}).
\end{itemize}
We also contribute an efficient \jax \citep{bradbury2018jax} implementation of
\openpiv and release it in \flowgym
\citep{banelli2026flowgym} alongside the already-ported \raftpiv and \dis.

At each operating point the baseline is the single best estimator that meets the
budget. For the \dis and \openpiv points, this baseline is the $\numEstimators=1$
solution of~\eqref{eq:library} over the budget-feasible pool. For a learned point,
the baseline is the corresponding pretrained network (\eg the pretrained \raftpiv at the accurate end).
\ourmethod draws on the same pool at the same budget, but with a library of
$\numEstimators>1$ specialists and selection for each image pair. The effect
of library size is studied in \cref{sec:appendix:ablations}.

\paragraph{Method}
We evaluate on the \gls{acr:piv} benchmark of \citet{cai2019dense}, following its
train/test split: every number in this subsection is on the held-out test set, while
model selection and the ablations of \cref{sec:appendix:ablations} use a
10\,\% validation split carved from the training set
(see \cref{sec:appendix:training:data}). We average over three training
seeds; hardware and training details are in
\cref{sec:appendix:training:hardware}. We build a separate library at each
operating point. The points span $\approx$\mainMinNominalRate{} (the pretrained \raftpiv)
to $\approx$\mainMaxNominalRate{} (\openpiv), with \dis filling the
range in between. The reported speeds include both the agent and the selected
estimator, measured with the sequential reference implementation. They provide a
conservative reference for speculative deployments that hide the agent's execution
time as in \cref{sec:method:execution}. We hold the measured pipeline to the same
target as the baseline. The classification
agent and greedy-versus-exact library selection are
ablated in \cref{sec:appendix:ablations,sec:appendix:library}.

\paragraph{Metrics}
We report the \gls{acr:aepe} (in px; lower is better), which averages
\gls{acr:epe} over all image pairs in the test set, and inference
throughput (\estimatorRateUnit, agent included, sequential reference implementation).
To compare a method $a$ against a baseline $b$ at matched speed, we report the
\gls{acr:raepe},
\begin{equation}\label{eq:raepe}
  \raepe{a}{b} = \frac{\aepe{b} - \aepe{a}}{\aepe{b}},
\end{equation}
where $\aepe{a}$ is the \gls{acr:aepe} of $a$ on the test set. Higher
$\raepe{a}{b}$ is better, and a value of 20\,\% means $a$ cuts the average
error by a fifth. Following common practice in flow quantification, we summarize
results using \gls{acr:aepe} or \gls{acr:raepe}; a statistical analysis of the
individual \gls{acr:epe} values is reported in \cref{sec:appendix:epe-statistics}.

\paragraph{Results}
At every operating point, selecting an estimator for each image pair beats the best single estimator
at matched speed, lowering the \gls{acr:aepe} by \mainRaepePctRange\,\%
(\cref{tab:main}). The gain holds across all three paradigms and both
specialization axes: whether the library members are per-class \raftpiv
weights or \dis and \openpiv
hyperparameter tunings, selection for each image pair recovers accuracy that no fixed
estimator attains on its own. The gap to
the oracle---the lowest \gls{acr:aepe} achievable by selecting the best estimator for each image pair
(\cref{sec:appendix:ablations})---measures the remaining improvement available
through more accurate selection, and we invite future work to close it by improving the agent
architecture.

\subsection{Active Flow Control}
\label{sec:afc}

We study the effectiveness of \ourmethod as the feedback source in two control
tasks: simulated local velocity regulation with a fixed
proportional--derivative controller (\cref{sec:afc:quieting}), and flow tracking
with a fluidic pinball of three rotating cylinders, using a \dreamerVThree
reinforcement learning controller trained directly in a physical water channel
(\cref{sec:afc:channel}).

\subsubsection{Flow regulation with distributed actuation}
\label{sec:afc:quieting}

\begin{figure}[!ht]\centering
  \href{https://github.com/antonioterpin/synthpix}{\includegraphics[width=\linewidth]{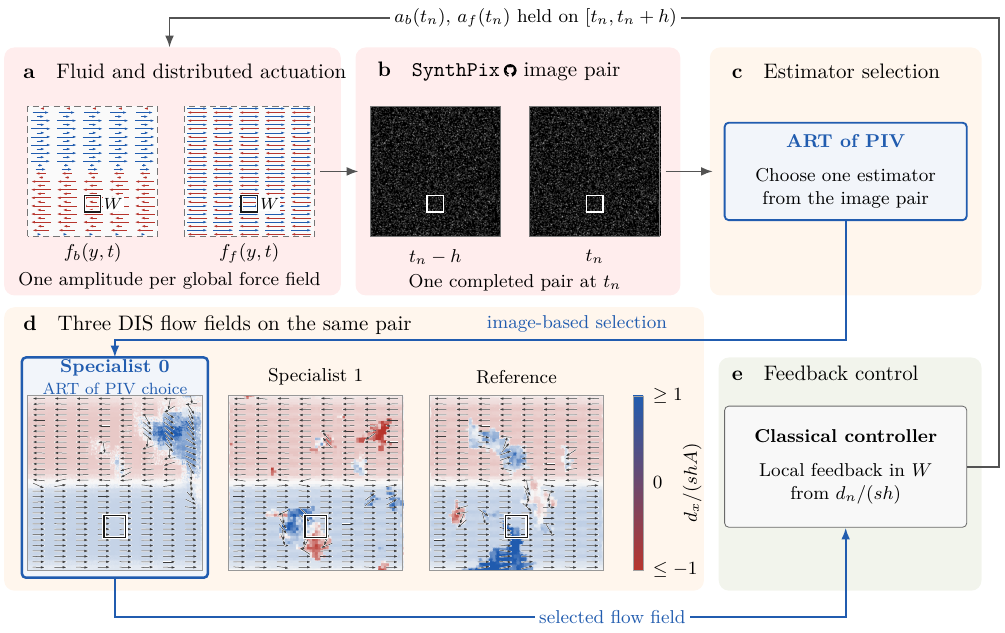}}
  \caption{Local velocity regulation setup and image-feedback loop.
    \textbf{(a)} The forces $\quietBroadForce(y,t),\quietFineForce(y,t)$ of
    the two actuators act throughout the periodic domain; $y$ is transverse
    position and $t$ is time. Arrows show negative commands, normalized per
    actuator. \textbf{(b)} \synthpix renders images at $t_n-h$ and $t_n$,
    where $t_n=nh$ is the $n$th update time and $h$ is the control period.
    \textbf{(c,d)} \ourmethod selects an estimator; the three actual \dis
    fields on the same pair are shown for comparison. The blue outline marks
    the recorded \ourmethod choice, specialist \quietArtSetupChoice. Images and fields show
    a qualitative example on its own trajectory at $t_n/T=\quietArtSetupTime$,
    where $T$ is the episode duration.
    Only specialists 0 and 1 are selectable; the reference is a fixed configuration,
    with lower mean closed-loop cost than either specialist used alone.
    \textbf{(e)} The selected displacement $d_n$ supplies local feedback to
    the controller, whose commands $a_b,a_f$ drive the fluid until the next update.
    Every map shows the full domain, with the work area $W$ outlined.
    Flow-field colors show streamwise displacement $d_x$ converted to velocity
    and normalized by nominal initial amplitude $A$: $d_x/(s h A)$, clipped at
    $\pm1$; their arrows indicate direction. Here $s$ is camera calibration in
    pixels per unit length.}
  \label{fig:quieting:forcing}
\end{figure}

Flow feedback can be exploited to create a work area usable for observation or
manipulation. For example, hydrodynamic traps use feedback to hold particles
near a stagnation point \citep{tanyeri2010hydrodynamic}. We isolate a related
requirement: regulate the velocity throughout a fixed work area
(\cref{fig:quieting:forcing}).

\begin{figure}[p]\centering
  \includegraphics[width=\linewidth]{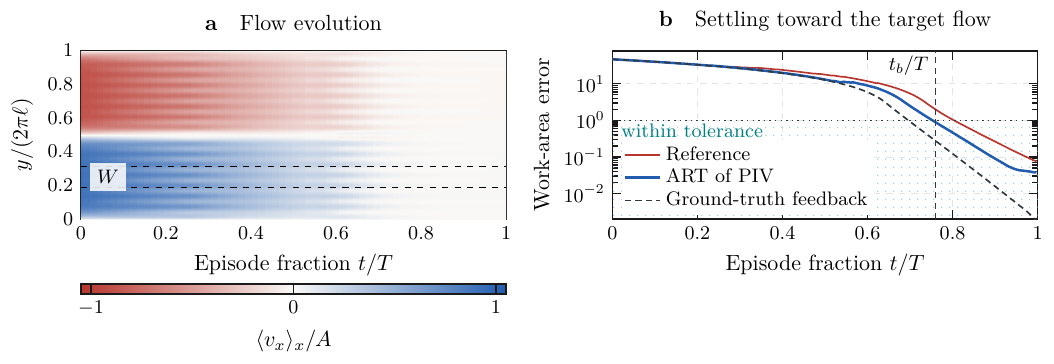}
  \par
  \includegraphics[width=\linewidth]{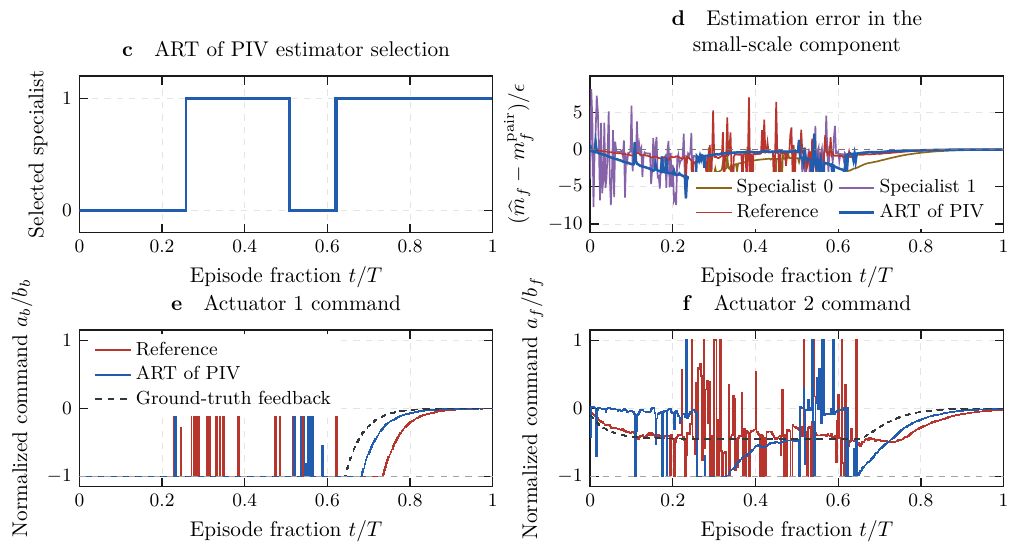}
  \par
  \includegraphics[width=\linewidth]{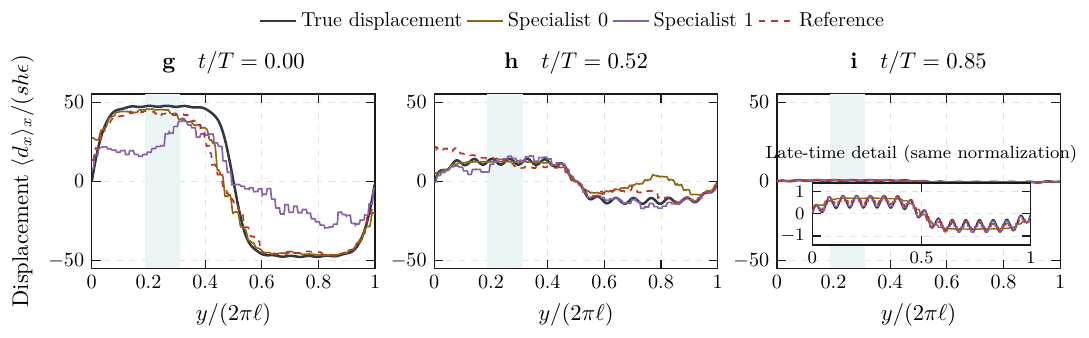}
  \caption{Flow regulation for one test initial condition.
    Time $t$ and transverse position $y$ are normalized by episode duration
    $T$ and domain width $2\pi\ell$, with length scale $\ell$;
    $\langle\cdot\rangle_x$ averages over the streamwise coordinate $x$.
    \textbf{(a)} Mean of instantaneous streamwise velocity $v_x$, normalized by
    nominal initial amplitude $A$; dashed lines bound the work area $W$ in $y$.
    \textbf{(b)} True root-mean-square velocity error over $W$ relative to the
    target, divided by tolerance $\quietTolerance$. The dotted region marks errors within
    tolerance; the vertical line marks nominal braking time
    $t_b/T\simeq\quietBrakingTimeRounded$.
    \textbf{(c,d)} Selected specialist and signed small-scale estimation error.
    The fitted velocity-error coefficients $\quietEstimatedError_f,m_f^{\rm pair}$
    within $W$ use estimated and exact pair displacements, respectively.
    \textbf{(e,f)} Actuator commands $a_b,a_f$ divided by magnitude bounds
    $b_b,b_f$; $\pm1$ indicates saturation.
    \textbf{(g--i)} Mean of streamwise pair displacement $d_x$ in pixels, divided
    by $s h\quietTolerance$, with camera calibration $s$ in pixels per unit
    length and image interval $h$. Shading marks $W$'s $y$ extent.
    Panels~\textbf{a, c, d, g--i} use the \ourmethod trajectory, with identical image
    pairs for all estimators; \textbf{b, e, f} follow each method's own trajectory.}
  \label{fig:quieting:mechanism}
\end{figure}

We consider a periodic square box of fluid flowing horizontally. The flow is
approximately the sum of two components. The \emph{large-scale component} forms two regions moving in
opposite directions; the \emph{small-scale component} alternates in direction
over much shorter vertical distances (\cref{fig:quieting:mechanism}, panel~\textbf{a}).
Initially, the flow is almost entirely large-scale, with nominal amplitude $A$;
the small-scale component starts negligibly weak.
The work area $W$ is a
relatively small
square within the domain, and the task is to reduce the flow there to a weak
prescribed target and maintain it at that target: \emph{local velocity regulation}.
A fixed background force maintains the
weak target flow, so holding it requires no control effort once it is reached.
Two actuators apply forces $\quietBroadForce$ and $\quietFineForce$
throughout the box. Their spatial patterns are fixed, and a controller sets
their strengths from camera images through the commands $a_b$ and $a_f$,
respectively.
Here, the subscripts $b$ and $f$ stand for broad and fine, referring to the
large-scale and small-scale components, respectively.
The broad actuator slows the large-scale component but, as a side effect, also
drives the small-scale component. The fine actuator acts only on the small-scale
component and can counteract this additional forcing. The images therefore change during an episode, from fast
large-scale motion to slow small-scale motion, so the estimator best suited
to them may change too.
Since this flow model serves only to create such a changing setting, we defer
its equations to \cref{sec:appendix:quieting}, together with the controller
and metric definitions, the training details, and a parameter sensitivity
analysis.

\paragraph{Estimators}
We compare \ourmethod with three fixed estimators and two idealized feedback
sources.

\ourmethod chooses, on each image pair, between two \dis configurations that
we call \emph{specialists}: specialist~0 uses larger patches at half
resolution, specialist~1 smaller patches at full resolution
(\cref{tab:quieting:library} in \cref{sec:appendix:quieting}). Two baselines
use a single specialist on every pair. The third, the \emph{reference}, is a
separately tuned \dis configuration that \ourmethod cannot select. We chose it because it
outperformed both specialists in closed loop on separate episodes
(\cref{tab:quieting:oracle-outcomes} in \cref{sec:appendix:quieting}); on the
test episodes, specialists~0 and~1 alone have mean cost differences of
$\quietArtFixedZeroCostDelta$ and
$\quietArtFixedOneCostDelta$ pp relative to it (\cref{tab:quieting:outcomes}).

The two idealized feedback sources are the \emph{whole-frame oracle} and
\emph{ground-truth feedback}. Both require access to the ground-truth flow
field and cannot be deployed in practice. The whole-frame oracle picks, on each pair,
the specialist
with lower whole-frame \gls{acr:epe}; it is what \ourmethod would do if its
selection were perfect according to its training criterion, which measures
image error, not closed-loop cost. Ground-truth feedback replaces the
estimate with the true velocity at each camera pixel, averaged over the
interval between the two images, \ie the measurement a perfect estimator
would provide.

\begin{table}[!t]\centering\small
  \caption{Local velocity regulation outcomes on \quietArtTestCount{} test episodes,
    starting from the same initial conditions for all feedback sources.
    Differences in cost and fraction of time within tolerance are in pp relative to the reference
    in each episode: negative $\quietCostDelta$ and positive
    $\quietUsableTimeDelta$ are better. \ourmethod intervals are min--max
    ranges of the episode means across \quietArtSeedCount{} training
    initializations. The number of successful episodes for \ourmethod
    is the same for every initialization. Bold marks the best deployable feedback
    for each metric; oracle and ground-truth feedback are reference results.}
  \label{tab:quieting:outcomes}
  \begin{tabular}{lrrr}\toprule
    Feedback & $\quietCostDelta$ (pp) & $\quietUsableTimeDelta$ (pp) & \shortstack{Successful\\episodes}\\\midrule
    \quietArtTableOutcomesRowAColA & \quietArtTableOutcomesRowAColB & \quietArtTableOutcomesRowAColC & \quietArtTableOutcomesRowAColD \\
    \quietArtTableOutcomesRowBColA & \quietArtTableOutcomesRowBColB & \quietArtTableOutcomesRowBColC & \quietArtTableOutcomesRowBColD \\
    Reference & \quietArtTableOutcomesRowCColB & \quietArtTableOutcomesRowCColC & \quietArtTableOutcomesRowCColD \\
    \ourmethod & \textbf{\boldmath \quietArtTableOutcomesRowDColB} & \textbf{\boldmath \quietArtTableOutcomesRowDColC} & \textbf{\boldmath \quietArtTableOutcomesRowDColD} \\
    Whole-frame oracle & \quietArtTableOutcomesRowEColB & \quietArtTableOutcomesRowEColC & \quietArtTableOutcomesRowEColD \\
    Ground-truth feedback & \quietArtTableOutcomesRowFColB & \quietArtTableOutcomesRowFColC & \quietArtTableOutcomesRowFColD \\
    \bottomrule
  \end{tabular}
\end{table}

\paragraph{Method}
All comparisons use the same controller gains, camera, initial conditions,
and tracer realizations, and differ only in how the displacement field is
obtained.

At each update time $t_n$, the estimator converts the latest image pair,
rendered with \synthpix \citep{terpin2025synthpix}, into a displacement field
$d_n$ in pixels. The images are separated by the control period $h$.
Dividing the displacement by camera calibration $s$ (pixels per unit length)
and by $h$ gives a velocity estimate. Within the work area, we measure errors in the
large-scale and small-scale components relative to the target flow. A fixed
proportional--derivative controller uses these two errors to command the
actuators, holding each command until the next update.

Even with perfect feedback, overcoming the flow's inertia can take a substantial
part of the episode. The \emph{nominal braking time} $t_b$ estimates the time needed
to slow the nominal initial flow to the target when the first actuator
is held at its actuation limit, with the command chosen to reduce the
large-scale velocity component.
Relative to the episode duration $T$, it is
$t_b/T\simeq\quietBrakingTimeRounded$ (\cref{fig:quieting:mechanism}, panel~\textbf{b}), leaving
the remaining time for settling near the target.

We train the \ourmethod estimator-selection agent by supervised learning
from simulated image pairs. We sample \quietArtTrainCount{} combinations of
initial flow conditions and tracer realizations for training and a disjoint set
of \quietArtValidationCount{} combinations for validation. For each combination,
we run the fixed controller with each of the three fixed estimators and both
idealized feedback sources, recording the rendered image pairs along every
trajectory.

As in \cref{sec:results:accuracy}, we label each pair with the specialist
whose displacement estimate has lower whole-frame \gls{acr:epe} against the
ground truth, and select the saved agent with the lowest validation
\gls{acr:aepe}. We deliberately do not restrict this
criterion to the work area $W$. Doing so would tailor the selector to this
particular controller, whereas we want to test whether the general-purpose
accuracy gains of \ourmethod carry over to closed-loop control unchanged.
The criterion also scores each image pair independently, rewarding accuracy
but not continuity over time. In feedback control, however, a state estimate
should be not only accurate but also free of abrupt changes, which propagate
to the commands and are amplified by derivative action
\citep{astrom2021feedback}; switching between estimators can introduce such
changes, and exploiting the temporal structure of the image sequence could
avoid them.

We repeat agent training with \quietArtSeedCount{} independent network
initializations and evaluate each selected agent in closed loop on the same
\quietArtTestCount{} test episodes, unused in training or validation.

\paragraph{Metrics}
We assess regulation using the true\footnote{Here, ``true'' means computed with access to the
ground-truth flow field. In a real-world experiment, this error may instead
be evaluated using a very accurate but slower estimator.}
root-mean-square velocity error relative to the target over $W$.
We denote by $\quietUsableTime$ the fraction of the episode for which this
error is within the prescribed tolerance $\quietTolerance$. The normalized
closed-loop cost $\quietCost$ adds a penalty for actuation effort to the
fraction of time outside tolerance. An episode counts as successful if the
error remains within tolerance throughout a prescribed final hold interval.

We report the cost difference $\quietCostDelta$ and the difference in fraction
of time within tolerance $\quietUsableTimeDelta$, in percentage points (pp)
relative to the fixed reference \dis configuration. Lower $\quietCostDelta$
and higher $\quietUsableTimeDelta$ are better; both differences are zero for
the reference itself.

\paragraph{Results}
Across \quietArtTestCount{} test episodes, \ourmethod gives mean
$\quietCostDelta\in[\quietArtCostDeltaMin,\quietArtCostDeltaMax]$ pp and
$\quietUsableTimeDelta\in[\quietArtUsableDeltaMin,\quietArtUsableDeltaMax]$ pp.
These ranges span the results from the \quietArtSeedCount{} training initializations
(\cref{tab:quieting:outcomes}). The cost improvement is consistent: it occurs
in every one of the \quietArtConsistentEpisodeCount{} episodes for every initialization.
The number of episodes that remain within tolerance throughout the final
hold interval rises from \quietArtReferenceHolds{} out of \quietArtTestCount{}
for the reference to \quietArtHoldsMin{} out of \quietArtTestCount{} for every initialization.
The benefit is predominantly additional time within tolerance. The whole-frame oracle
gives $\quietCostDelta=\quietArtOracleCostDelta$ pp.

\Cref{fig:quieting:mechanism} illustrates how estimator selection adapts as
the flow slows down. Panel~\textbf{a} shows the streamwise velocity $\quietVelocity_x$:
as the large-scale velocity component decreases, the small-scale variations become relatively
more prominent. \ourmethod initially selects
specialist~0, then shifts to specialist~1 as the flow slows and the
displacements decrease, with a brief return to specialist~0 during the transition
(panel~\textbf{c}). The profiles of streamwise displacement $d_x$ show why both
specialists are useful:
specialist~1 substantially underestimates the large initial displacements
(panel~\textbf{g}), whereas specialist~0 smooths out the small-scale variations late
in the episode, which specialist~1 resolves more accurately (panel~\textbf{i}).

The reference captures the initial large-scale profile reasonably well, but
shows larger relative deviations from the true displacement in the middle
snapshot (panel~\textbf{h}). These profiles compare all estimators on the same
\ourmethod image pairs.
On these pairs, panel~\textbf{d} compares the image-based estimate $\quietEstimatedError_f$
of the small-scale velocity error relative to the target with the
corresponding value $m_f^{\rm pair}$ obtained from ground-truth displacements
between the two image times. Their
difference, divided by the velocity tolerance $\quietTolerance$, measures the
feedback estimation error. This comparison also shows that
\ourmethod does not always select the specialist that most accurately estimates
the small-scale velocity error within the work area, consistent with training
on whole-frame \gls{acr:epe}.
Once it settles on specialist~1, however, its feedback estimation error is
several times smaller than that of specialist~0 on the same pairs.

Panels~\textbf{e, f} show each method's commands $a_b$ and $a_f$ during its own control run,
normalized by their respective magnitude limits $b_b$ and $b_f$; values of
$\pm1$ mean that the command has reached its actuator limit.
With reference feedback, the first actuator's command repeatedly changes sign
in the middle of the episode, interrupting its action to slow the large-scale
flow. \ourmethod keeps the command directed against this flow more consistently
over the same interval. Both
image-based feedback sources produce fluctuations in the commands, especially
for the second actuator, but the \ourmethod commands decrease in magnitude
earlier as the flow approaches the target. Correspondingly, \ourmethod brings
the work-area error below tolerance earlier than the reference and keeps it
there (panel~\textbf{b}), while ground-truth feedback brings the error below
tolerance earlier still. This example connects the
changing displacement scales and estimator choices to the additional time
within tolerance observed across the test episodes.

\subsubsection{Flow tracking with a fluidic pinball}
\label{sec:afc:channel}

\begin{figure}[p]\centering
  \includegraphics[width=\linewidth]{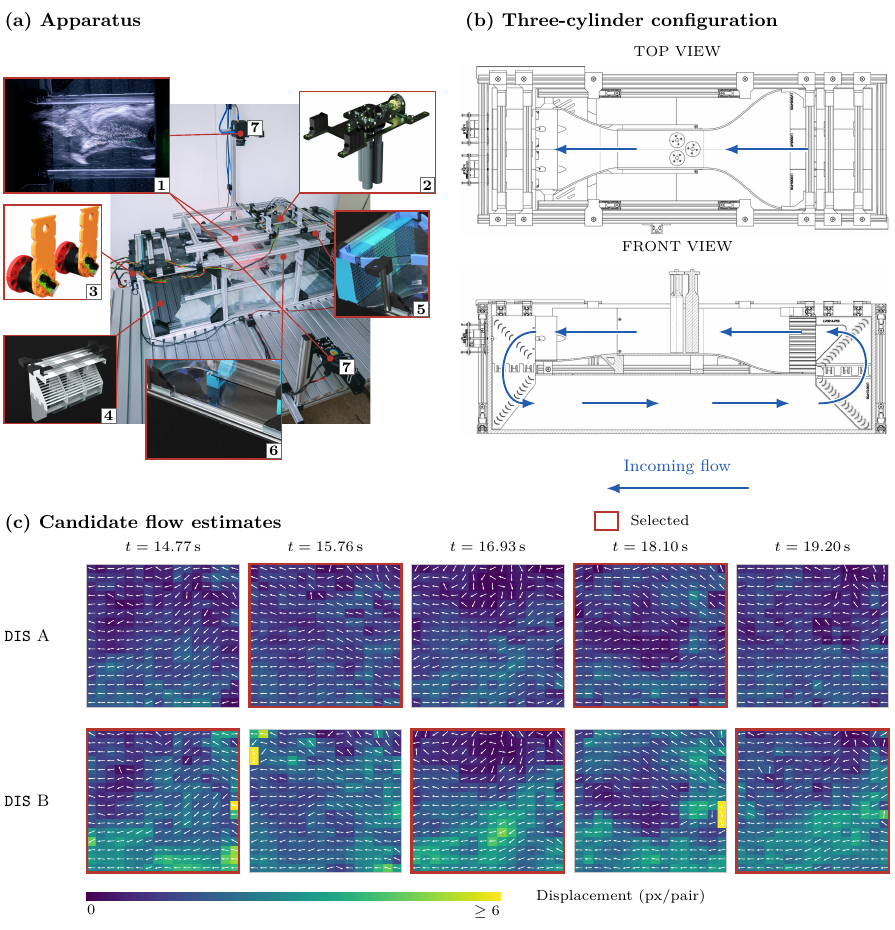}
  \caption{Water-channel apparatus and flow-estimator selection.
    \textbf{(a)} Updated tabletop channel based on \citet{terpin2025ff}, with
    component photographs and renders from
    \href{https://activefluidcontrol.com}{Active Fluid Control}.
    Insets identify the illuminated test section (\textbf{1}), the three-cylinder actuation assembly
    (\textbf{2}), external propulsion motors (\textbf{3}), turning vanes
    (\textbf{4}), honeycomb (\textbf{5}), and contraction (\textbf{6}).
    The camera and laser are labelled \textbf{7} directly on the photograph;
    connectors link inset \textbf{1} to both devices and the test section.
    Each of the \channelCylinderCount{} cylinders is driven
    independently; the \channelPropellerCount{} propulsion motors circulate
    water through magnetic couplings across the glass.
    \textbf{(b)} Updated top and front CAD views of the upstream three-cylinder
    arrangement. Blue arrows indicate the water circulation direction.
    \textbf{(c)} Both \dis tunings evaluated on five saved image
    pairs from training episode~\pinballTrainingEpisode{}, at the displayed
    episode times (approximately one second apart). Red outlines mark the
    recorded \ourmethod choices, switching B--A--B--A--B. Fields use the
    $16\times16$ grid supplied to the controller; colour encodes displacement
    magnitude and equal-length arrows show direction.}
  \label{fig:tasks}
\end{figure}

We next study \emph{flow tracking}: controlling a velocity field to follow a
prescribed spatial pattern that changes over time. Spatial control of fluid
motion is useful for localized reagent delivery \citep{juncker2005multipurpose},
cell culture under distinct flow conditions \citep{zhang2014microfluidic}, and
hydrodynamic trapping of particles \citep{tanyeri2010hydrodynamic}.

For this example, we consider an idealized flushing task followed by local
quieting. We first request leftward flow throughout the observed region,
representing a flushing phase. We then request a stationary upper half while
maintaining the same leftward flow in the lower half. This represents creating
a quiet work area while preserving fluid transport alongside it.

We implement this task on an updated version of the tabletop circulating
water channel developed by \citet{terpin2025ff}.
The test section contains a fluidic pinball, a configuration of three
independently rotating cylinders
(\citealp{cornejomaceda2021stabilization}; see also \cref{fig:tasks}\textbf{(a,b)}).
Here, \ourmethod selects between complementary \dis tunings and supplies
flow feedback to a \dreamerVThree controller trained directly on the physical
setup for flow tracking.
Training directly on the apparatus addresses both the computational cost and
the modeling difficulty of this setting. Repeated high-fidelity,
three-dimensional control rollouts are expensive, and the predicted wake
depends on how channel confinement and boundary conditions are represented
\citep[Section~2.1 and Appendix~A.2]{terpin2025ff}.
Physical interaction supplies training data that include these boundary
effects and the actual hardware dynamics, without first constructing and
calibrating a fluid simulator. This also places \ourmethod inside the learning
loop, where its estimates must remain useful as the controller changes the flow.

\begin{remark}
  We also use this same physical setup, with the limited data available from
  physical experiments, to illustrate a different use of \ourmethod:
  selecting regions of interest with coherent rotation or shear while
  retaining a common flow estimator (\cref{fig:channel:regions}).
  We describe the library and our labels
  in \cref{sec:appendix:water-channel:tuning}, and report the training and
  validation of the region-selection agent in
  \cref{sec:appendix:water-channel:selector}.
\end{remark}

\paragraph{Method}
\providecommand{\pinballTrackingControlHz}{30}
\providecommand{\pinballTrackingEpisodeS}{60}
\providecommand{\pinballTrackingPhaseS}{30}
\providecommand{\pinballTrackingResetS}{60}
\providecommand{\pinballTrackingAverageCount}{3}
\providecommand{\pinballTrackingTrainEpisodes}{20}
\providecommand{\pinballTrackingEvalEpisodes}{10}
\providecommand{\pinballTrackingFlowScale}{4}
\providecommand{\pinballTrackingCylinderLimit}{150}
\providecommand{\pinballTrackingEvaluationEpisode}{1}
\providecommand{\pinballTrackingFlushingTime}{15.00}
\providecommand{\pinballTrackingFlushingReward}{-0.398}
\providecommand{\pinballTrackingQuietingTime}{45.00}
\providecommand{\pinballTrackingQuietingReward}{-0.414}

\begin{figure}[p]\centering
  \includegraphics[width=\linewidth]{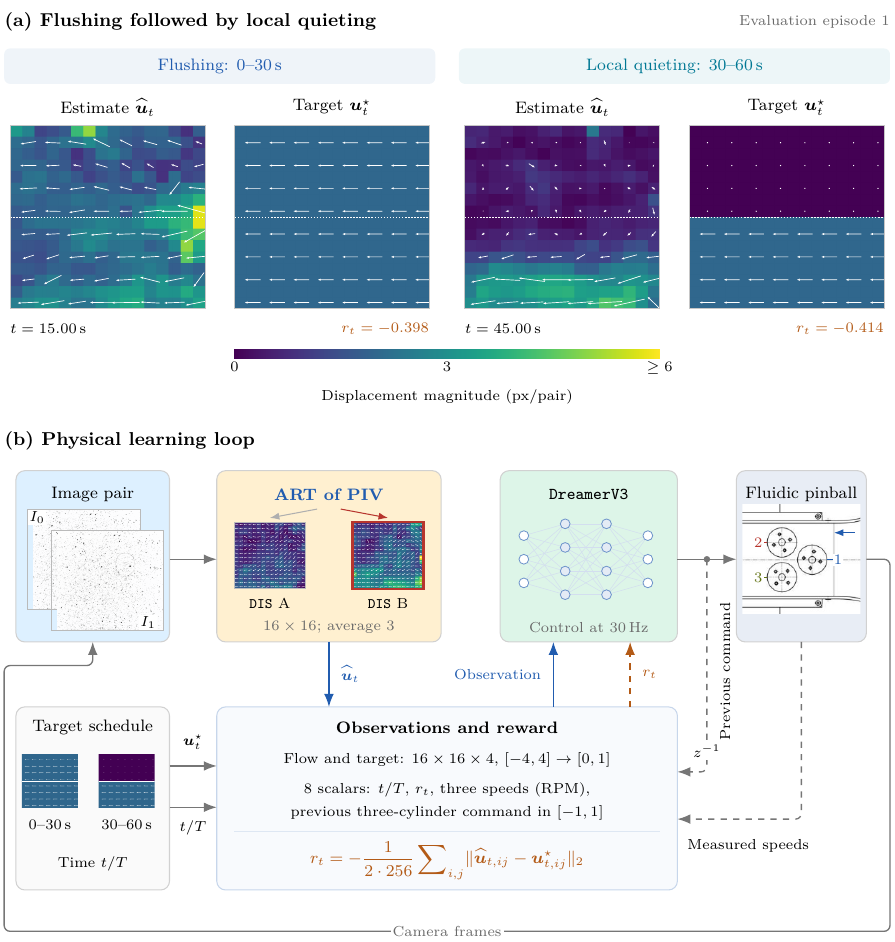}
  \caption{Physical flow tracking with \ourmethod feedback.
    \textbf{(a)} Targets and recorded $16\times16$ estimates from the first
    evaluation rollout, nearest the phase midpoints
    $t=\pinballTrackingFlushingTime$ and
    $\pinballTrackingQuietingTime\,\mathrm{s}$.
    Flushing requests $(-2,0)$ pixels per pair throughout; local quieting
    changes only the upper half to $(0,0)$.
    Estimates and targets share the displacement-magnitude colour scale
    of \cref{fig:tasks}; colours saturate outside the displayed range.
    Arrows retain a common displacement scale and show every second grid
    point in each direction; dots denote zero vectors and dashed lines
    separate the halves.
    \textbf{(b)} The fixed \ourmethod selector chooses \dis~A or \dis~B;
    downsampling and a \pinballTrackingAverageCount{}-field running average
    produce $\channelEstimatedFlow_t$. This estimate and the target supply
    both the observations and reward. The controller receives four scaled
    image channels and the eight scalars listed in the box. The reward uses
    the fields before observation scaling.
    Dashed grey paths carry measured speeds and the previous command
    ($z^{-1}$ denotes a one-step delay); the dashed orange arrow supplies
    the learning reward, also included in the observations.
    \dreamerVThree commands the three cylinder speeds at
    \pinballTrackingControlHz{}\,Hz. The image pair and candidate thumbnails
    correspond to the first snapshot in \cref{fig:tasks}; image intensities
    are inverted for display, and the red outline identifies the selected
    tuning. The network is schematic; the pinball inset crops the top view
    in \cref{fig:tasks}\textbf{(b)}, retaining the cylinder numbering and
    incoming flow from the right.
    These snapshots illustrate the spatial task and its measured error.}
  \label{fig:channel:loop}
\end{figure}

We control the three cylinder rotation rates continuously within
$\pm\pinballTrackingCylinderLimit{}\,\mathrm{RPM}$, while holding the
propellers at $-200$ and $200\,\mathrm{RPM}$.
At each \pinballTrackingControlHz{}\,Hz update, \ourmethod selects a
\dis tuning for the latest image pair $\imagePair$.
\Cref{fig:tasks}\textbf{(c)} shows both candidate estimates and the recorded
selections during training.
We reduce the selected displacement field to $16\times16$ and average the latest
\pinballTrackingAverageCount{} fields to obtain the feedback
$\channelEstimatedFlow_t$.
Each \pinballTrackingEpisodeS{}\,s episode requests
$(u,v)=(-2,0)$ throughout the grid for
\pinballTrackingPhaseS{}\,s, followed by $(0,0)$ in the displayed upper
half while retaining $(-2,0)$ below (\cref{fig:channel:loop}\textbf{(a)}).
Displacements remain in source-image pixels per pair.
Between episodes, we stop the cylinders and allow
\pinballTrackingResetS{}\,s of settling with the propellers running.

We construct a library of two complementary tunings
(\cref{fig:channel:tuning}) under a common
\channelDisLatencyBudgetMs{}\,ms budget: \dis~A searches at native
resolution to retain fine spatial detail, whereas \dis~B uses a
coarse-to-fine pyramid to accommodate larger particle displacements.
We assign our preferred tuning to recordings from prescribed motor settings
and train \ourmethod on the resulting image-pair labels, using
cross-entropy and photometric augmentation as in \cref{sec:method:learning}.
The classifier receives only the images; separate validation recordings
select its checkpoint by agreement with our labels.
The library, the rationale for our choices and its quantitative consistency
check, and the collection and training procedures are detailed in
\cref{sec:appendix:water-channel:tuning,sec:appendix:water-channel:selection}.

\newcommand{\pinballValidationColourMax}{6}
\newcommand{\pinballValidationVorticityMax}{0.5}
\newcommand{\pinballValidationSelectedPairOrdinal}{900}
\newcommand{\pinballValidationPairsPerRecording}{1799}
\newcommand{\pinballValidationDisSequenceCount}{5}
\newcommand{\pinballValidationDisSequenceFirst}{898}
\newcommand{\pinballValidationDisSequenceLast}{902}
\newcommand{\pinballValidationRoiSequenceCount}{3}
\newcommand{\pinballValidationRoiSequenceFirst}{899}
\newcommand{\pinballValidationRoiSequenceLast}{901}

\begin{figure}[p]\centering
  \includegraphics[width=\linewidth]{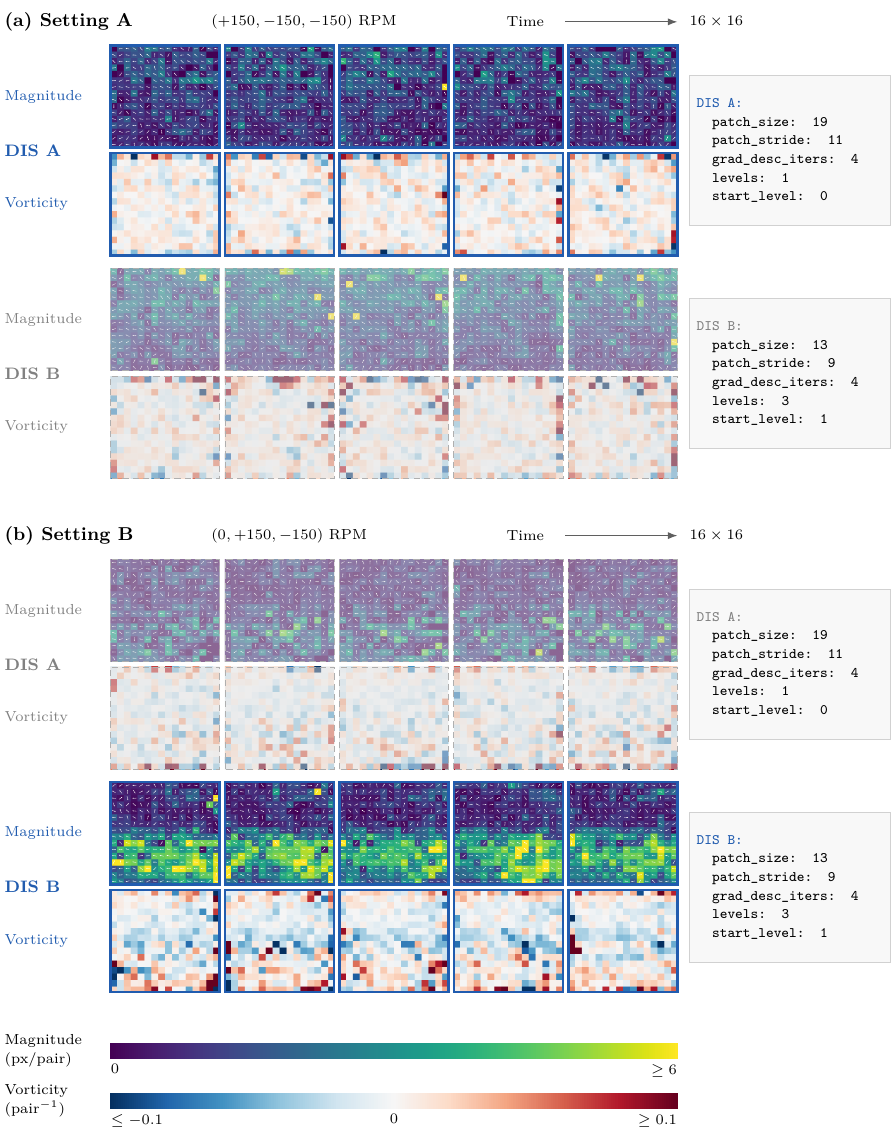}
  \caption{\dis tuning-selection pool on separate validation recordings.
    \textbf{(a,b)} Settings A and B; headers give
    signed commands for cylinders $(1,2,3)$. Columns advance through the
    consecutive pairs \pinballValidationDisSequenceFirst--\pinballValidationDisSequenceLast{}
    of \pinballValidationPairsPerRecording{}, chosen by temporal position alone.
    Each tuning has magnitude and vorticity rows. Fields are bilinearly reduced
    to $16\times16$ without rescaling displacement. Blue outlines identify the
    \ourmethod choice at checkpoint $\pinballDisSelectedStep$; other members
    are greyed and dashed. Colour scales are shared across settings and tunings.
    Magnitude is in source-image pixels per pair; vorticity is displacement
    curl per pair, positive counterclockwise in the displayed view.
    Equal-length arrows indicate direction; fields are reflected vertically
    for display.}
  \label{fig:channel:tuning}
\end{figure}

\begin{remark}[Library and task design]
  \label{rem:channel:tracking-tasks}
  This library encodes our choices for the prescribed settings; it
  need not cover every flow attainable with the apparatus, and it may be suboptimal
  even for this task. We keep this example deliberately simple to illustrate
  the method and its integration with a learning-based controller, rather than to optimize performance.
  In fact, future work may choose the collection settings and estimator tunings to match the
  intended control task. The regimes in \cref{fig:channel:tuning} suggest that
  the present library may be better suited to maintaining a quiet observed
  area and intermittently flushing only its lower half. Other cylinder
  settings and expert preferences could guide the construction of a library
  for different targets. More broadly, this two-phase example motivates
  richer tracking tasks on the same apparatus: repeated target switches,
  flushing different regions, dynamic target selection, flushing on demand to reject
  disturbances, more complex tracking scenarios.
  Larger libraries combining additional estimators and tunings
  could cover the corresponding range of flow regimes.
  Overall, this task design flexibility renders this setup and \ourmethod
  a useful testbed for future work on learning-based active flow control.
\end{remark}

With the library and selector fixed, we train a \dreamerVThree controller
\citep{hafner2025mastering} directly on the apparatus, using the feedback
loop in \cref{fig:channel:loop}\textbf{(b)}.
Its convolutional encoder receives a $16\times16\times4$ tensor containing
both components of the estimated and target fields, each mapped from
$[-\pinballTrackingFlowScale,\pinballTrackingFlowScale]$ pixels per pair to
$[0,1]$ with clipping. Its vector input contains eight scalars: episode time
$t/T$ ($T=\pinballTrackingEpisodeS\,\mathrm{s}$), the current tracking reward,
the three measured cylinder speeds (RPM), and the three components of the
previous command in $[-1,1]$.
The reward is the negative mean pointwise Euclidean distance
between $\channelEstimatedFlow_t$ and $\channelTargetFlow_t$, divided by
the target displacement span of $2$ pixels per pair, as in
\cref{sec:appendix:water-channel:training}. Thus both feedback and reward use the same
averaged field. We train for \pinballTrackingTrainEpisodes{} episodes
with one initialization, then evaluate the final controller in
\pinballTrackingEvalEpisodes{} rollouts without further learning.

\paragraph{Results}
We first evaluate whether \ourmethod reproduces our preferences.
On the separate validation recordings used for checkpoint selection
(\cref{sec:appendix:water-channel:selector}), the
selected tuning classifier reproduces all \pinballDisValPairs{} labels
(\pinballDisValidationAccuracyPct{}\,\%); the region classifier reaches
\pinballRoiValidationAccuracyPct{}\,\%
(\cref{tab:channel:selector-validation}).
For tuning selection, \pinballAugmentationWithCompletedSeeds{} paired training
seeds yield test classification accuracies of
$\pinballAugmentationWithAccuracyPct\pm\pinballAugmentationWithStdPct$\,\%
with photometric augmentation and
$\pinballAugmentationWithoutAccuracyPct\pm\pinballAugmentationWithoutStdPct$\,\%
without it (mean $\pm$ population standard deviation;
\cref{tab:channel:augmentation}). The \pinballDisTestPairs{} test pairs come
from additional recordings collected \pinballTestSeparationMinutes{} minutes
after the initial collection, under nominally unchanged conditions.
Both variants can therefore learn our choices.

For control, our aim is to illustrate how \ourmethod's observations support
learning of a spatially and temporally structured behavior. We therefore
examine the evolution of the feedback fields, using the requested patterns
as a visual reference. \Cref{fig:channel:training-sequences} follows training
at regular episode intervals. Early rollouts either retain motion in the upper half
when quieting is requested or quiet that half before the target changes.
A response aligned with the phase change becomes visible during training.

\Cref{fig:channel:evaluation-sequences} shows all ten rollouts of the final
controller. Across these repetitions, the fields show the same progression:
leftward motion throughout the observed area, followed by a relatively quiet
upper half while transport continues below. The snapshots around the target
switch expose the transition. The corresponding cylinder commands and \ourmethod selections are shown
in \cref{fig:channel:training-actions,fig:channel:evaluation-actions}, using
the same episode order.

Although the \ourmethod selector remains fixed, its use of \dis~B becomes
more frequent overall during control training, and nearly exclusive in the
final rollouts (\cref{fig:channel:training-actions,fig:channel:evaluation-actions}).
Under the preferences encoded by our labels, B therefore appears better
suited to the flows visited by the learned controller, a choice that was not
obvious a priori from the prescribed targets. Exploratory actions during
training also produce flows for which the selector chooses A. This suggests
that adaptive estimation can be useful throughout controller learning, even
when the eventual policy largely visits flow regimes served by a single
estimator.

These qualitative results demonstrate that \ourmethod can provide flow
feedback for training a controller directly on the physical apparatus.
Our deliberately simple configuration leaves room for future work to refine
both estimation and control for closer tracking of the prescribed fields
(\cref{rem:channel:tracking-tasks}).

\providecommand{\pinballSequenceTrainingEpisodeList}{1, 3, 5, 7, 9, 11, 13, 15, 17, and 19}
\providecommand{\pinballSequenceTrainingRowCount}{10}
\providecommand{\pinballSequenceEvaluationRowCount}{10}

\begin{figure}[p]\centering
  \includegraphics[width=\linewidth]{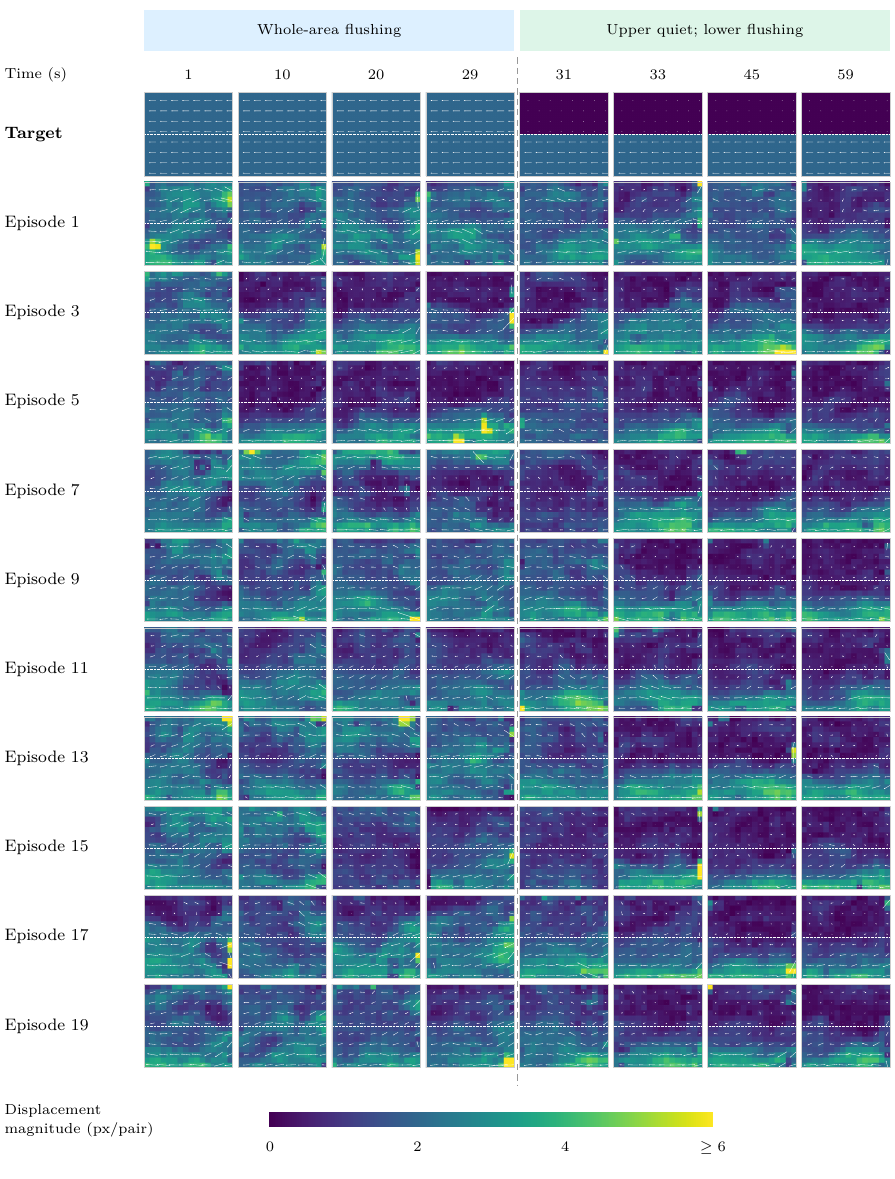}
  \caption{Development of flow tracking with \ourmethod feedback.
    Rows show episodes \pinballSequenceTrainingEpisodeList{} in chronological
    order; the top row gives the target. Each column uses the recorded field
    nearest the displayed time. The vertical dashed line separates the two
    target phases. Tiles show the $16\times16$ feedback after the online
    \pinballTrackingAverageCount{}-field average; colours, arrow scales,
    and display orientation match \cref{fig:channel:loop}.
    Horizontal dotted lines separate the upper and lower halves.}
  \label{fig:channel:training-sequences}
\end{figure}

\begin{figure}[p]\centering
  \includegraphics[width=\linewidth]{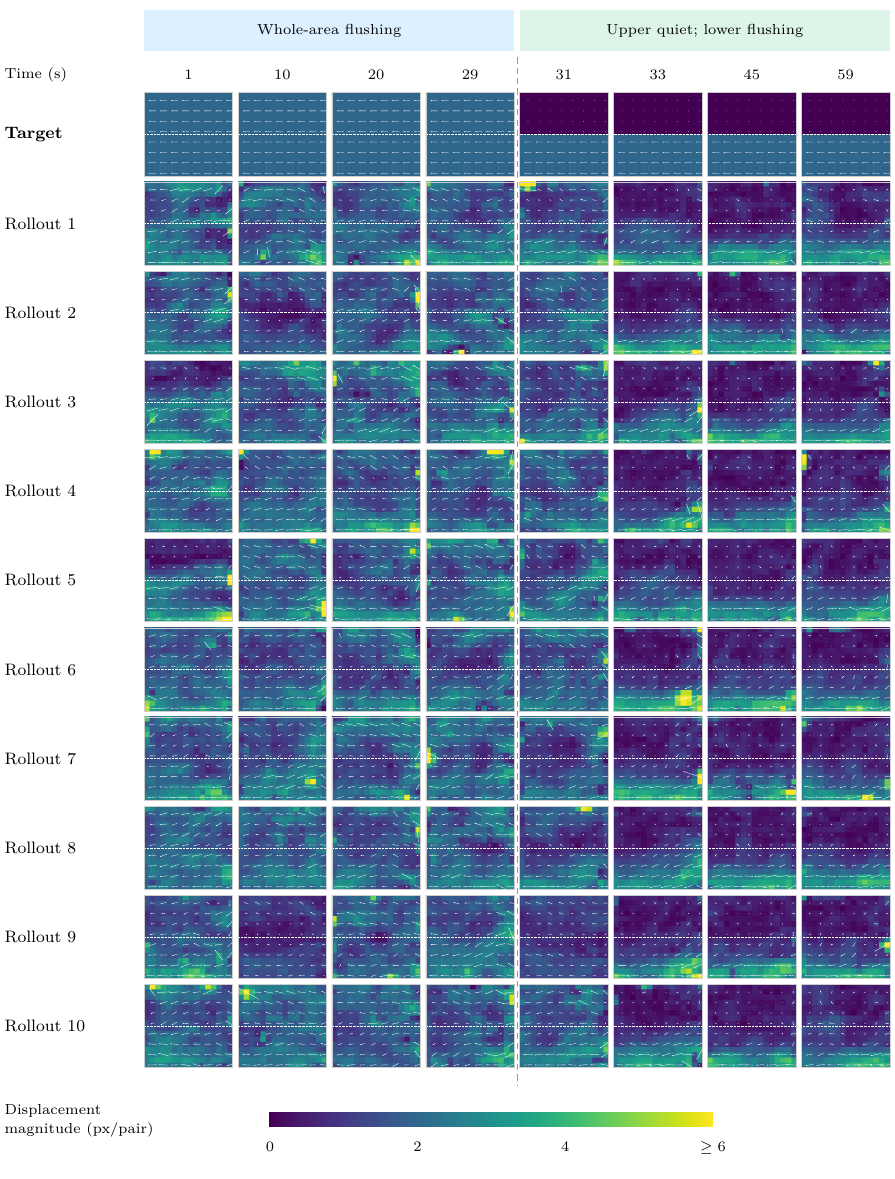}
  \caption{Repeatability of the final controller with \ourmethod feedback.
    We show all \pinballSequenceEvaluationRowCount{} evaluation rollouts in
    acquisition order, one per row, using the final checkpoint without
    learning updates. The target row, snapshot times, and rendering are
    shared with \cref{fig:channel:training-sequences}.
    The columns around the target switch show the transition; subsequent
    snapshots repeatedly show a relatively quiet upper half alongside
    continued transport below.}
  \label{fig:channel:evaluation-sequences}
\end{figure}

\begin{figure}[p]\centering
  \includegraphics[width=\linewidth]{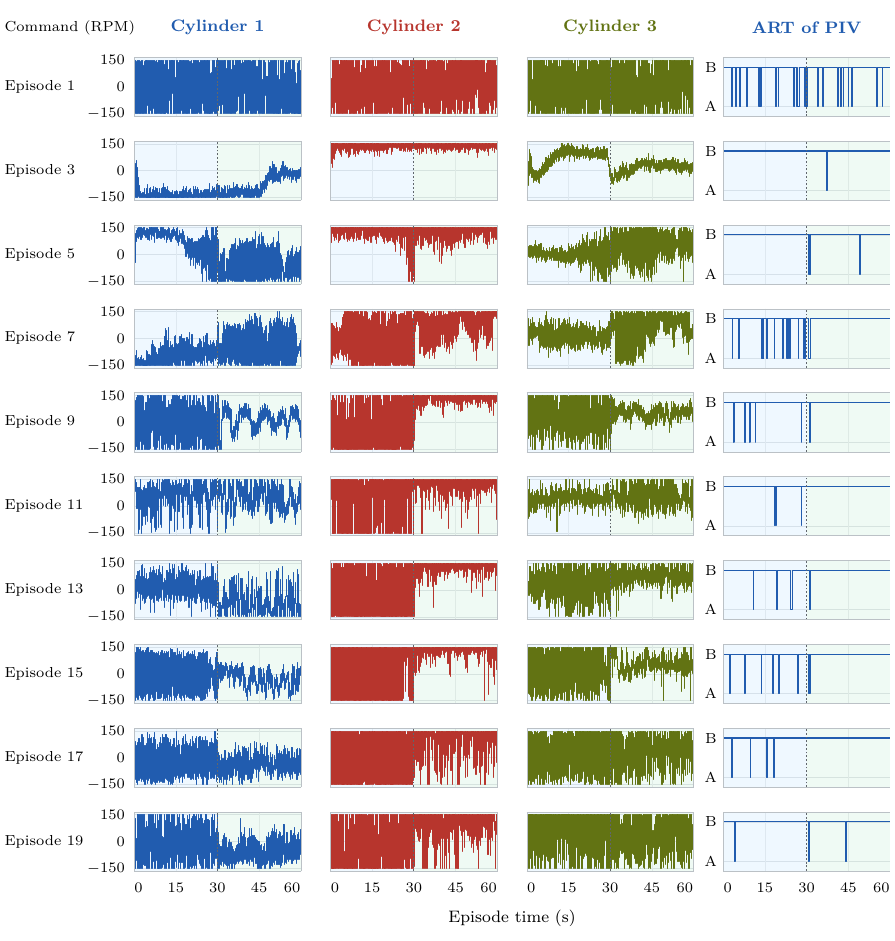}
  \caption{Cylinder commands and \ourmethod selections during flow-tracking training.
    Rows match episodes \pinballSequenceTrainingEpisodeList{} in
    \cref{fig:channel:training-sequences}. The first three columns follow the
    cylinder numbering and colours of \cref{fig:channel:loop}, showing
    \dreamerVThree's commands in RPM on a common vertical scale.
    The final column shows the DIS tuning selected by \ourmethod for each
    recorded image pair: A or B. All traces use their recorded episode times,
    without temporal smoothing. Pale blue and green
    backgrounds indicate whole-area flushing and upper-half quieting with
    continued lower-half flushing, respectively; dashed lines mark the target
    switch at \pinballTrackingPhaseS{}\,s. Step traces hold values between
    logged updates, leaving gaps longer than two nominal update periods
    unconnected.}
  \label{fig:channel:training-actions}
\end{figure}

\begin{figure}[p]\centering
  \includegraphics[width=\linewidth]{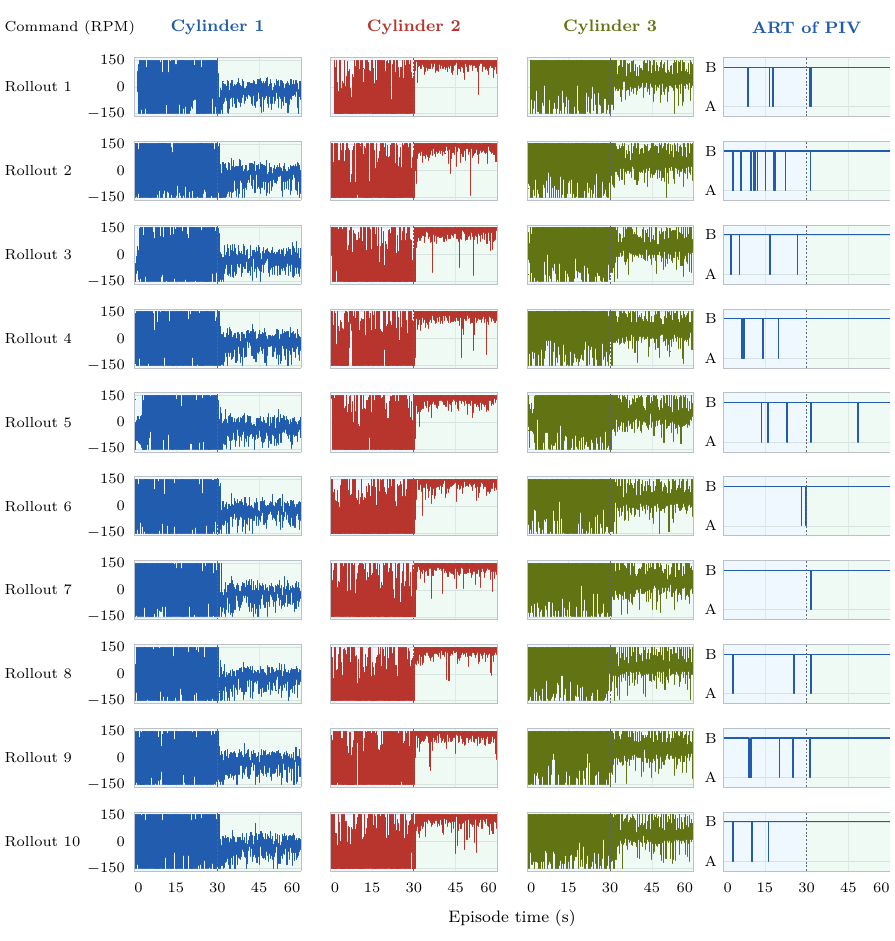}
  \caption{Cylinder commands and \ourmethod selections in the final controller's
    \pinballSequenceEvaluationRowCount{} evaluation rollouts.
    Rows correspond to \cref{fig:channel:evaluation-sequences}, in acquisition
    order. The last column reports the recorded DIS A/B choices.
    Cylinder colours, command scales, phase shading, and plotting conventions match \cref{fig:channel:training-actions}.
    The repeated change in command patterns around the target switch
    accompanies the repeated flow transition shown in
    \cref{fig:channel:evaluation-sequences}.}
  \label{fig:channel:evaluation-actions}
\end{figure}

\section{Conclusion}
\label{sec:conclusion}

We presented \ourmethod, which learns to select a configured flow-measurement
pipeline for each image pair within a specified speed budget.
Across specialized \dis and \openpiv tunings and fine-tuned \raftpiv variants we reduce
\gls{acr:aepe} by \mainRaepePctRange\,\% relative to the best single configuration at matched
speed. In simulated local velocity regulation, dynamic estimator selection also
reduces closed-loop cost and increases time within tolerance relative to a tuned fixed
\dis reference, with the same classical controller (\cref{sec:afc:quieting}).
We also deploy \ourmethod in the tabletop water channel of
\citet{terpin2025ff}, using its flow estimates to train a
\dreamerVThree reinforcement learning controller directly on the fluidic
pinball and demonstrate flow tracking (\cref{sec:afc:channel}).

\paragraph{Limitations}
The accuracy of \ourmethod is bounded by its library: it is only as good as the
best specialist available at the chosen budget, as the oracle lower bound in
\cref{fig:dis-num-models} of \cref{sec:appendix:ablations} makes explicit.
Given this dependence on the library, selection and consensus are
complementary rather than opposed. Where we commit to a single estimator, the
consensus refinement of \citet{bonomi2025consensus} reconciles the outputs of
several, and the two could be combined: one could select a small, complementary
subset and then reconcile its outputs within the chosen speed budget.
The selector also treats image pairs independently and makes one choice for
the whole image. Finally, training requires numerical scores or expert labels
(\cref{rem:method:objectives}), which may be costly to obtain on a new setup.

\paragraph{Outlook}
Two directions follow directly. First, \emph{recurrence}: because a control loop
observes the same flow over time, the agent should reason over the temporal
stream, carrying its choice and the flow estimate across image pairs rather than
deciding afresh each time. Second, \emph{automatic, self-supervised fine-tuning}:
from signals available online, such as photometric and temporal consistency and a
control performance index, the agent could keep improving \emph{after}
deployment, specializing to the specific setup and flow regime it actually faces,
without new labels or human intervention.
\synthpix \citep{terpin2025synthpix} could complement photometric augmentation by
regenerating particle-image pairs from the same flow fields with new particle
positions and seeding densities. Selection labels would need to be checked or
recomputed.
Alongside these developments, future work could expand the estimator library of
\ourmethod to include additional estimation methods and specialist configurations,
such as \lima \citep{manickathan2023lima}, consensus ADMM refinement
\citep{bonomi2025consensus}, and possibly learned refinements
with enforced constraints \citep{grontas2026pinet}.
The modular hardware, estimator library, and control algorithms supported by
\flowgames invite \emph{co-design}: jointly choosing components and control
rewards to improve closed-loop performance under resource constraints
\citep{bonomi2026codesign}.
The fluidic-pinball extensions in \cref{rem:channel:tracking-tasks} offer a
concrete setting for this co-design: adapting the library to repeated spatial
target changes, flushing on demand, and disturbance rejection.
Pretrained generalist agents could also guide pipeline selection using prior
knowledge, visual observations, and a description of the measurement objective,
potentially reducing the need for labels from each new setup.
\citet{lopez2026vision} demonstrate general-purpose vision--language agents
making measurement and control decisions in physical optics experiments.
Applying this approach to \gls{acr:piv} would require assessing both selection
reliability and compatibility with the real-time latency budget.
More broadly, we view estimator selection as a first, deliberately simple
instance of \emph{agency in the measurement loop}. The same budget-aware,
performance-driven formulation that here decides \emph{how} to estimate could
also decide \emph{what to measure} and \emph{where to look}, including where to
spend the compute budget, how to point the camera, and how to seed the flow. We believe
that treating fluid quantification as an active, agentic process, rather than a
fixed sensing front-end, is a fertile direction for future work.

\paragraph{Data availability}
The hardware designs and the data collected in this study are available at
\url{https://activefluidcontrol.com}.

\paragraph{Code availability}
The method and evaluation infrastructure are available in \flowgym, and the
deployment code is available in \flowgames.

\ifbool{anonmode}{}{\paragraph{Acknowledgements}
  The authors thank Filippo Coletti and Claudio Mucignat for insightful
  discussions, and Fabian Weber and Alexander Przybysz
  for their support with the hardware setup.

  \paragraph{Funding}
  This work was supported by the RobotX Research Program 2026.

  \paragraph{Author contributions}
  Conceptualization, visualization, and project administration: A.T.
  Methodology and formal analysis: A.T. and F.B.
  Data curation, software, validation, and investigation: A.T., F.B., and A.B.
  Resources, supervision, and funding acquisition: A.T. and R.D.
  Writing---original draft: A.T.
  Writing---review and editing: all authors.
}

\paragraph{Competing interests}
The authors declare no competing interests.

\clearpage
\appendix
\crefalias{section}{appendix}
\crefname{appendix}{Appendix}{Appendices}\Crefname{appendix}{Appendix}{Appendices}
\hypersetup{linkcolor=accentdark}
\setcounter{tocdepth}{2}
{\small\tableofcontents}
\bigskip
\bigskip

\section{Ablations}
\label{sec:appendix:ablations}

We compare training signals and agent sizes for the experiments in
\cref{sec:results:accuracy}. All ablations use the validation split of
\citet{cai2019dense}; the training-signal comparisons use \dis libraries.

\subsection{Learning signal: classification, regression, and bandit}
\label{sec:appendix:ablations:signal}
\begin{figure}[!ht]\centering
  \includegraphics[width=\linewidth]{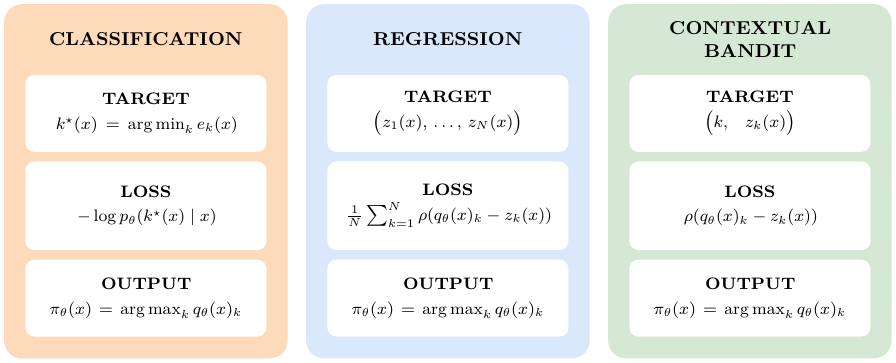}
  \caption{The three training signals for the agent. \emph{Classification}
    uses the best estimator for each image pair as its label;
    \emph{regression} fits scores to all $\numEstimators$ error-derived targets
    $\scoreTarget_k(x)$; the \emph{contextual bandit} fits only the score of
    the selected estimator. Targets negate the transformed errors, so larger
    scores indicate better estimators. Here $\huberLoss$ is the Huber loss and
    $p_{\selectorParams}$ is the softmax of the scores. Each card lists the
    training target, per-pair loss, and deployed selection rule.}
  \label{fig:training-signals}
\end{figure}

\Cref{fig:training-signals} compares classification, regression, and
contextual bandit training. We write
$e_k(x)$ for the \gls{acr:epe} of library estimator $\estimator{k}$ on image pair
$x=\imagePair$, with $k\in\nrange{\numEstimators}$. As in
\cref{sec:method:learning}, the network produces one score $\score(x)_k$
per estimator, and the deployed policy chooses
$\action=\selector(x)=\argmax_k\score(x)_k$. The three training signals differ in
how they train these scores; all use this same selection rule.

\paragraph{Classification}
We label each image pair with its best estimator,
$k^\star(x)=\argmin_k e_k(x)$, using the precomputed errors. We train with
cross-entropy $-\log p_{\selectorParams}(k^\star(x)\mid x)$, where
$p_{\selectorParams}(k\mid x)=\mathrm{softmax}(\score(x))_k$.
Maximizing this probability gives the same estimator as maximizing the score.
Labels depend only on the error ordering.

\paragraph{Regression}
The agent fits \emph{every} score $\score(x)_k$ to a target
$\scoreTarget_k(x)$ obtained by negating the estimator's error after the
chosen target transformation. With raw errors,
$\scoreTarget_k(x)=-e_k(x)$; target transformations are compared in
\cref{sec:appendix:ablations:params}. These transformations preserve the
within-pair error ordering. The per-pair regression loss used in our experiments is
$\frac{1}{\numEstimators}\sum_{k=1}^{\numEstimators}
\huberLoss\!\left(\score(x)_k-\scoreTarget_k(x)\right)$.
Here $\huberLoss$ is the Huber loss: $\huberLoss(d)=d^2/2$ for $|d|\leq 1$
and $|d|-1/2$ otherwise. We average the loss over image pairs in each training batch.
Regression must fit both the ordering and scale of the transformed targets.

\paragraph{Bandit}
Both signals above require every $e_k(x)$ at training time. The bandit variant
needs only the error of the estimator it \emph{actually ran}: after choosing
$k$, it computes $e_k(x)$ and the corresponding target $\scoreTarget_k(x)$.
It fits that score with
$\huberLoss(\score(x)_k-\scoreTarget_k(x))$, averaged over the training batch.
The raw-error target equals the reward $\reward=-e_k(x)$ defined in
\cref{sec:method:formulation}; a transformed target changes its scale while
preserving the estimator ordering.

During training, the bandit selects a uniformly random estimator with
probability $\explore$ and the highest-scoring estimator otherwise.

\paragraph{Method}
We compare validation \gls{acr:aepe} using the same greedily constructed
libraries (\cref{sec:method:library}) and image pairs
(\cref{sec:appendix:training:data}). Greedy libraries are nested, so each
size adds one estimator to the previous library.

\begin{figure}[t]\centering
  \includegraphics[width=0.7\linewidth]{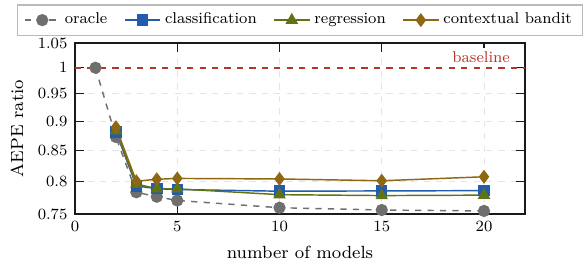}
  \caption{\gls{acr:aepe} (relative to the single-\dis baseline) versus the
    number of models in the library, on the validation split. The oracle is the
    best achievable by any selection. Regression leads at the largest
    library sizes tested.}
  \label{fig:dis-num-models}
\end{figure}

\paragraph{Results}
All three signals improve over the single-model baseline
(\cref{tab:ablation-signal,fig:dis-num-models}). The better of classification
and regression depends on library size; regression leads for the largest
libraries tested, where the bandit trails both.

\begin{table}[t]\centering
  \caption{\gls{acr:aepe} (in px; lower is better) versus the
    number of \dis models in the library, on the validation split, for each
    learning signal and the oracle (the best achievable by any selection). The
    baseline is the single \dis model; the bandit is trained only on the error
    of the estimator it ran. These are the data plotted in
    \cref{fig:dis-num-models}. Bold marks the best learned selector for each
    library size.}
  \label{tab:ablation-signal}
  \small
  \begin{tabular}{rccccc}
    \toprule
    \# models & Baseline (px) & Oracle (px) & Classification (px) & Regression (px) & Bandit (px) \\
    \midrule
    \ablationSignalRows
    \bottomrule
  \end{tabular}
\end{table}

\subsection{Parameters of the regression and bandit agents}
\label{sec:appendix:ablations:params}
\paragraph{Method}
For the bandit we sweep the initial exploration rate $\explore$, which
decays linearly to 0 over the first $4{,}000$ of the $5{,}000$ training steps
(the deployed agent does not explore); for the regression agent we vary the
target normalization. We report validation \gls{acr:aepe}.

For image pair $x$, let $\meanPairError(x)=\frac{1}{\numEstimators}
\sum_{j=1}^{\numEstimators}e_j(x)$ be the mean error across the library.
The raw, per-pair-mean, and log targets are, respectively,
$\scoreTarget_k(x)=-e_k(x)$,
$\scoreTarget_k(x)=-e_k(x)/\meanPairError(x)$, and
$\scoreTarget_k(x)=-\log(e_k(x)+\targetLogOffset)$.
Here $\targetLogOffset=10^{-8}$ is a small offset that keeps the logarithm finite
when an estimator has zero error.

\begin{figure}[t]\centering
  \includegraphics[width=0.62\linewidth]{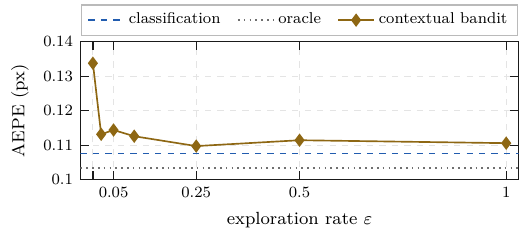}
  \caption{Exploration rate of the bandit agent on the \dis library (validation
    split, 20-model library, 3-seed mean). The bandit's \gls{acr:aepe} is flat
    for every nonzero initial exploration rate $\explore$ and degrades sharply
    without exploration. Classification (dashed) and oracle (dotted) at 20 models for reference;
    the bandit trails both, consistent with its partial-information training.}
  \label{fig:ablation-params}
\end{figure}

\paragraph{Results}
\Cref{fig:ablation-params,tab:ablation-regression-target} summarize the sweeps.
The bandit is essentially insensitive to $\explore$ once it is nonzero, and
clearly worse without exploration (\cref{fig:ablation-params}). Both per-pair-mean and log normalization
improve regression over the raw target and outperform classification at
this library size; the log target gives the lowest regression \gls{acr:aepe}
(\cref{tab:ablation-regression-target}).

\begin{table}[t]\centering
  \caption{Regression target normalization on the \dis 20-model library
    (validation split, 3-seed mean). The log target gives the lowest regression
    \gls{acr:aepe}. Classification and oracle at 20 models are included for comparison
    (\gls{acr:aepe} in px; lower is better). Bold marks the best regression target.}
  \label{tab:ablation-regression-target}
  \small
  \begin{tabular}{lccccc}
    \toprule
    & \multicolumn{3}{c}{Regression target} & \multicolumn{2}{c}{Reference} \\
    \cmidrule(lr){2-4}\cmidrule(lr){5-6}
    & None (raw \gls{acr:epe}) & Per-pair mean & Log & Classification & Oracle \\
    \midrule
    \gls{acr:aepe} (px) & \ablationRegressionRawAepeTable & \ablationRegressionMeanAepeTable
      & \ablationRegressionLogAepeTable & \ablationClassificationRefAepe
      & \ablationOracleRefAepe \\
    \bottomrule
  \end{tabular}
\end{table}

\subsection{Agent architecture}
\label{sec:appendix:ablations:arch}
At the high-rate operating points
(\mainHighRateRange\,\estimatorRateUnit), the deployed agent uses a single
convolutional block (width 32) feeding a shared \gls{acr:mlp} head of width
512, with one set of head weights across all arms. For \dis at
\disTwentyHzNominalRate{} and for \raftpiv, where the agent's execution time
is negligible next to the estimator's, it uses three convolutional blocks
(widths $32/64/64$) and a separate head per arm; so do the ablations of
\cref{sec:appendix:ablations:signal,sec:appendix:ablations:params}. At the
high-rate operating points we compare the two, retraining the larger agent
on the same libraries (\cref{tab:ablation-agent}). Accuracy changes little, while standalone
agent execution time increases
(\dis $\agentOverheadDisDeployedMs\!\to\!\agentOverheadDisLargeMs$\,ms,
\openpiv $\agentOverheadOpenpivDeployedMs\!\to\!\agentOverheadOpenpivLargeMs$\,ms).
The smaller agent is more likely to finish before the selected estimator,
allowing speculation to hide its execution time
(\cref{sec:method:execution}).

\begin{table}[H]\centering
  \caption{Agent-size comparison using the same libraries
    (3-seed mean, validation split) at the three high-rate operating points.
    \gls{acr:aepe} in px; $\raepe{\ourmethod}{b}$ in \%; standalone agent
    execution time in ms, measured on an RTX 5090. The larger agent increases
    execution time with little change in accuracy. Bold marks the better value
    for each metric at each operating point.}
  \label{tab:ablation-agent}
  \small
  \begin{tabular}{lclccc}
    \toprule
    Estimator & Speed (\estimatorRateUnit) & Agent & \gls{acr:aepe} (px) & \acrshort{acr:raepe} (\%) & Agent time (ms) \\
    \midrule
    \dis & \disOneKHzNominalRateValue & deployed & \archDisOneKHzArtofpivAepeTable & \archDisOneKHzRaepePctTable & \agentOverheadDisDeployedMsTable \\
         & & larger & \archDisOneKHzLargeArtofpivAepeTable & \archDisOneKHzLargeRaepePctTable & \agentOverheadDisLargeMsTable \\
    \openpiv & \openpivOneKHzNominalRateValue & deployed & \archOpenpivOneKHzArtofpivAepeTable & \archOpenpivOneKHzRaepePctTable & \agentOverheadOpenpivDeployedMsTable \\
             & & larger & \archOpenpivOneKHzLargeArtofpivAepeTable & \archOpenpivOneKHzLargeRaepePctTable & \agentOverheadOpenpivLargeMsTable \\
    \openpiv & \openpivFourKHzNominalRateValue & deployed & \archOpenpivFourKHzArtofpivAepeTable & \archOpenpivFourKHzRaepePctTable & \agentOverheadOpenpivDeployedMsTable \\
             & & larger & \archOpenpivFourKHzLargeArtofpivAepeTable & \archOpenpivFourKHzLargeRaepePctTable & \agentOverheadOpenpivLargeMsTable \\
    \bottomrule
  \end{tabular}
\end{table}

\section{Library construction: why greedy is near-optimal}
\label{sec:appendix:library}

\Cref{sec:method:library} builds the library by greedily solving~\eqref{eq:library}:
starting from the empty set, it repeatedly adds the candidate with the largest
expected gain until the cap $\numEstimators$ is reached, at a cost of
$\landau{\abs{\pool_{\speedBudget}}\,\numEstimators}$ evaluations against the
cached rewards. We show this is near-optimal for a fixed reward and compare
it to the exact optimum in the negative-\gls{acr:epe} experiments.

\paragraph{The gain is monotone and submodular}
Fix finite rewards on the cached image pairs and budget-feasible candidates,
and choose a constant $\rewardBaseline$ no larger than any of these rewards.
For an image pair $\imagePair$, define the \emph{gain} $g_{\imagePair}(\varnothing)=0$ and
$g_{\imagePair}(\library)=\max_{\estimator{k}\in\library}\reward(\framePrev,\frameNext,k)-\rewardBaseline\ge 0$
for nonempty $\library$. Adding an estimator $\estimator{j}$ to a nonempty
set $\library$ raises this gain by
$\max\set{0,\ \reward(\framePrev,\frameNext,j)-\max_{\estimator{k}\in\library}\reward(\framePrev,\frameNext,k)}$,
which is non-negative and non-increasing as $\library$ grows. From the empty
set, the gain is $\reward(\framePrev,\frameNext,j)-\rewardBaseline$, which is no smaller. Thus
$g_{\imagePair}$ is monotone and submodular. These properties are preserved under
non-negative combinations, so
$g(\library)=\expectedValue[\imagePair]{g_{\imagePair}(\library)}$ is monotone and submodular too,
where the expectation is over the cached pairs. Maximizing the expected
oracle reward in \eqref{eq:library} is equivalent to maximizing $g$ under
the same cardinality cap. Greedy therefore attains at least $(1-1/e)$ of
the optimal gain \citep{nemhauser1978}.
For $\reward(\framePrev,\frameNext,k)=-e_k\imagePair$, the gain is the reduction in oracle error relative
to the common upper error bound $-\rewardBaseline$.

\paragraph{Greedy versus exact}
The exact optimum is obtained by enumerating subsets for small pools or by a
mixed-integer program for larger ones. On the \dis pool, for libraries of one to
20 models, the gap between the greedy and the optimal \gls{acr:aepe} is largest
at two models (\librarySelectionGreedyGapTwoPct\,\%) and at most
\librarySelectionGreedyGapFromThreeMaxPct\,\% from three models on. The exact
mixed-integer program solves in
\librarySelectionMilpMinS--\librarySelectionMilpMaxS\,s. Since refining
is this cheap, the libraries in \cref{tab:main} are the exact optima. The
library-size ablation (\cref{sec:appendix:ablations:signal}) uses the nested
greedy libraries instead.

\section{Training details}
\label{sec:appendix:training}

\paragraph{Data and splits}
\label{sec:appendix:training:data}
The offline comparisons in \cref{sec:results:accuracy} use the \gls{acr:piv} dataset of \citet{cai2019dense}:
$256\times256$ px image pairs with ground-truth flow, spanning a range of flow
regimes. We follow the split of \citet{cai2019dense}:
97\,\% of the pairs form the training set and the remaining 3\,\%
the held-out test set. For
model selection and the ablations of
\cref{sec:appendix:ablations} we further hold out a 10\,\% validation split
from the training set, and a separate 10\,\% tuning split on which the
libraries are selected (\cref{sec:method:library,sec:appendix:library}).

\paragraph{Optimization and hardware}
\label{sec:appendix:training:hardware}
For these comparisons, we train agents in \jax \citep{bradbury2018jax} with Adam
\citep{kingma2014adam} (batch size 16, learning rate $10^{-3}$ decayed
exponentially to $10^{-4}$ over $4{,}000$ steps) for $5{,}000$ steps on an
RTX 5090, and keep the checkpoint with the lowest validation \gls{acr:aepe}.
The agents of the high-rate operating points are smaller
(\cref{sec:appendix:ablations:arch}) and use batch size 64 and learning
rate $5\times10^{-3}$ decayed to $5\times10^{-4}$. We average over 3 seeds.

\paragraph{\texorpdfstring{\raft}{RAFT} fine-tuning: flow-class specialists}
\label{sec:appendix:training:raft}
Starting from a pretrained \raftpiv \citep{lagemann2021raft}, we fine-tune one specialist on
each flow subcategory of \citet{cai2019dense} for $50{,}000$ steps, varying only
the fine-tuning data; this yields a small library with one member per flow class.
We use Adam (batch size 25, gradient clipping at global norm 0.5) with a
one-cycle learning rate peaking at $10^{-5}$, and keep the checkpoint with the
lowest validation \gls{acr:aepe}.

\paragraph{Classical estimator pools}
\label{sec:appendix:training:dis}
\label{sec:appendix:training:openpiv}
We form the \dis candidate pool of 565 configurations by varying patch size,
patch stride, number of gradient-descent iterations, and pyramid levels. In
our experiments these span patch sizes $5$--$37$ px, strides $1$--$31$ px,
$1$--$21$ gradient-descent iterations, $1$--$3$ pyramid levels, and a
finest level of full or half resolution. We select the subset meeting the
speed budget as in \cref{sec:method:library}.

For \openpiv \citep{liberzon2021openpiv}, we form a pool of 1080
configurations by varying interrogation-window size ($12$--$32$ px), overlap,
search-area size, the number of passes ($1$--$3$), the correlation (circular or
linear), the sub-pixel estimator, and whether signal-to-noise validation is
applied, then select the subset meeting the speed budget as in
\cref{sec:method:library}.

\section{Statistical analysis of end-point errors}
\label{sec:appendix:epe-statistics}

We report the distribution of the \gls{acr:epe} over the test image pairs, from
the runs of \cref{tab:main} (3-seed mean).

\paragraph{Spread and tails}
\Cref{tab:epe-spread} gives the spread and tails of the \gls{acr:epe}.
\ourmethod lowers every statistic at every operating point. For \openpiv, the
largest reduction is in the tails: at $\approx$\openpivOneKHzNominalRate, the
95th percentile drops from \epeStatsOpenpivOneKHzBestNinetyFivePx{} to
\epeStatsOpenpivOneKHzRouterNinetyFivePx\,px and the \gls{acr:aepe} from
\epeStatsOpenpivOneKHzBestMeanPx{} to \epeStatsOpenpivOneKHzRouterMeanPx\,px.

\begin{table}[!ht]\centering
  \caption{Per-image-pair \gls{acr:epe} (px) on the test set (3-seed mean).
    The mean is the \gls{acr:aepe}; p90--p99 are percentiles. Bold marks the
    lower mean, median, and tail errors within each operating point.}
  \label{tab:epe-spread}
  \footnotesize
  \setlength{\tabcolsep}{3pt}
  \begin{tabular}{llcccccccc}
    \toprule
    Estimator & \shortstack{Speed\\(\estimatorRateUnit)} & Method & Mean & Std & Median & p90 & p95 & p99 & Max \\
    \midrule
    \multirow{2}{*}{\raftpiv} & \multirow{2}{*}{$\approx$\raftThirtyTwoNominalRateValue}
      & Best single & \epeStatsRaftThirtyTwoBestMeanPxTable & \epeStatsRaftThirtyTwoBestStdPx & \epeStatsRaftThirtyTwoBestMedianPxTable & \epeStatsRaftThirtyTwoBestNinetyPxTable & \epeStatsRaftThirtyTwoBestNinetyFivePxTable & \epeStatsRaftThirtyTwoBestNinetyNinePxTable & \epeStatsRaftThirtyTwoBestMaxPxTable \\
    & & \ourmethod & \epeStatsRaftThirtyTwoRouterMeanPxTable & \epeStatsRaftThirtyTwoRouterStdPx & \epeStatsRaftThirtyTwoRouterMedianPxTable & \epeStatsRaftThirtyTwoRouterNinetyPxTable & \epeStatsRaftThirtyTwoRouterNinetyFivePxTable & \epeStatsRaftThirtyTwoRouterNinetyNinePxTable & \epeStatsRaftThirtyTwoRouterMaxPxTable \\
    \addlinespace
    \multirow{2}{*}{\dis} & \multirow{2}{*}{$\approx$\disTwentyHzNominalRateValue}
      & Best single & \epeStatsDisTwentyHzBestMeanPxTable & \epeStatsDisTwentyHzBestStdPx & \epeStatsDisTwentyHzBestMedianPxTable & \epeStatsDisTwentyHzBestNinetyPxTable & \epeStatsDisTwentyHzBestNinetyFivePxTable & \epeStatsDisTwentyHzBestNinetyNinePxTable & \epeStatsDisTwentyHzBestMaxPxTable \\
    & & \ourmethod & \epeStatsDisTwentyHzRouterMeanPxTable & \epeStatsDisTwentyHzRouterStdPx & \epeStatsDisTwentyHzRouterMedianPxTable & \epeStatsDisTwentyHzRouterNinetyPxTable & \epeStatsDisTwentyHzRouterNinetyFivePxTable & \epeStatsDisTwentyHzRouterNinetyNinePxTable & \epeStatsDisTwentyHzRouterMaxPxTable \\
    \addlinespace
    \multirow{2}{*}{\dis} & \multirow{2}{*}{$\approx$\disOneKHzNominalRateValue}
      & Best single & \epeStatsDisOneKHzBestMeanPxTable & \epeStatsDisOneKHzBestStdPx & \epeStatsDisOneKHzBestMedianPxTable & \epeStatsDisOneKHzBestNinetyPxTable & \epeStatsDisOneKHzBestNinetyFivePxTable & \epeStatsDisOneKHzBestNinetyNinePxTable & \epeStatsDisOneKHzBestMaxPxTable \\
    & & \ourmethod & \epeStatsDisOneKHzRouterMeanPxTable & \epeStatsDisOneKHzRouterStdPx & \epeStatsDisOneKHzRouterMedianPxTable & \epeStatsDisOneKHzRouterNinetyPxTable & \epeStatsDisOneKHzRouterNinetyFivePxTable & \epeStatsDisOneKHzRouterNinetyNinePxTable & \epeStatsDisOneKHzRouterMaxPxTable \\
    \addlinespace
    \multirow{2}{*}{\openpiv} & \multirow{2}{*}{$\approx$\openpivOneKHzNominalRateValue}
      & Best single & \epeStatsOpenpivOneKHzBestMeanPxTable & \epeStatsOpenpivOneKHzBestStdPx & \epeStatsOpenpivOneKHzBestMedianPxTable & \epeStatsOpenpivOneKHzBestNinetyPxTable & \epeStatsOpenpivOneKHzBestNinetyFivePxTable & \epeStatsOpenpivOneKHzBestNinetyNinePxTable & \epeStatsOpenpivOneKHzBestMaxPxTable \\
    & & \ourmethod & \epeStatsOpenpivOneKHzRouterMeanPxTable & \epeStatsOpenpivOneKHzRouterStdPx & \epeStatsOpenpivOneKHzRouterMedianPxTable & \epeStatsOpenpivOneKHzRouterNinetyPxTable & \epeStatsOpenpivOneKHzRouterNinetyFivePxTable & \epeStatsOpenpivOneKHzRouterNinetyNinePxTable & \epeStatsOpenpivOneKHzRouterMaxPxTable \\
    \addlinespace
    \multirow{2}{*}{\openpiv} & \multirow{2}{*}{$\approx$\openpivFourKHzNominalRateValue}
      & Best single & \epeStatsOpenpivFourKHzBestMeanPxTable & \epeStatsOpenpivFourKHzBestStdPx & \epeStatsOpenpivFourKHzBestMedianPxTable & \epeStatsOpenpivFourKHzBestNinetyPxTable & \epeStatsOpenpivFourKHzBestNinetyFivePxTable & \epeStatsOpenpivFourKHzBestNinetyNinePxTable & \epeStatsOpenpivFourKHzBestMaxPxTable \\
    & & \ourmethod & \epeStatsOpenpivFourKHzRouterMeanPxTable & \epeStatsOpenpivFourKHzRouterStdPx & \epeStatsOpenpivFourKHzRouterMedianPxTable & \epeStatsOpenpivFourKHzRouterNinetyPxTable & \epeStatsOpenpivFourKHzRouterNinetyFivePxTable & \epeStatsOpenpivFourKHzRouterNinetyNinePxTable & \epeStatsOpenpivFourKHzRouterMaxPxTable \\
    \bottomrule
  \end{tabular}
\end{table}

\paragraph{Per image pair}
\Cref{tab:epe-pairs} compares the two on each test image pair. \ourmethod is
better on most image pairs at every operating point; ties are image pairs where
the agent selects the best single estimator.

\begin{table}[!ht]\centering
  \caption{Share of test image pairs (\%, 3-seed mean) on which \ourmethod has
    a lower (better), equal (tie), or higher (worse) \gls{acr:epe} than the
    best single estimator.}
  \label{tab:epe-pairs}
  \small
  \begin{tabular}{lcccc}
    \toprule
    Estimator & Speed (\estimatorRateUnit) & Better & Tie & Worse \\
    \midrule
    \raftpiv & $\approx$\raftThirtyTwoNominalRateValue & \epeStatsRaftThirtyTwoBetterPct & \epeStatsRaftThirtyTwoTiePct & \epeStatsRaftThirtyTwoWorsePct \\
    \dis & $\approx$\disTwentyHzNominalRateValue & \epeStatsDisTwentyHzBetterPct & \epeStatsDisTwentyHzTiePct & \epeStatsDisTwentyHzWorsePct \\
    \dis & $\approx$\disOneKHzNominalRateValue & \epeStatsDisOneKHzBetterPct & \epeStatsDisOneKHzTiePct & \epeStatsDisOneKHzWorsePct \\
    \openpiv & $\approx$\openpivOneKHzNominalRateValue & \epeStatsOpenpivOneKHzBetterPct & \epeStatsOpenpivOneKHzTiePct & \epeStatsOpenpivOneKHzWorsePct \\
    \openpiv & $\approx$\openpivFourKHzNominalRateValue & \epeStatsOpenpivFourKHzBetterPct & \epeStatsOpenpivFourKHzTiePct & \epeStatsOpenpivFourKHzWorsePct \\
    \bottomrule
  \end{tabular}
\end{table}

\section{Local velocity regulation: parameters and sensitivity}
\label{sec:appendix:quieting}

We specify the local velocity regulation experiment and assess its parameter sensitivity.

\subsection{Model and parameters}
\label{sec:appendix:quieting:model}

Let $\quietPosition=(x,y)\in[0,2\pi\ell)^2$ denote position in a periodic
square domain, where $\ell$ is a length scale, and let $t$ denote time. The
work area $W$ is a fixed square in which we measure feedback and velocity error.
We model two idealized actuators: each applies a fixed spatial
force pattern throughout the domain, and a single command sets its strength.
The actuator forces per unit mass are
\[
  \quietBroadForce(y,t)=a_b(t)\bigl(\quietBroad(y)-\kappa\quietFine(y)\bigr),
  \qquad \quietFineForce(y,t)=a_f(t)\quietFine(y),
\]
with the fixed spatial profiles
\begin{equation}\label{eq:quieting:profiles}
  \quietBroad(y)=c_b\tanh\!\left(\frac{\sin(y/\ell)}{w}\right)e_x,
  \qquad
  \quietFine(y)=c_f\sin(k_f y/\ell)e_x.
\end{equation}
Both profiles point along the streamwise unit vector $e_x$ and vary only
across $y$. The large-scale profile $\quietBroad$ pushes two wide bands in opposite
directions; $w\ell$ approximately sets the thickness of the transition layers
where the profile changes sign. The small-scale profile $\quietFine$ varies sinusoidally
with dimensionless wavenumber $k_f$. The constants $c_b,c_f$ normalize both profiles to unit
root-mean-square magnitude over $W$ (\cref{fig:quieting:forcing}).
Actuator~1 combines both patterns with fixed coupling $\kappa>0$; actuator~2
applies the small-scale profile alone. The signed commands $a_b(t)$ and $a_f(t)$ set
their respective strengths, with $|a_b(t)|\le b_b$ and $|a_f(t)|\le b_f$,
where $b_b,b_f>0$ are the maximum command magnitudes.

The initial velocity is
$\quietVelocity(\quietPosition,-h)=A\quietBroad(y)+F\quietFine(y)$,
where $A$ and $F$ are the amplitudes of the initial large-scale and small-scale
velocity components, respectively. For these initial conditions,
the velocity and all prescribed forces
point along $e_x$ and vary only across $y$, so the flow retains the form
$\quietVelocity(\quietPosition,t)=\quietVelocity_x(y,t)e_x$.
Incompressibility then holds automatically, and both nonlinear advection and
the pressure gradient vanish. The forced incompressible Navier--Stokes
equations therefore reduce to a one-dimensional diffusion equation with forcing:
\begin{equation}\label{eq:quieting:dynamics}
  \partial_t\quietVelocity(\quietPosition,t)
  =\nu\partial_{yy}\quietVelocity(\quietPosition,t)
  +\quietBroadForce(y,t)+\quietFineForce(y,t)+\quietReferenceForce(y),
\end{equation}
where $\nu$ is kinematic viscosity. The fixed background force
$\quietReferenceForce(y)=-\nu\partial_{yy}\quietReference(y)$
keeps the weak target flow $\quietReference(y)=U_r\quietBroad(y)$, of prescribed
amplitude $U_r$, steady without actuation. The controller therefore only has to
bring the flow to the target; maintaining it requires no actuator effort.

We compute the normalization constants $c_b,c_f$ of the force profiles
\eqref{eq:quieting:profiles} by numerical integration over $W$ on a sampling grid
independent of the fluid grid. For numerical integration in time, we retain
the full two-dimensional vorticity equation, with a two-thirds Fourier cutoff
and fourth-order Runge--Kutta steps of duration $\delta t$.
The ratio
$\ell^2/(\nu k_f^2T)=\quietDiffusionRatio$, where $T$ is episode duration,
means that viscosity alone damps the small-scale velocity component only slowly.
\Cref{tab:quieting:parameters,tab:quieting:library} give the parameters.

\paragraph{Coupled actuation}
Together, the two actuators apply the small-scale force profile with net amplitude
$a_f(t)-\kappa a_b(t)$. When actuator~1 slows the large-scale flow ($a_b(t)<0$),
it also applies small-scale forcing, which actuator~2 can cancel by setting
$a_f(t)=\kappa a_b(t)$.
We call $t_b=(A-U_r)/b_b$ the \emph{nominal braking time}: an estimate of the time
actuator~1 needs, with its command held at the negative actuation limit
$a_b=-b_b$, to slow the flow in the work area
from the nominal initial amplitude $A$ to the target amplitude $U_r$,
neglecting viscosity and the fixed background force.
For the nominal parameters, $t_b/T=\quietBrakingTimePrecise$.

\begin{remark}[Relation to laboratory forcing]
  Electromagnetic shallow-layer experiments apply fixed spatial force
  patterns whose strengths are set by electric current
  \citep[Sec.~2]{tithof2017bifurcations}; magnet geometry can also shape forcing
  at multiple length scales \citep{rossi2009control}. This motivates our
  separation of spatial profiles and actuator commands. The mixture of spatial scales,
  the coupling $\kappa$ between them, and the second independent command
  are numerical design choices that define a simple, controlled setting
  for studying estimator selection in closed loop.
  \Cref{sec:appendix:quieting:screens} assesses the sensitivity of the closed-loop
  results to the parameter values. Quantitatively reproducing those
  experiments requires modelling bottom friction, depth
  dependence, and wall boundaries \citep{tithof2017bifurcations}.
\end{remark}

\paragraph{Feedback and controller}
For numerical convenience, adjacent pairs $[t_n-h,t_n]$ and $[t_n,t_n+h]$
share the image at $t_n=nh$, so $h$ is both the time between images in a pair and
the time between control updates. Laboratory \gls{acr:piv} can acquire separate pairs
at a different rate, as in \cref{sec:afc:channel}
\citep[Fig.~8]{terpin2025ff}.
The flow evolves without actuation over $[-h,0]$ to supply the first image pair.
We track the same tracers throughout each episode and render them as Gaussian spots.
Copies across the periodic boundaries fill a padded image border (the halo),
which we crop after rendering.
At time $t_n$, the estimator sees only the completed pair $[t_n-h,t_n]$
and returns a displacement field $d_n$. Dividing by camera calibration $s$
(in pixels per unit length) and by $h$ gives the velocity estimate
$\widehat v_n=d_n/(s h)$. We fit the difference from the target over $W$
by least squares as a weighted sum of the large-scale and small-scale profiles.
The fitted weights $\quietEstimatedError_n=G^{-1}r_n$ are the large-scale and small-scale velocity
errors, with
\[
  G_{ij}=\frac{1}{|W|}\int_W\phi_i(y)\cdot\phi_j(y)\,d\quietPosition,
  \qquad
  r_{n,i}=\frac{1}{|W|}\int_W
    \phi_i(y)\cdot\bigl(\widehat v_n(\quietPosition)-\quietReference(y)\bigr)
    \,d\quietPosition,
\]
where $|W|$ is the area of $W$ and $i,j\in\{b,f\}$.
For ground-truth feedback, $d_n$ is instead the exact displacement: at
$t_n-h$, we place a virtual tracer at every camera pixel,
advect it with the simulated flow to $t_n$, and take its displacement in
pixels. The rest of the feedback path is unchanged. For the parallel flows of
the nominal setting, a tracer keeps its $y$ coordinate, so $d_n/(s h)$ equals
the true velocity at the pixel averaged over $[t_n-h,t_n]$; with the
non-parallel perturbations of the sensitivity analysis, this holds
approximately.
When $d_n$ is the exact displacement, we denote the resulting
coefficient $G^{-1}r_n$ by $m_n^{\mathrm{pair}}$.
Its fine component is $m_{n,f}^{\mathrm{pair}}$, abbreviated as
$m_f^{\mathrm{pair}}$ in plots where the time index is implicit.
The small-scale coefficient $\quietEstimatedError_{n,f}$ is written
$\quietEstimatedError_f$ in \cref{fig:quieting:mechanism}, with time given by the horizontal axis.
The controller smooths the rate of change of these errors as
$D_n=(1-\alpha)D_{n-1}+\alpha(\quietEstimatedError_n-\quietEstimatedError_{n-1})/h$, with
$\alpha=h/(\tau_D+h)$, and sets the commands to
$a_{n,i}=\operatorname{clip}(-K_p\quietEstimatedError_{n,i}-K_dD_{n,i},-b_i,b_i)$.
Here $K_p$ and $K_d$ are the proportional and derivative gains, $\tau_D$ sets
the smoothing time, and $\operatorname{clip}$ limits each command to $[-b_i,b_i]$.
We initialize $D_0=0$ and keep $a_{n,i}$ constant over $[t_n,t_n+h)$.
The nominal integral gain $K_i$ is zero. For the nonzero-$K_i$ sensitivity tests,
we initialize the integral contribution $\quietIntegral_{-1,i}=0$ and compute
$\widetilde{\quietIntegral}_{n,i}=\quietIntegral_{n-1,i}-K_i h\quietEstimatedError_{n,i}$.
The command before clipping is
$\quietUnclippedCommand_{n,i}=-K_p\quietEstimatedError_{n,i}-K_dD_{n,i}+\widetilde{\quietIntegral}_{n,i}$,
and $a_{n,i}=\operatorname{clip}(\quietUnclippedCommand_{n,i},-b_i,b_i)$.
We retain the candidate integral contribution as $\quietIntegral_{n,i}$ if the command
is not clipped or if
$(\quietUnclippedCommand_{n,i}-a_{n,i})\quietEstimatedError_{n,i}\ge0$;
otherwise, $\quietIntegral_{n,i}=\quietIntegral_{n-1,i}$.
This prevents accumulation when the integral update would push the command
further beyond an actuator limit (conditional anti-windup). We neglect computation time, so
the commands take effect immediately at $t_n$.

\paragraph{Metrics}
For episode duration $T$, velocity tolerance $\quietTolerance$, and effort
weight $\lambda$, we define the fraction of time within tolerance
$\quietUsableTime$ and normalized closed-loop cost $\quietCost$ as
\[
  \begin{aligned}
    \quietUsableTime&=\frac{1}{T}\int_0^T
    \mathbf{1}\!\left\{
      \left(\frac{1}{|W|}\int_W
      |\quietVelocity(\quietPosition,t)-\quietReference(y)|^2\,d\quietPosition
      \right)^{1/2}
      \le\quietTolerance\right\}\,dt,\\
    \quietCost&=1-\quietUsableTime+\frac{\lambda}{T}\int_0^T
    \sum_{i\in\{b,f\}}(a_i(t)/b_i)^2\,dt.
  \end{aligned}
\]
Here the error uses the true velocity $\quietVelocity(\quietPosition,t)$;
the effort penalty normalizes each actuator command $a_i(t)$ by its
magnitude limit $b_i$. Success additionally requires that the true
root-mean-square velocity error remain within tolerance throughout the
final interval $[T-d_h,T]$ of prescribed duration $d_h$.

For each episode, $\quietCost_{\mathrm{ref}}$ and
$\quietUsableTime_{\mathrm{ref}}$ denote the values obtained with the fixed
reference \dis configuration, starting from the same flow and tracer positions.
The reported differences in percentage points are
$\quietCostDelta=100(\quietCost-\quietCost_{\mathrm{ref}})$ and
$\quietUsableTimeDelta=100(\quietUsableTime-\quietUsableTime_{\mathrm{ref}})$.
Lower $\quietCostDelta$ and higher $\quietUsableTimeDelta$ are better.

\begin{table}[!htb]\centering\small
  \caption{Nominal local velocity regulation parameters. Lengths use the simulation
    unit $\ell$ and times use seconds. Spot intensity is a peak value in
    camera gray levels.}
  \label{tab:quieting:parameters}
  \begin{tabular}{p{.28\linewidth}p{.29\linewidth}p{.33\linewidth}}\toprule
    Component & Symbol or quantity & Nominal value\\\midrule
    Fluid and forces & Viscosity $\nu$ & $\quietParamViscosity\,\ell^2/\mathrm{s}$\\
    & \raggedright Width $w$; dimensionless wavenumber $k_f$; coupling $\kappa$ & $\quietParamWidth$; $\quietParamFineMode$; $\quietParamCoupling$\\
    Initial condition & $A$; $F$ & $\quietParamAmplitude\,\ell/\mathrm{s}$; $\quietParamFineAmplitude\,\ell/\mathrm{s}$\\
    & Episode variation & Independent uniform draws: amplitudes $\pm\quietArtAmplitudeVariationPercent\,\%$, viscosity $\pm\quietArtViscosityVariationPercent\,\%$\\
    Target and work area & $U_r$; side of $W$ & $\quietParamReference\,\ell/\mathrm{s}$; $\quietWindowSide$\\
    & Work-area center & $x/\ell\in\{\quietWindowCentersX\}$, $y/\ell=\quietWindowCenterY$\\
    & Feedback and force integration & $\quietParamQuadrature\times\quietParamQuadrature$ sample points over $W$\\
    Controller & $K_p$; $K_i$; $K_d$ & $\quietParamKp\,\mathrm{s}^{-1}$; $\quietParamKi\,\mathrm{s}^{-2}$; $\quietParamKd$\\
    & $\tau_D$; $b_b$; $b_f$ & $\quietParamDerivativeTau\,\mathrm{s}$; $\quietParamBroadBound\,\ell/\mathrm{s}^2$; $\quietParamFineBound\,\ell/\mathrm{s}^2$\\
    Timing and integration & $h$; $T$; $\delta t$ & $\quietParamInterval\,\mathrm{s}$; $\quietParamHorizon\,\mathrm{s}$; $\quietParamStep\,\mathrm{s}$\\
    & Fluid grid & $\quietParamGrid^2$\\
    Imaging & Image side; calibration $s$ & $\quietParamImageSize$ px; $\quietParamImageSize/(2\pi)$ px$/\ell$\\
    & Particle density; diameter & $\quietParamDensity$ particles/px; $\quietParamDiameter$ px\\
    & Spot intensity; bit depth & $\quietParamIntensity$ gray levels; $\quietParamBitDepth$ bits\\
    & Halo & $\quietParamHalo$ px\\
    & Added image noise; changes in tracer appearance; tracer loss & All disabled\\
    Objective & $\quietTolerance$; $\lambda$; $d_h$ & $\quietParamTolerance\,\ell/\mathrm{s}$; $\quietParamEffortWeight$; $\quietParamHold\,\mathrm{s}$\\
    \bottomrule
  \end{tabular}
\end{table}

\begin{table}[!htb]\centering\small
  \caption{\dis configurations. Variational refinement and reuse of the
    previous displacement estimate (temporal propagation) are disabled.
    Patch side lengths and strides are in pixels at the image resolution
    used for estimation. Finest level 0 uses full resolution; level 1 estimates at half resolution
    and upsamples to the full image size.}
  \label{tab:quieting:library}
  \begin{tabular}{lrrrrr}\toprule
    Estimator & Patch (px) & Stride (px) & Search iterations & \shortstack{Pyramid\\levels} & \shortstack{Finest\\level}\\\midrule
    \quietTableLibraryRowAColA & \quietTableLibraryRowAColB & \quietTableLibraryRowAColC & \quietTableLibraryRowAColD & \quietTableLibraryRowAColE & \quietTableLibraryRowAColF \\
\quietTableLibraryRowBColA & \quietTableLibraryRowBColB & \quietTableLibraryRowBColC & \quietTableLibraryRowBColD & \quietTableLibraryRowBColE & \quietTableLibraryRowBColF \\
Reference & \quietTableLibraryRowCColB & \quietTableLibraryRowCColC & \quietTableLibraryRowCColD & \quietTableLibraryRowCColE & \quietTableLibraryRowCColF \\
\bottomrule

  \end{tabular}
\end{table}

\subsection{Training and evaluation of \ourmethod}
\label{sec:appendix:quieting:training}

We train the \ourmethod estimator-selection agent on image pairs collected
from simulated closed-loop episodes. Training and validation use disjoint
sets of \quietArtTrainCount{} and \quietArtValidationCount{} combinations of
initial flow condition and tracer positions, respectively. For each
combination, we run the fixed controller in \quietArtTeacherCount{} separate
episodes, using ground-truth feedback, each of the three fixed estimators,
and the whole-frame oracle. We record the image pairs rendered by \synthpix
along these trajectories, yielding \quietArtTrainPairs{} training and
\quietArtValidationPairs{} validation pairs. We evaluate both specialists
on every pair and label it with the one whose displacement estimate has lower
whole-frame \gls{acr:epe} against the ground truth.

The selector uses convolutional layers with $(\quietArtChannels)$ channels and
initial stride $\quietArtStride$ to produce a $\quietArtEmbedding$-dimensional
feature vector. Its classifier has a hidden layer with $\quietArtHidden$ units
and rectified linear activations.
We train with Adam at learning rate $\quietArtLearningRate$ for
\quietArtSteps{} steps, using batches of \quietArtBatch{} uniformly sampled pairs.
For each of \quietArtSeedCount{} independent network initializations, we select
the saved network with the lowest validation \gls{acr:aepe} and evaluate the agent
in closed loop on the same
\quietArtTestCount{} test episodes, unused in training or validation.
All feedback sources start from the same flow and tracer positions in each test episode.
Qualitative figures use the
model with lowest validation \gls{acr:aepe}.

\subsection{Sensitivity analysis}
\label{sec:appendix:quieting:screens}
\label{sec:appendix:quieting:statistics}

\begin{samepage}
With the estimator library fixed, we vary one parameter at a time and compare
the whole-frame oracle with the reference, starting each episode from the same
flow and tracer positions.
\Cref{tab:quieting:sensitivity-model,tab:quieting:sensitivity-numerics} give
the nominal and tested values. They also report the min--max ranges of the
setting-wise mean $\quietCostDelta$ and $\quietUsableTimeDelta$.
\Cref{tab:quieting:oracle-outcomes} reports nominal comparisons
on \quietOriginalCount{} episodes.
\par
\end{samepage}

\begin{table}[H]\centering\small
  \caption{Nominal comparisons on \quietOriginalCount{} episodes used in the sensitivity analysis.
    Differences are in pp relative to the reference;
    successful episodes satisfy the final-interval criterion. The window oracle minimizes
    \gls{acr:epe} in $W$. Bold marks the best deployable feedback
    for each metric; oracle and ground-truth feedback are reference results.}
  \label{tab:quieting:oracle-outcomes}
  \begin{tabular}{lrrr}\toprule
    Feedback & $\quietCostDelta$ (pp) & $\quietUsableTimeDelta$ (pp) & \shortstack{Successful\\episodes}\\\midrule
    No control & \quietTableOutcomesRowAColB & \quietTableOutcomesRowAColC & \quietTableOutcomesRowAColD \\
    \quietTableOutcomesRowBColA & \quietTableOutcomesRowBColB & \quietTableOutcomesRowBColC & \quietTableOutcomesRowBColD \\
    \quietTableOutcomesRowCColA & \quietTableOutcomesRowCColB & \quietTableOutcomesRowCColC & \quietTableOutcomesRowCColD \\
    Reference & \textbf{\boldmath \quietTableOutcomesRowDColB} & \textbf{\boldmath \quietTableOutcomesRowDColC} & \textbf{\boldmath \quietTableOutcomesRowDColD} \\
    Whole-frame oracle & \quietTableOutcomesRowEColB & \quietTableOutcomesRowEColC & \quietTableOutcomesRowEColD \\
    Window oracle & \quietTableOutcomesRowFColB & \quietTableOutcomesRowFColC & \quietTableOutcomesRowFColD \\
    Ground-truth feedback & \quietTableOutcomesRowGColB & \quietTableOutcomesRowGColC & \quietTableOutcomesRowGColD \\
    \bottomrule
  \end{tabular}
\end{table}

\paragraph{Physical and numerical parameters}
Varying the fluid grid, integration time step, and image padding yields
mean $\quietCostDelta\in[\quietNumericsCostDeltaMin,\quietNumericsCostDeltaMax]$ pp,
indicating low sensitivity to numerical resolution. Slower imaging
and lower camera resolution can prevent the error from reaching tolerance;
larger initial motion or weaker large-scale actuation can leave too little time
to reach tolerance, even with ground-truth feedback.

\begin{table}[H]\centering\scriptsize
  \caption{Sensitivity to physical and controller parameters, \quietScreenCount{}
    episodes per setting. Ranges span mean oracle differences from the
    reference across nominal and tested settings; nominal differences are
    $\quietCostDelta=\quietScreenCostDelta$ pp and
    $\quietUsableTimeDelta=\quietScreenUsableDelta$ pp.
    \quietSensitivityExclusions{}
    The variation multiplier scales the amplitude and viscosity ranges.}
  \label{tab:quieting:sensitivity-model}
  \setlength{\tabcolsep}{3.5pt}
  \begin{tabular}{lrrrr}\toprule
    Parameter & Nominal & Tested values & $\quietCostDelta$ (pp) & $\quietUsableTimeDelta$ (pp)\\\midrule
    \multicolumn{5}{l}{\emph{Fluid and initial condition}}\\
    $\kappa$ & \quietTableSensitivityPhysicsRowAColB & \quietTableSensitivityPhysicsRowAColC & \quietTableSensitivityPhysicsRowAColD & \quietTableSensitivityPhysicsRowAColE \\
$k_f$ & \quietTableSensitivityPhysicsRowBColB & \quietTableSensitivityPhysicsRowBColC & \quietTableSensitivityPhysicsRowBColD & \quietTableSensitivityPhysicsRowBColE \\
$w$ & \quietTableSensitivityPhysicsRowCColB & \quietTableSensitivityPhysicsRowCColC & \quietTableSensitivityPhysicsRowCColD & \quietTableSensitivityPhysicsRowCColE \\
\quietTableSensitivityPhysicsRowDColA & \quietTableSensitivityPhysicsRowDColB & \quietTableSensitivityPhysicsRowDColC & \quietTableSensitivityPhysicsRowDColD & \quietTableSensitivityPhysicsRowDColE \\
$A$ ($\ell/\mathrm{s}$) & \quietTableSensitivityPhysicsRowEColB & \quietTableSensitivityPhysicsRowEColC & \quietTableSensitivityPhysicsRowEColD & \quietTableSensitivityPhysicsRowEColE \\
$F$ ($\ell/\mathrm{s}$) & \quietTableSensitivityPhysicsRowFColB & \quietTableSensitivityPhysicsRowFColC & \quietTableSensitivityPhysicsRowFColD & \quietTableSensitivityPhysicsRowFColE \\
$U_r$ ($\ell/\mathrm{s}$) & \quietTableSensitivityPhysicsRowGColB & \quietTableSensitivityPhysicsRowGColC & \quietTableSensitivityPhysicsRowGColD & \quietTableSensitivityPhysicsRowGColE \\
Variation multiplier & \quietTableSensitivityPhysicsRowHColB & \quietTableSensitivityPhysicsRowHColC & \quietTableSensitivityPhysicsRowHColD & \quietTableSensitivityPhysicsRowHColE \\

    \midrule
    \multicolumn{5}{l}{\emph{Controller and actuation}}\\
    \quietTableSensitivityControllerRowAColA & \quietTableSensitivityControllerRowAColB & \quietTableSensitivityControllerRowAColC & \quietTableSensitivityControllerRowAColD & \quietTableSensitivityControllerRowAColE \\
\quietTableSensitivityControllerRowBColA & \quietTableSensitivityControllerRowBColB & \quietTableSensitivityControllerRowBColC & \quietTableSensitivityControllerRowBColD & \quietTableSensitivityControllerRowBColE \\
$K_d$ & \quietTableSensitivityControllerRowCColB & \quietTableSensitivityControllerRowCColC & \quietTableSensitivityControllerRowCColD & \quietTableSensitivityControllerRowCColE \\
$\tau_D$ (s) & \quietTableSensitivityControllerRowDColB & \quietTableSensitivityControllerRowDColC & \quietTableSensitivityControllerRowDColD & \quietTableSensitivityControllerRowDColE \\
\quietTableSensitivityControllerRowEColA & \quietTableSensitivityControllerRowEColB & \quietTableSensitivityControllerRowEColC & \quietTableSensitivityControllerRowEColD & \quietTableSensitivityControllerRowEColE \\
\quietTableSensitivityControllerRowFColA & \quietTableSensitivityControllerRowFColB & \quietTableSensitivityControllerRowFColC & \quietTableSensitivityControllerRowFColD & \quietTableSensitivityControllerRowFColE \\

    \bottomrule
  \end{tabular}
\end{table}

\begin{table}[H]\centering\scriptsize
  \caption{Sensitivity to observation, imaging, and numerical settings.
    Format follows \cref{tab:quieting:sensitivity-model}; pixel lengths use
    the nominal camera. Resolution changes preserve force normalization and
    physical work-area size, with tracer density and diameter fixed in pixels.
    Work-area dimensions for feedback and evaluation coincide nominally;
    the indicated feedback or evaluation dimension varies separately in these tests.
    Large-scale force pulses have amplitude $\quietPulseAmplitude b_b$, duration
    $\quietPulseDuration h$, and timing variation $\quietPulseJitter h$.
    The longer episode uses its own duration for normalization.}
  \label{tab:quieting:sensitivity-numerics}
  \setlength{\tabcolsep}{3.5pt}
  \begin{tabular}{lrrrr}\toprule
    Parameter & Nominal & Tested values & $\quietCostDelta$ (pp) & $\quietUsableTimeDelta$ (pp)\\\midrule
    \multicolumn{5}{l}{\emph{Observation and imaging}}\\
    Work-area side for feedback (px) & \quietTableSensitivityObservationRowAColB & \quietTableSensitivityObservationRowAColC & \quietTableSensitivityObservationRowAColD & \quietTableSensitivityObservationRowAColE \\
Work-area center for feedback $y/\ell$ & \quietTableSensitivityObservationRowBColB & \quietTableSensitivityObservationRowBColC & \quietTableSensitivityObservationRowBColD & \quietTableSensitivityObservationRowBColE \\
$h$ (s) & \quietTableSensitivityObservationRowCColB & \quietTableSensitivityObservationRowCColC & \quietTableSensitivityObservationRowCColD & \quietTableSensitivityObservationRowCColE \\
Particle density (particles/px) & \quietTableSensitivityObservationRowDColB & \quietTableSensitivityObservationRowDColC & \quietTableSensitivityObservationRowDColD & \quietTableSensitivityObservationRowDColE \\
Spot diameter (px) & \quietTableSensitivityObservationRowEColB & \quietTableSensitivityObservationRowEColC & \quietTableSensitivityObservationRowEColD & \quietTableSensitivityObservationRowEColE \\
Spot intensity (gray levels) & \quietTableSensitivityObservationRowFColB & \quietTableSensitivityObservationRowFColC & \quietTableSensitivityObservationRowFColD & \quietTableSensitivityObservationRowFColE \\
Image side (px) & \quietTableSensitivityObservationRowGColB & \quietTableSensitivityObservationRowGColC & \quietTableSensitivityObservationRowGColD & \quietTableSensitivityObservationRowGColE \\
Work-area side for evaluation (px) & \quietTableSensitivityObservationRowHColB & \quietTableSensitivityObservationRowHColC & \quietTableSensitivityObservationRowHColD & \quietTableSensitivityObservationRowHColE \\

    \midrule
    \multicolumn{5}{l}{\emph{Numerics and disturbances}}\\
    Grid side (nodes) & \quietTableSensitivityNumericsRowAColB & \quietTableSensitivityNumericsRowAColC & \quietTableSensitivityNumericsRowAColD & \quietTableSensitivityNumericsRowAColE \\
$\delta t$ (s) & \quietTableSensitivityNumericsRowBColB & \quietTableSensitivityNumericsRowBColC & \quietTableSensitivityNumericsRowBColD & \quietTableSensitivityNumericsRowBColE \\
$T$ (s) & \quietTableSensitivityNumericsRowCColB & \quietTableSensitivityNumericsRowCColC & \quietTableSensitivityNumericsRowCColD & \quietTableSensitivityNumericsRowCColE \\
Halo (px) & \quietTableSensitivityNumericsRowDColB & \quietTableSensitivityNumericsRowDColC & \quietTableSensitivityNumericsRowDColD & \quietTableSensitivityNumericsRowDColE \\
Pulse timing $t/T$ & $\text{none}$ & \quietTableSensitivityNumericsRowEColC & \quietTableSensitivityNumericsRowEColD & \quietTableSensitivityNumericsRowEColE \\

    \bottomrule
  \end{tabular}
\end{table}

\paragraph{Evaluation sensitivity}
\label{sec:appendix:quieting:objective}

\begin{figure}[H]\centering
  \includegraphics[width=\linewidth]{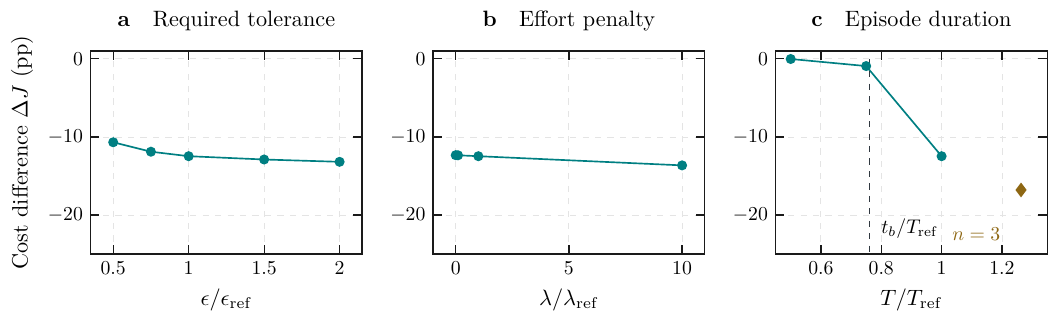}
  \caption{Sensitivity to the metric definitions and episode duration. Teal points
    use metrics recomputed from \quietOriginalCount{} recorded episodes;
    the diamond uses \quietScreenCount{} additional, longer episodes.
    Points give mean oracle $\quietCostDelta$ in pp; the dashed line marks
    nominal braking time. The subscript $\mathrm{ref}$ denotes the nominal parameter values
    in \cref{tab:quieting:parameters}.}
  \label{fig:quieting:sensitivity}
\end{figure}

We recompute the metrics from these \quietOriginalCount{} episodes under
\quietObjectiveSettings{} evaluation settings, without rerunning the simulations
(\cref{tab:quieting:objective,fig:quieting:sensitivity}).
Removing the effort penalty gives $\quietCostDelta=\quietNoEffortCostDelta$ pp,
and the oracle retains its advantage across the tested tolerances.
Ending the evaluation earlier, before the error reaches tolerance, gives
differences near zero.

\begin{table}[H]\centering\scriptsize
  \caption{Evaluation sensitivity on \quietOriginalCount{} episodes. Format
    follows \cref{tab:quieting:sensitivity-model}, with nominal outcomes
    $\quietCostDelta=\quietOriginalCostDelta$ pp and
    $\quietUsableTimeDelta=\quietOriginalUsableDelta$ pp.
    Each evaluation uses its own duration for normalization; the required
    final interval within tolerance $d_h$ affects success but not cost.}
  \label{tab:quieting:objective}
  \setlength{\tabcolsep}{3.5pt}
  \begin{tabular}{lrrrr}\toprule
    Parameter & Nominal & Tested values & $\quietCostDelta$ (pp) & $\quietUsableTimeDelta$ (pp)\\\midrule
    $\quietTolerance$ ($\ell/\mathrm{s}$) & \quietTableSensitivityObjectiveRowAColB & \quietTableSensitivityObjectiveRowAColC & \quietTableSensitivityObjectiveRowAColD & \quietTableSensitivityObjectiveRowAColE \\
$\lambda$ & \quietTableSensitivityObjectiveRowBColB & \quietTableSensitivityObjectiveRowBColC & \quietTableSensitivityObjectiveRowBColD & \quietTableSensitivityObjectiveRowBColE \\
$d_h$ (s) & \quietTableSensitivityObjectiveRowCColB & \quietTableSensitivityObjectiveRowCColC & \quietTableSensitivityObjectiveRowCColD & \quietTableSensitivityObjectiveRowCColE \\
$T$ (s) & \quietTableSensitivityObjectiveRowDColB & \quietTableSensitivityObjectiveRowDColC & \quietTableSensitivityObjectiveRowDColD & \quietTableSensitivityObjectiveRowDColE \\

    \bottomrule
  \end{tabular}
\end{table}

\section{Fluidic pinball experiment}
\label{sec:appendix:water-channel}

We describe the experimental apparatus and measurement pipeline, the procedure
used to construct the tuning-selection and region-selection libraries, and
the training of their selection agents. The physical controller uses tuning
selection, as in \cref{sec:afc:channel}.

\subsection{Experimental apparatus and measurement pipeline}
\label{sec:appendix:water-channel:apparatus}
\providecommand{\pinballTrackingControlHz}{30}
\providecommand{\pinballTrackingEpisodeS}{60}
\providecommand{\pinballTrackingPhaseS}{30}
\providecommand{\pinballTrackingResetS}{60}
\providecommand{\pinballTrackingAverageCount}{3}
\providecommand{\pinballTrackingTrainEpisodes}{20}
\providecommand{\pinballTrackingEvalEpisodes}{10}
\providecommand{\pinballTrackingFlowScale}{4}
\providecommand{\pinballTrackingCylinderLimit}{150}
\providecommand{\pinballTrackingEvaluationEpisode}{1}
\providecommand{\pinballTrackingFlushingTime}{15.00}
\providecommand{\pinballTrackingFlushingReward}{-0.398}
\providecommand{\pinballTrackingQuietingTime}{45.00}
\providecommand{\pinballTrackingQuietingReward}{-0.414}

We collect data using a workstation running \channelOperatingSystem{}, equipped
with an \channelCpuModel{} processor and \channelGpuCount{} \channelGpuModel{} GPUs.
Following \citet[Appendix~A.1]{terpin2025ff}, we run flow estimation, policy
inference, and policy training on separate dedicated GPUs.

We use an updated version of the tabletop circulating water channel designed
by \citet{terpin2025ff}. The design and fabrication files are available at
\url{https://activefluidcontrol.com}. \Cref{fig:tasks} shows the channel and
labelled close-ups of its components. The mean incoming flow speed
at the entrance to the test section is \channelInletSpeedMS{}\,m/s. The apparatus combines
off-the-shelf components with 3D-printed parts. It is housed in an
\channelAquariumModel{} (\channelAquariumCapacityL{}\,L) measuring
$\channelAquariumLengthM\times\channelAquariumWidthM\times\channelAquariumHeightM\,\mathrm{m}^{3}$.

The channel has an upper test branch and a lower return branch.
The test section (\textbf{1}) contains the fluidic pinball (\textbf{2}).
The circulation propellers (\textbf{3}) are driven by
\channelPropellerCount{} external \channelMotorModel{} motors
through magnetic couplings across the glass wall, producing a right-to-left
flow through the test section. Throughout selection-data collection and control training,
we hold the two propellers at $-\channelPropellerSpeedRpm$ and
$+\channelPropellerSpeedRpm$\,RPM (hardware identifiers 8 and 20,
respectively). Turning vanes (\textbf{4}) guide the water
between the two branches. A honeycomb (\textbf{5}) aligns the flow and
attenuates large vortical structures, while the flow contractions
(\textbf{6}) accelerate the stream and elongate the remaining smaller vortices.
The \gls{acr:piv} camera (\textbf{7}) records
$\channelAcquisitionWidthPx\times\channelAcquisitionHeightPx$-pixel images
(width $\times$ height) at \channelCameraRateHz{}\,Hz.

Each of the \channelCylinderCount{} cylinders is driven independently by a
\channelMotorModel{} motor, of the same type used for the circulation propellers.
All \channelMotorCount{} motors share a controller area network (CAN) bus
connected to the host through a \channelCanInterface{} interface.
\flowgames sends individually addressed
rotation-rate commands and receives motor-speed feedback over this bus.
The cylinders have radius $R=\channelCylinderRadiusM\,\mathrm{m}$. We limit their angular
velocities to $|\omega_i|\leq\channelActuationLimit$,
$i=1,\dots,\channelCylinderCount$, with
$\channelActuationLimit=\channelRotationRateLimitRadS\,\mathrm{rad/s}$.
The resulting maximum surface speed is
$R\channelActuationLimit=\channelSurfaceSpeedMS\,\mathrm{m/s}$, corresponding to
\channelSurfaceSpeedRatio{} times the mean
incoming flow speed.
We update the cylinder rotation-rate commands at \pinballTrackingControlHz{}\,Hz.

\paragraph{Modular motor mount}
All \channelCylinderCount{} cylinder motors are mounted on a common aluminum
plate. Exchanging the plate allows the apparatus to switch between the fluidic
pinball and a single-cylinder configuration.

\paragraph{Image processing and flow observations}
We correct the acquired
$\channelAcquisitionWidthPx\times\channelAcquisitionHeightPx$-pixel frames
(width $\times$ height) using the homography calibrated from AprilTags. From
each corrected frame, we retain a
$\channelCropWidthPx\times\channelCropHeightPx$-pixel crop, starting at column
\channelCropLeftPx{} and row \channelCropTopPx{} in the native array. The crop
excludes the channel boundary and is passed to the estimator without image
resizing. The displayed camera view is reflected vertically: the same crop
starts at row
$\channelAcquisitionHeightPx-\channelCropTopPx-\channelCropHeightPx=20$
in that view. We cap each cropped image at its median intensity plus
\channelPreprocessIntensityCappingN{} standard deviations, then linearly
stretch its intensity range to $[0,\channelPreprocessStretchContrastScale]$
(\cref{tab:channel:preprocessing}).

We form pairs from consecutive frames separated by the short laser-pulse
interval of \channelPairIntervalMs{}\,ms, giving approximately
\channelFlowAcquisitionRateHz{} image pairs and flow estimates per second.
The tuning-selection library estimates flow over the whole retained crop;
the region-selection library estimates flow within one of the subregions in
\cref{tab:channel:regions}. After estimation, we reduce the selected vector
field to $\channelObservationHeight\times\channelObservationWidth$ grid
points by antialiased linear interpolation, without rescaling its components:
flow remains a displacement in source-image pixels per image pair. The grid
covers the parent crop for tuning selection and the selected subregion for
region selection.

For the physical flow-tracking controller, we then take a causal running
average of the latest \pinballTrackingAverageCount{} reduced fields.
The averaging history is reset at each episode; the first outputs average
only the samples available so far. This averaged field, in source-image
pixels per pair, supplies the reward measurement in \cref{sec:appendix:water-channel:training}.
For the controller observation, we map both the averaged field and the current
target componentwise from
$[-\pinballTrackingFlowScale,\pinballTrackingFlowScale]$ pixels per pair to
$[0,1]$, clipping values outside this range. We stack their two displacement
components into a
$\channelObservationHeight\times\channelObservationWidth\times4$ tensor,
which \dreamerVThree receives through its convolutional encoder at
\pinballTrackingControlHz{}\,Hz.

\begin{table}[htbp]\centering\small
  \caption{Processing in the physical flow-tracking pipeline, in execution
    order. Steps 1--2 act on each cropped image; steps 3--4 act on the selected
    flow; steps 5--6 prepare the averaged flow and target for the controller.
    Intensity-routine names follow \flowgym; temporal averaging uses
    the \texttt{AverageFilter} provided by \goggles*.}
  \label{tab:channel:preprocessing}
  \begin{tabular}{cll}
    \toprule
    Step & Operation & Parameters\\\midrule
    1 & \texttt{intensity\_capping} & $n=\channelPreprocessIntensityCappingN$\\
    2 & \texttt{stretch\_contrast} & $\texttt{scale}=\channelPreprocessStretchContrastScale$\\
    3 & Spatial reduction & $\channelObservationHeight\times\channelObservationWidth$, antialiased linear\\
    4 & Temporal averaging & \pinballTrackingAverageCount{} fields at \channelFlowAcquisitionRateHz{}\,Hz\\
    5 & Observation scaling & $[-\pinballTrackingFlowScale,\pinballTrackingFlowScale]\to[0,1]$, clipped\\
    6 & Controller input & $\channelObservationHeight\times\channelObservationWidth\times4$, CNN at \pinballTrackingControlHz{}\,Hz\\
    \bottomrule
  \end{tabular}
\end{table}

\subsection{Constructing the two libraries}
\label{sec:appendix:water-channel:tuning}

We manually tune the estimator parameters for both libraries, then fix the
libraries, preprocessing, and our labels before collecting the
selection-agent training recordings. Both libraries use the same parent crop
and intensity processing of \cref{sec:appendix:water-channel:apparatus}.
\Cref{tab:channel:configured-dis} gives the \dis parameters;
\cref{tab:channel:labelled-settings} associates each of our choices with
the constant cylinder commands used to collect its examples.
We number the cylinders as in \cref{fig:channel:loop}: 1 is upstream,
2 is downstream right, and 3 is downstream left when looking downstream.
All command tuples and table columns follow the order $(1,2,3)$.

\begin{table}[htbp]\centering\small
  \caption{Manually tuned \dis configurations. \dis~A and \dis~B form the
    tuning-selection library; \dis~C is shared by all region-selection members. Level zero is
    native resolution, and each successive pyramid level halves each image
    dimension. All configurations disable variational refinement and temporal
    flow propagation.}
  \label{tab:channel:configured-dis}
  \begin{tabular}{lrrrrr}
    \toprule
    Configuration & \shortstack{Finest\\level} & Levels & \shortstack{Patch\\size (px)} & \shortstack{Patch\\stride (px)} & Iterations\\\midrule
    \channelFrozenDisRows
    \bottomrule
  \end{tabular}
\end{table}

\begin{table}[htbp]\centering\small
  \caption{Recording labels and constant cylinder commands. Columns give
    cylinder numbers 1, 2, and 3; the circulation-propeller commands remain fixed.}
  \label{tab:channel:labelled-settings}
  \begin{tabular}{llrrr}
    \toprule
    Variant & Label & Cylinder 1 (RPM) & Cylinder 2 (RPM) & Cylinder 3 (RPM)\\\midrule
    \channelFrozenRegimeRows
    \bottomrule
  \end{tabular}
\end{table}

\subsubsection{DIS tuning selection}
\label{sec:appendix:water-channel:tuning-library}
We retain two configurations with different allocations of computation,
each applied to the same preprocessed parent image pair.
\dis~A uses one native-resolution search level; \dis~B searches from
one-eighth to one-half resolution and interpolates its output to the native
grid. \dis~A retains particle-image detail that \dis~B discards; \dis~B
can initialize its search on smaller pixel displacements at the coarsest
level. We manually tune both configurations to meet a common
\channelDisLatencyBudgetMs{}\,ms budget for warmed estimator execution on
resident, already-preprocessed image pairs.

We check our tuning choices against a slower reference in
\cref{sec:appendix:water-channel:self-supervised}.

\subsubsection{Region selection}
\label{sec:appendix:water-channel:regions}
Here, the library varies where to measure the flow while retaining a common
estimator. We choose the three
$\channelRoiSidePx\times\channelRoiSidePx$-pixel regions in
\cref{tab:channel:regions}. Each member extracts its region from the
preprocessed parent image pair and applies \dis~C, with the identical
numerical parameters listed in \cref{tab:channel:configured-dis}.
We downsample the selected field to
$\channelObservationHeight\times\channelObservationWidth$ grid points while
preserving source-pixel displacement units. Equal crop sizes allocate the
same number of output grid points to each region. They also limit the area
on which dense flow estimation is performed.

The measurement objective is to observe coherent spatial variation, where
neighboring parts of the fluid move at different speeds or in different
directions, as in rotation or shear. We identify the preferred
region under each of the three prescribed actuation settings in
\cref{tab:channel:labelled-settings}. The selection agent sees the common
parent image pair and learns where to direct the measurement.

In exploratory checks, we inspect the coherent spatial variation by
removing each region's mean displacement and fitting an affine
displacement field on particle-supported locations. The trace-free spatial gradient
of this fit summarizes rotation and anisotropic stretching, with uniform
drift and isotropic expansion removed. We inspect fields before and after
downsampling, using the same source-image support when comparing resolutions,
so a change of sampling locations does not itself change the comparison.
These checks support our choice of a region; they are not labels
recomputed during classifier training, nor evidence that the preferred region
is necessarily the most useful one for control.

\begin{table}[htbp]\centering
  \caption{Region offsets within the common parent crop. Coordinates are
    native array coordinates before the vertical display reflection: columns
    increase rightward and rows downward. Every region has side length
    \channelRoiSidePx{} pixels.}
  \label{tab:channel:regions}
  \begin{tabular}{lrr}
    \toprule
    Region & Left column (px) & Top row (px)\\\midrule
    \channelRoiRows
    \bottomrule
  \end{tabular}
\end{table}

\begin{figure}[p]\centering
  \includegraphics[width=\linewidth]{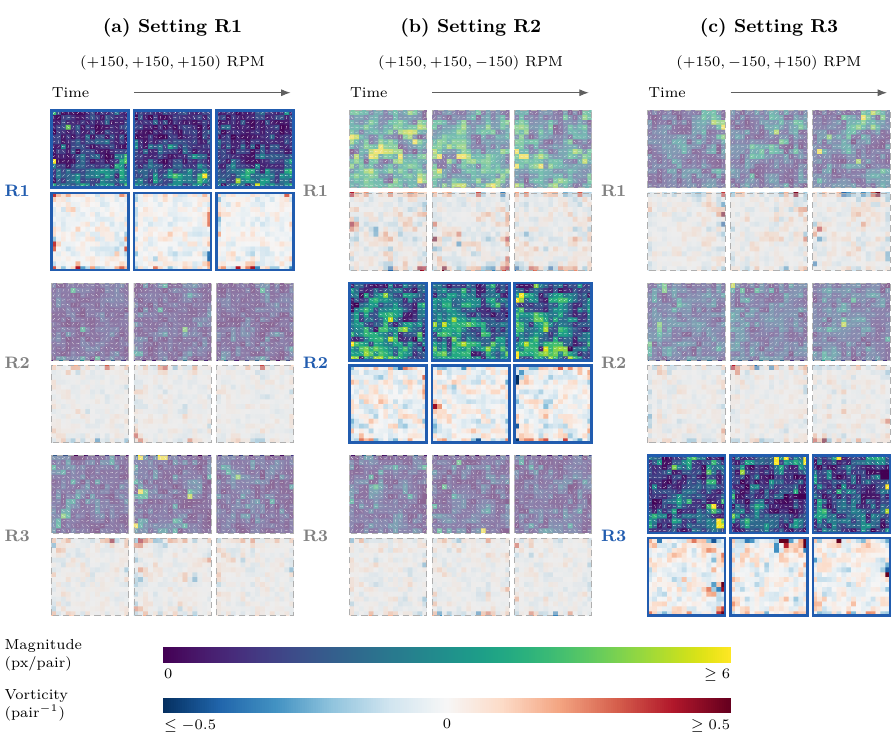}
  \caption{Region-selection pool on separate validation recordings.
    \textbf{(a--c)} Settings R1--R3; headers
    give signed cylinder commands in the same order as \cref{fig:channel:tuning}.
    Within each motor panel, columns advance through consecutive pairs
    \pinballValidationRoiSequenceFirst--\pinballValidationRoiSequenceLast{}.
    For each region, the upper row shows displacement magnitude and the
    lower row shows vorticity, both obtained with \dis~C on its
    $\channelRoiSidePx\times\channelRoiSidePx$-pixel support.
    The displayed grid is $16\times16$; an otherwise identical native-resolution
    counterpart is retained in the figure sources. Blue outlines identify the
    actual archived classifier choice at checkpoint $\pinballRoiSelectedStep$;
    unselected tiles use the same grey and dashed treatment as
    \cref{fig:channel:tuning}. Pair selection, arrows and display orientation
    follow that figure; scalar colour scales are shared within each figure.
    The regions have different spatial supports: their fields illustrate what each choice measures, rather than
    alternative estimates of one common velocity field.}
  \label{fig:channel:regions}
\end{figure}

\paragraph{Visualizing the libraries}
We visualize the two pools of estimators in
\cref{fig:channel:tuning,fig:channel:regions}. Blue outlines indicate
\ourmethod's choices on separate validation recordings, using the selectors
trained in \cref{sec:appendix:water-channel:selector}.

\subsection{Data collection, training, and validation}
\label{sec:appendix:water-channel:selection}

We first describe the labelled recordings, then the shared classifier-training
procedure and validation agreement. We subsequently check the \dis tuning
preferences against a slower reference and assess photometric augmentation
on later test recordings.

\subsubsection{Labelled recordings}
\label{sec:appendix:water-channel:recordings}

Each recording inherits the preferred library member assigned to its actuation
setting in \cref{tab:channel:labelled-settings}. For every setting, we collect
\channelSelectorTrainingRecordings{} training recordings and
\channelSelectorValidationRecordings{} separate validation recording, each
containing approximately \channelSelectorRecordingDurationS{}\,s of data.
We allow \channelSelectorSettlingDurationS{}\,s of settling before each recording.
We collect all tuning-selection training recordings first, then all
region-selection training recordings, followed by validation recordings for
the two variants in the same order. This increases the time separation
between training and validation for each library, helping reduce temporal
correlation between the splits.

Each image pair contains the frames illuminated by the two pulses of the same
laser cycle, separated by \channelPairIntervalMs{}\,ms, giving approximately
\channelFlowAcquisitionRateHz{} pairs per second. We verify the pulse phase
in each recording before forming pairs. Each recording provides
$\channelSelectorPairsPerRecording$ complete image pairs.
This gives $\channelSelectorTrainPairsPerClass$ training pairs and
$\channelSelectorValidationPairsPerClass$ validation pairs per class. The tuning-selection
dataset therefore contains $\channelDisTrainPairs$ training and
$\channelDisValidationPairs$ validation pairs; the region-selection dataset
contains $\channelRoiTrainPairs$ and $\channelRoiValidationPairs$, respectively.
Splitting by recording keeps adjacent frames out of different splits, although
pairs within a recording remain temporally correlated.

For the augmentation ablation in \cref{sec:appendix:water-channel:augmentation},
we collect one additional test recording of approximately
\channelSelectorRecordingDurationS{}\,s per tuning-selection setting, after
\channelSelectorSettlingDurationS{}\,s of settling, giving
$\pinballDisTestPairs$ test pairs. The additional collection began
\pinballTestSeparationMinutes{} minutes after the original collection ended.
We kept the camera, optics, and illumination settings unchanged and made no
deliberate change to seeding density. Thus this evaluates transfer to a later
acquisition block under nominally unchanged conditions.
Test recordings are excluded from training and checkpoint selection.

\subsubsection{Selection-agent training and validation}
\label{sec:appendix:water-channel:selector}

We train a separate \ourmethod classifier for each library. Each classifier
receives the parent image pair and learns the supplied library-member label.
We use the small convolutional architecture of the main experiments,
cross-entropy, and Adam, with batch size \channelSelectorBatchSize{} and
\channelSelectorTrainingSteps{} steps per run. The tuning selector uses a
constant learning rate of \channelSelectorLearningRate{}.

\paragraph{Photometric augmentation}
We vary the illumination of every training pair. For images with intensities in
$[0,255]$, we apply a gain drawn uniformly from $[0.4,1.6]$ and an offset from
$[-20,20]$, both shared by the two frames, and an extra gain drawn
independently per frame from $[0.9,1.1]$, as the two laser pulses are never
equally bright. Intensities are clipped to $[0,255]$ after augmentation.
We apply no random spatial flips or crops. Validation uses unaugmented image
pairs.

\paragraph{Region-selector training}
For the region selector, we reduce the learning rate to
\channelRoiSelectorLearningRate{}.

\paragraph{Checkpoint selection and validation agreement}
We select checkpoints by the lowest classification error over every pair in
the validation recordings. The selected tuning and region classifiers are
saved at steps \pinballDisSelectedStep{} and \pinballRoiSelectedStep{},
respectively. \Cref{tab:channel:selector-validation} gives their agreement
with our labels by setting. These recordings are separate from training but
also select the checkpoint, so the results measure agreement within the
collection session. Generalization to later recordings is assessed separately
for tuning selection in \cref{sec:appendix:water-channel:augmentation}.

\begin{table}[htbp]\centering
  \caption{Agreement with our labels for the selected tuning and region
    classifiers. Each setting contributes one validation recording, held out
    from training and used for checkpoint selection.}
  \label{tab:channel:selector-validation}
  \begin{tabular}{llrr}
    \toprule
    Library & Our choice & Correct/total & Agreement (\%)\\\midrule
    \pinballSelectorValidationRows
    \bottomrule
  \end{tabular}
\end{table}

\subsubsection{Reference-based check of the DIS tunings}
\label{sec:appendix:water-channel:self-supervised}

We compare both library members with a slower \dis reference on all
$\pinballDisValPairs$ image pairs in the tuning-selection validation
recordings defined in \cref{sec:appendix:water-channel:recordings}.
This checks the flow estimates associated with our tuning preferences.
The classifier learns the fixed labels; reference agreement supplies
supporting evidence for those choices.

\paragraph{Common spatial support and output scales}
We score only particle-rich locations, using a mask computed independently
of the candidate flow from the mean of the two preprocessed images. For mask
construction only, we smooth this mean with Gaussian standard deviation five
pixels, threshold at intensity 12 on the $[0,255]$ scale, and discard a
24-pixel border. The estimator images themselves remain unblurred.
We compare native fields and their bilinear reductions to $64\times64$ and
$16\times16$ grids, retaining displacement in source-image pixels per pair.
For each reduced grid, we reconstruct the field on the native grid before
scoring. The same particle-supported locations are therefore used at every
scale.

\paragraph{Agreement with a slower reference}
We use OpenCV \dis with native finest scale, patch size 12,
stride three, 64 gradient-descent iterations, and 20 variational-refinement
iterations, with mean normalization and spatial propagation enabled (R12).
For each scale, we apply the same reduction and reconstruction to both the
candidate and reference fields, then average their Euclidean discrepancy on
the fixed particle support. \Cref{tab:channel:self-supervised-epe} reports its
mean over image pairs, in source-image pixels. This reference discrepancy
has the form of \gls{acr:epe}, but uses an estimated field in place of ground
truth. Agreement can inherit shared \dis errors, so we use it as a consistency
check alongside our assessment of the observed flow structure.

\begin{table}[htbp]\centering\small
  \caption{Mean discrepancy from the slower R12 reference on the separate
    DIS validation recordings (px; lower is better). Both fields undergo the
    same resampling, and all scales use the same native particle support.
    These are reference-agreement errors, not ground-truth \gls{acr:aepe}.
    Bold marks the lower discrepancy between A and B for each setting and scale.}
  \label{tab:channel:self-supervised-epe}
  \begin{tabular}{llrrr}
    \toprule
    Setting & Configuration & Native & $64\times64$ & $16\times16$\\\midrule
    \pinballPreferenceReferenceRows
    \bottomrule
  \end{tabular}
\end{table}

\paragraph{Results}
The comparison with R12 supports the preference reversal at all three output
scales: \dis~A has lower mean reference discrepancy in setting A, while \dis~B has lower
discrepancy in setting B. At $16\times16$, the discrepancies in setting A are
\pinballSettingAASixteenEpe{}\,px for \dis~A and
\pinballSettingABSixteenEpe{}\,px for \dis~B; in setting B they are
\pinballSettingBASixteenEpe{}\,px and \pinballSettingBBSixteenEpe{}\,px,
respectively. The preferred tuning has lower discrepancy on
\pinballSettingASixteenPreferredWinPct{}\,\% and
\pinballSettingBSixteenPreferredWinPct{}\,\% of pairs in the respective
recordings.

\subsubsection{Photometric augmentation ablation}
\label{sec:appendix:water-channel:augmentation}

We assess photometric augmentation for the tuning-selection library using the
additional test recordings defined in
\cref{sec:appendix:water-channel:recordings}.
We train with and without it over five paired seeds, holding the architecture,
optimizer, training examples, and validation procedure fixed. Each run uses
\channelSelectorTrainingSteps{} updates; the lowest validation label error
selects its checkpoint,
with ties resolved in favor of the earliest step.

We evaluate every test pair exactly once for each selected checkpoint and
report mean label agreement and its population standard deviation across
completed seeds, with the completion count shown explicitly
(\cref{tab:channel:augmentation}). The variation across seeds reflects
training variability on the same test recordings.

\begin{table}[htbp]\centering
  \caption{DIS-selector augmentation ablation on the additional test
    recordings. Agreement with our labels is reported as mean $\pm$
    population standard deviation over the completed seeds, out of five
    planned paired seeds per variant. Bold marks the higher mean agreement.
    \pinballAugmentationCompletionNote}
  \label{tab:channel:augmentation}
  \begin{tabular}{lcr}
    \toprule
    Training & Completed seeds & Test agreement (\%)\\\midrule
    \pinballAugmentationRows
    \bottomrule
  \end{tabular}
\end{table}

Across all five paired seeds, test agreement is
$\pinballAugmentationWithAccuracyPct \pm \pinballAugmentationWithStdPct$\,\%
with augmentation and
$\pinballAugmentationWithoutAccuracyPct \pm \pinballAugmentationWithoutStdPct$\,\%
without it.

\subsection{DreamerV3 control-agent training}
\label{sec:appendix:water-channel:training}

We give the reward and training details for the physical flow-tracking
experiment of \cref{sec:afc:channel}. The target schedule, observations, and
episode protocol are specified there; image processing and flow normalization
are detailed in \cref{sec:appendix:water-channel:apparatus}.

\paragraph{Reward and actions}
The reward penalizes the spatial mean of the pointwise vector error between
the averaged estimate $\channelEstimatedFlow_t$ and target
$\channelTargetFlow_t$:
\begin{equation*}
  \reward_t = -\frac{1}{\channelTrackingVelocitySpanPx\cdot\channelObservationCellCount}
    \sum_{i=1}^{\channelObservationHeight}\sum_{j=1}^{\channelObservationWidth}
    \left\lVert\channelEstimatedFlow_{t,ij}
      -\channelTargetFlow_{t,ij}\right\rVert_2.
\end{equation*}
Here, $i$ and $j$ index the \channelObservationCellCount{} grid points, and
both fields retain source-image pixels per pair. Dividing by the horizontal
target span of \channelTrackingVelocitySpanPx{} pixels per pair makes the
reward dimensionless; zero indicates perfect tracking according to the
deployed measurement pipeline. Taking the norm before spatial averaging
prevents opposite local errors from cancelling. Each of the three continuous
actions lies in $[-1,1]$ and maps linearly to its cylinder's rotation-rate
limits.

\paragraph{Model and optimization}
We use \dreamerVThree* \citep{hafner2025mastering} with the
\texttt{\pinballControlModelPreset} preset.
Relative to this preset's defaults, we reduce the convolutional encoder and
decoder to two stages (\pinballControlConvChannels{} channels) for the
$16\times16$ fields, reduce replay capacity to
\pinballControlReplayCapacity{} sequence starts, and set
\texttt{train\_ratio=\pinballControlTrainRatio}.
The remaining model and optimizer parameters retain their default values.

\paragraph{Training and evaluation}
The selector and library remain fixed while the controller learns from
scratch. The training schedule in \cref{sec:afc:channel}
provides \pinballControlTrainMinutes{} minutes of controlled interaction
(\pinballControlTrainSteps{} nominal control steps), excluding settling.
Evaluation uses the final checkpoint.
The training and evaluation sequences are shown in
\cref{fig:channel:training-sequences,fig:channel:evaluation-sequences}.

\end{document}